\documentclass[cernpreprint,texlive=2011,txfonts,UKenglish]{latex/atlasdoc}
\pdfoutput=1

\usepackage[subfigure=true]{latex/atlaspackage}
\usepackage{latex/atlasphysics}

\graphicspath{{logos/}{figures/}}

\usepackage{latex/atlasbiblatex}
\AtlasTitle{Luminosity determination in $\boldsymbol{pp}$\ collisions at \mbox{$\sqrt{s}$ = 8 TeV} using the ATLAS detector at the LHC}

\PreprintIdNumber{CERN-PH-EP-2016-117}

\AtlasRefCode{DAPR-2013-01}

\AtlasJournalRef{Eur. Phys. J. C (2016) 76: 653}
\AtlasDOI{10.1140/epjc/s10052-016-4466-1}

\AtlasAbstract{
The luminosity determination for the ATLAS detector at the LHC during $pp$ collisions  
at $\sqrt s =$~8~\textrm{Te\kern -0.1em V} in 2012 is presented.
The evaluation of the luminosity scale is performed using several luminometers, and comparisons between these 
luminosity detectors are made to assess the accuracy, consistency and long-term stability of the results.
A luminosity uncertainty of $\delta {\cal L}/ {\cal L} = \pm 1.9\%$ is obtained for the $22.7\,\mathrm{fb}^{-1}$ 
of $pp$ collision data delivered to ATLAS at $\sqrt s =$~8~\textrm{Te\kern -0.1em V} in 2012.
}

\usepackage{rotating}
\usepackage{multirow}
\usepackage{units}

\input{lumipaper-defs.sty}

\begin{document}

\title{Luminosity determination in $pp$ collisions at \mbox{$\sqrt{s}$ = 8 TeV} using the ATLAS detector at the LHC}
\author{The ATLAS Collaboration}

\maketitle



\section{Introduction}

An accurate measurement of the delivered luminosity is a key component of the 
ATLAS~\cite{bib:ATLASDetectorPaper} physics programme.
For cross-section measurements, the uncertainty in the delivered luminosity is often one 
of the major systematic uncertainties.
Searches for, and eventual discoveries of,  physical phenomena beyond the 
Standard Model also rely on accurate information about the delivered luminosity to 
evaluate background levels and determine sensitivity to the signatures of new 
phenomena.

\par
This paper describes the measurement of the luminosity delivered to the ATLAS detector 
at the LHC in $pp$ collisions at a centre-of-mass energy of $\sqrt{s}=8$~\TeV\ during 2012. It is structured as follows.
The strategy for measuring and calibrating the luminosity is outlined  in Sect.~\ref{sec:overview}, followed in Sect.~\ref{sec:detectors} by a brief description of the detectors and algorithms used for luminosity determination.
The absolute calibration of these algorithms by the van der Meer (\vdM) method~\cite{bib:vdm}, which must be carried out under specially tailored beam conditions, is described in Sect.~\ref{sec:calibration}; the associated systematic uncertainties are detailed in Sect.~\ref{sec:vdMerrors}.  
The comparison of the relative response of several independent luminometers during physics running  reveals that significant time- and rate-dependent effects impacted the performance of the  ATLAS bunch-by-bunch luminometers during the 2012 run (Sect.~\ref{sec:stability}). Therefore this absolute \vdM\ calibration cannot be invoked as is. 
Instead, it must be transferred, at one point in time and using an independent relative-luminosity monitor, from the low-luminosity regime of \vdM scans to the high-luminosity conditions typical of routine physics running. Additional corrections must be applied over the course of the 2012 data-taking period to compensate for 
detector aging (Sect.~\ref{sec:LDtmPhys}). The various contributions to the systematic uncertainty affecting the integrated luminosity delivered to ATLAS in 2012 are recapitulated in Sect.~\ref{sec:totalSyst}, and the final results are summarized in Sect.~\ref{sec:conclusions}.

\section{Luminosity-determination methodology}
\label{sec:overview}

The analysis presented in this paper closely parallels, and where necessary expands, the one used to determine the luminosity in $pp$ collisions at  $\sqrt{s}=7$~\TeV~\cite{Aad:2013ucp}.

\par
The bunch luminosity ${\mathcal L}_{\mathrm b}$ produced by a single pair of colliding bunches can be expressed as
\begin{equation}
\mathcal{L} _{\mathrm b}= \frac{{\mu} f_{\mathrm{r} }}{{\sigma _{\mathrm{inel}} }}\,,
\end{equation}
where the pile-up parameter $\mu$ is the average number of inelastic interactions per bunch crossing, $f_{\mathrm{r}}$ is the bunch revolution frequency, and $\sigma_{\mathrm{inel}}$ is the $pp$ inelastic cross-section. The total instantaneous luminosity is given by
\begin{equation*}
{\mathcal L}= \sum_{\mathrm b\,=\,1}^{n_{\mathrm{b}}} {\mathcal L}_{\mathrm b}
                  = n_{\mathrm{b}}\, \langle {\mathcal L}_{\mathrm b} \rangle \,
                 =  n_{\mathrm{b}}     \frac{{\langle \mu \rangle} f_{\mathrm{r} }}{{\sigma _{\mathrm{inel}} }}         \,.
\end{equation*}
Here the sum runs over the $n_{\mathrm{b}}$ bunch pairs colliding at the interaction point (IP), $\langle {\mathcal L}_{\mathrm b} \rangle $ is the mean bunch luminosity and $\langle \mu \rangle $ is the bunch-averaged pile-up parameter.
Table~\ref{tab:LHC} highlights the operational conditions of the LHC during Run 1 from 2010 to 2012.  
Compared to previous years, operating conditions did not vary significantly during 2012, with typically 1368 bunches colliding and a peak instantaneous luminosity delivered by the LHC at the start of a fill of ${\cal L}_{\mathrm{peak}} \approx  6$--$8 \times 10^{33}\, \mathrm{cm}^{-2}\, \mathrm{s}^{-1}$, on the average three times higher than in 2011.

\begin{table}[h]
\centering
\vspace{2mm}
\begin{tabular*}{\columnwidth}{@{\extracolsep{\fill}}lccc@{}}
      	\hline
      	Parameter & 2010 & 2011 & 2012 \\
      	\hline
	Number of bunch pairs colliding ($n_{\mathrm{b}} $) & 348 & 1331 & 1380 \\
	Bunch spacing [ns] & 150 & 50 & 50 \\
	Typical bunch population [$10^{11}$ protons] & $0.9$ & $1.2$ & 1.7 \\
	Peak luminosity ${\cal L}_{\mathrm{peak}}$~[$10^{33}\, \mathrm{cm}^{-2}\, \mathrm{s}^{-1}$] & 0.2 & 3.6 & 7.7 \\
	Peak number of inelastic interactions per crossing & $\sim 5$ & $\sim 20$ & $\sim 40$ \\
          Average number of  interactions per crossing (luminosity weighted) & $\sim 2$ & $\sim 9$ & $\sim 21$ \\

	Total integrated luminosity delivered & $47\, \mathrm{pb}^{-1}$ & $5.5\, \mathrm{fb}^{-1}$ & $23\, \mathrm{fb}^{-1}$\\ 
	\hline
   \end{tabular*}
   \caption{Selected LHC parameters for $pp$ collisions at $\sqrt{s} = 7$~\TeV\ in 2010 and 2011, and at $\sqrt{s} = 8$~\TeV\ in 2012. 
Values shown are representative of the best accelerator performance during normal physics operation.
}
\label{tab:LHC}
\end{table}

\par
ATLAS monitors the delivered luminosity by measuring $\mu_{\mathrm{vis}}$, the visible interaction rate per bunch crossing, with a variety of independent detectors and using several different algorithms (Sect. 3). The bunch luminosity can then be written as
\begin{equation}
\mathcal{L}_{\mathrm b} = \frac{\muvis \, f_{\mathrm r} } {\sigmavis}\,,
\label{eqn:defmu}
\end{equation}
where $\muvis = \varepsilon \, \mu$, $\varepsilon$ is the efficiency of the detector and algorithm under consideration, and the visible cross-section for that same detector and algorithm is defined by $\sigmavis \equiv \varepsilon \, \sigma _{\mathrm{inel}}$.
Since $\mu_{\mathrm{vis}}$ is a directly measurable quantity, 
the calibration of the luminosity scale for a particular detector and algorithm
amounts to determining the visible cross-section \sigmavis.
This calibration, described in detail in Sect.~\ref{sec:calibration}, is performed using 
dedicated beam-separation scans, where the absolute luminosity can be inferred from direct 
measurements of the beam parameters~\cite{bib:vdm, bib:Rubbia}. This known luminosity is then combined with the simultaneously measured interaction rate $\muvis$ to extract $\sigmavis$.

\par
A fundamental ingredient of the ATLAS strategy to assess and control the 
systematic uncertainties affecting the absolute luminosity determination is to 
compare the measurements of several luminometers, most of which use 
more than one algorithm to determine the luminosity. 
These multiple detectors and algorithms are characterized by 
significantly different acceptance, response to pile-up, 
and sensitivity to instrumental effects and to beam-induced backgrounds. 
Since the calibration of the absolute luminosity scale is carried out only two or three times per year, this calibration must either remain constant over extended  periods of time and under different machine conditions, or be corrected for long-term drifts.
The level of consistency across the various methods, over the full range of 
luminosities and beam conditions, and across many months of LHC operation,
provides a direct test of the accuracy and stability of the results.
A full discussion of the systematic uncertainties is presented in Sects.~\ref{sec:vdMerrors}--\ref{sec:totalSyst}.

\par
The information needed for physics analyses is the integrated luminosity for some well-defined data samples.
The basic time unit for storing ATLAS luminosity information for physics use is the luminosity block (LB). 
The boundaries of each LB are defined by the ATLAS central trigger processor (CTP), and in general the duration of each LB is approximately one minute.
Configuration changes, such as a trigger prescale adjustment, prompt a luminosity-block transition, and data are analysed assuming that each luminosity block contains data taken under uniform conditions, including luminosity.
For each LB, the instantaneous luminosity from each detector and algorithm, averaged over the luminosity block, is stored in a relational database along with a variety of general ATLAS data-quality information.
To define a data sample for physics, quality criteria are applied to select LBs where conditions are acceptable; then the instantaneous luminosity in that LB is multiplied by the LB duration to provide the integrated luminosity delivered in that LB.
Additional corrections can be made for trigger deadtime and trigger prescale factors, which are also recorded on a per-LB basis.
Adding up the integrated luminosity delivered in a specific set of luminosity blocks provides the integrated luminosity of the entire data sample.

\section{Luminosity detectors and algorithms}
\label{sec:detectors}

The ATLAS detector is discussed in detail in Ref.~\cite{bib:ATLASDetectorPaper}.  
The two primary luminometers, the BCM (Beam Conditions Monitor) and LUCID (LUminosity measurement using a Cherenkov Integrating Detector), both make deadtime-free, bunch-by-bunch luminosity measurements (Sect.~\ref{subsec:bunchCapDets}).
These are compared with the results of the track-counting method (Sect.~\ref{subsec:IDalgos}), a new approach developed by ATLAS which monitors the multiplicity of charged particles produced in randomly selected colliding-bunch crossings, and is essential to assess the calibration-transfer correction from the \vdM\ to the high-luminosity regime.
Additional methods have been developed to disentangle the relative long-term drifts and run-to-run variations between the BCM, LUCID and track-counting measurements during high-luminosity running, thereby reducing the associated systematic uncertainties to the sub-percent level. These techniques measure the total instantaneous luminosity, summed over all bunches, by monitoring detector currents sensitive to average particle fluxes through the ATLAS calorimeters, or by reporting fluences observed in radiation-monitoring equipment; they are described in Sect.~\ref{subsec:bunchIntegDets}.

\subsection{Dedicated bunch-by-bunch luminometers}
\label{subsec:bunchCapDets}

The BCM consists of four $8 \times 8$\,mm$^2$ diamond sensors arranged around the beampipe in a cross pattern at $z = \pm 1.84$\,m on each side of the ATLAS IP.\footnote{ATLAS uses a right-handed coordinate system with its origin at the nominal interaction point in the centre of the detector, and the $z$-axis along the beam line.  The $x$-axis points from the IP to the centre of the LHC ring, and the $y$-axis points upwards. Cylindrical coordinates $(r, \phi)$ are used in the transverse plane, $\phi$ being the azimuthal angle around the beam line.  The pseudorapidity is defined in terms of the polar angle $\theta$ as $\eta = -\ln \tan(\theta/2)$.}
If one of the sensors produces a signal over a preset threshold, a {\em hit} is recorded for that bunch crossing, thereby providing a low-acceptance bunch-by-bunch luminosity signal at $|\eta| = 4.2$ with sub-nanosecond time resolution. The horizontal and vertical pairs of BCM sensors are read out separately, leading to two luminosity measurements labelled BCMH and BCMV respectively. Because the thresholds, efficiencies and noise levels may exhibit small differences between BCMH and BCMV, these two measurements are treated for calibration and monitoring purposes as being produced by independent devices, although the overall response of the two devices is expected to be very similar. 

\par
LUCID is a Cherenkov detector specifically designed to measure the luminosity in ATLAS.  Sixteen aluminium tubes originally filled  with ${\mathrm C}_4{\mathrm F}_{10}$ gas surround the beampipe on each side of the IP at a distance of 17~m, covering the pseudorapidity range $5.6 < |\eta| < 6.0$.
For most of 2012, the LUCID tubes were operated under vacuum to reduce the sensitivity of the device, thereby mitigating pile-up effects and providing a wider operational dynamic range. In this configuration, Cherenkov photons are produced only in the quartz windows that separate the gas volumes from the photomultiplier tubes (PMTs) situated at the back of the detector. If one of the LUCID PMTs produces a signal over a preset threshold, that tube records a hit for that bunch crossing.

\par
Each colliding-bunch pair is identified numerically by a bunch-crossing identifier  (BCID) which labels each of the 3564 possible 25~ns slots in one full revolution of the nominal LHC fill pattern. Both BCM and LUCID are fast detectors with electronics capable of reading out the diamond-sensor and PMT hit patterns separately for each bunch crossing, thereby making full use of the available statistics. These FPGA-based front-end electronics run autonomously from the main data acquisition system, and are not affected by any deadtime imposed by 
the CTP.\footnote{The CTP inhibits triggers (causing deadtime) for a variety of reasons, 
but especially for several bunch crossings after a triggered event to allow time for the 
detector readout to conclude.  
Any new triggers which occur during this time are ignored.}
They execute  in real time several different online algorithms, characterized by diverse efficiencies, background sensitivities, and linearity characteristics~\cite{bib:ATLASLumiPaper}.

\par
The BCM and LUCID detectors consist of two symmetric arms placed in the forward (``A'') and backward (``C'') direction from the IP, which can also be treated as independent devices. 
The baseline luminosity algorithm is an inclusive hit requirement, known as the EventOR algorithm, which requires that at least one hit be recorded anywhere in the detector considered. Assuming that the number of interactions in a bunch crossing obeys a Poisson distribution, the probability of observing an event which satisfies the EventOR criteria can be computed as
\begin{equation}
P_\textsf{\tiny EventOR}\,(\mu_{\mathrm{vis}}^{\mathrm{OR}})  =  N_{\mathrm{OR}} / N_{\mathrm{BC}} =  1-\mathrm{e}^{-\mu_{\mathrm{vis}}^{\mathrm{OR}}}. 
\label{eqn:muor}
\end{equation}
Here the raw event count $N_{\mathrm{OR}}$ is the number of bunch crossings, during a given time interval, in which at least one $pp$ interaction satisfies the event-selection 
criteria of the OR algorithm under consideration, and $N_{\mathrm{BC}}$ is the total 
number of bunch crossings during the same interval.
Solving for $\mu_{\mathrm{vis}}$ in terms of the event-counting rate yields
\begin{equation}
\begin{array}{ll}
\mu_{\mathrm{vis}}^{\mathrm{OR}} = - \ln \left( 1- \frac{N_{\mathrm{OR}}}{N_{\mathrm{BC}}} \right ).\\
\end{array}
\label{eqn:muand}
\end{equation}

\par
When $\mu_{\mathrm{vis}} \gg 1$, event counting algorithms lose sensitivity as fewer and fewer bunch crossings in a given time interval report zero observed interactions. In the limit where $N_{\mathrm{OR}}/N_{\mathrm{BC}} = 1$, event counting algorithms can no longer be used to determine the interaction rate $\mu_{\mathrm{vis}}$: this is referred to as {\em saturation}.
The sensitivity of the LUCID detector is high enough (even without gas in the tubes) that the LUCID\_EventOR algorithm saturates in a one-minute interval at around 20 interactions per crossing, while the single-arm inclusive LUCID\_EventA and LUCID\_EventC algorithms can be used up to around 30 interactions per crossing. The lower acceptance of the BCM detector allowed event counting to remain viable for all of 2012.

\subsection{Tracker-based luminosity algorithms}
\label{subsec:IDalgos}

The ATLAS inner detector (ID) measures the trajectories of charged particles over the pseudorapidity range $|\eta|<2.5$ and the full azimuth. It consists~\cite{bib:ATLASDetectorPaper} of a silicon pixel detector (Pixel), a silicon micro-strip detector (SCT) and a straw-tube transition-radiation detector (TRT). Charged particles are reconstructed as tracks using an inside-out algorithm, which starts with three-point seeds from the silicon detectors and then adds hits using a combinatoric Kalman filter~\cite{bib:ATLAS_NEWT}. 

\par
The luminosity is assumed to be proportional to the number of reconstructed charged-particle tracks, with the visible interaction rate \muvis taken as the number of tracks per bunch crossing averaged over a given time window (typically a luminosity block). In standard physics operation, silicon-detector data are recorded in a dedicated partial-event stream using a random trigger at a typical rate of 100\,Hz, sampling each colliding-bunch pair with equal probability. Although a bunch-by-bunch luminosity measurement is possible in principle, over 1300 bunches were colliding in ATLAS for most of 2012, so that in practice only the bunch-integrated luminosity can be determined with percent-level statistical precision in a given luminosity block. During \vdM\ scans, Pixel and SCT data are similarly routed to a dedicated data stream for a subset of the colliding-bunch pairs at a typical rate of 5 kHz per BCID, thereby allowing the bunch-by-bunch determination of \sigmavis.

\par
For the luminosity measurements presented in this paper, charged-particle track reconstruction uses hits from the silicon detectors only. 
Reconstructed tracks are required to have at least nine silicon hits, zero holes\footnote{In this context, a hole is counted when a hit is expected in an active sensor located on the track trajectory between the first and the last hit associated with this track, but no such hit is found. If the corresponding sensor is known to be inactive and therefore not expected to provide a hit, no hole is counted.} 
in the Pixel detector and transverse momentum in excess of 0.9\,GeV.  Furthermore, the absolute transverse impact parameter with respect to the luminous centroid~\cite{bib:LumR} is required to be no larger than seven times its uncertainty, as determined from the covariance matrix of the fit. 

\par
This default track selection makes no attempt to distinguish tracks originating from primary vertices from those produced in secondary interactions, as the yields of both are expected to be proportional to the luminosity. Previous studies of track reconstruction in ATLAS show that in low pile-up conditions ($\mu \leq 1$) and with a track selection looser than the above-described default, single-beam backgrounds remain well below the per-mille level~\cite{bib:Aad:2010ac}.  However, for pile-up parameters typical of 2012 physics running, tracks formed from random hit combinations, known as {\em fake tracks}, can become significant~\cite{bib:CONF2012-042}. The track selection above is expected to be robust against such non-linearities, as demonstrated by analysing simulated events of overlaid inelastic \pp\ interactions produced using the PYTHIA 8 Monte Carlo event generator \cite{bib:pythia8}. 
In the simulation, the fraction of fake tracks per event can be parameterized as a function of the true pile-up parameter, yielding a fake-track fraction of less than 0.2\% at $\mu = 20$ for the default track selection. In data, this fake-track contamination is subtracted from the measured track multiplicity using the simulation-based parameterization with, as input, the \muav value reported by the BCMH\_EventOR luminosity algorithm.
An uncertainty equal to half the correction is assigned to the measured track multiplicity to account for possible systematic differences between data and simulation.

\par
Biases in the track-counting luminosity measurement can arise from $\mu$-dependent effects in the track reconstruction or selection requirements, which would change the reported track-counting yield per collision between the low pile-up \vdM-calibration regime and the high-$\mu$ regime typical of physics data-taking. Short- and long-term variations in the track reconstruction and selection efficiency can also arise from changing ID conditions, for example because of temporarily disabled silicon readout modules. In general, looser track selections are less sensitive to such fluctuations in instrumental coverage; however, they typically suffer from larger fake-track contamination. 

\par
To assess the impact of such potential biases, several looser track selections, or {\em working points} (WP), are investigated. Most are found to be consistent with the default working point once the uncertainty affecting the simulation-based fake-track subtraction is accounted for. In the case where the Pixel-hole requirement is relaxed from zero to no more than one, a moderate difference in excess of the fake-subtraction uncertainty is observed in the data. This working point, labelled ``Pixel holes $\le 1$'', is used as an alternative algorithm when evaluating the systematic uncertainties associated with track-counting luminosity measurements.

\par
In order to all but eliminate fake-track backgrounds and minimize the associated $\mu$-dependence, another alternative is to remove the impact-parameter requirement and use the resulting superset of tracks as input to the primary-vertex reconstruction algorithm. Those tracks which, after the vertex-reconstruction fit, have a non-negligible probability of being associated to any primary vertex are counted to provide an alternative luminosity measurement. 
In the simulation, the performance of this ``vertex-associated'' working point is comparable, in terms of fake-track fraction and other residual non-linearities, to that of the default and ``Pixel holes $\le 1$'' track selections discussed above. 

\subsection{Bunch-integrating detectors}
\label{subsec:bunchIntegDets}

Additional algorithms, sensitive to the instantaneous luminosity summed over all bunches, provide relative-luminosity monitoring on time scales of a few seconds rather than of a bunch crossing, allowing independent checks of the linearity and long-term stability of the BCM, LUCID and track-counting algorithms. The first technique measures the particle flux from \pp\ collisions as reflected in the current drawn by the PMTs of the 
hadronic calorimeter (TileCal). This flux, which is proportional to the instantaneous luminosity, is also monitored by the total ionization current flowing through a well-chosen set of liquid-argon (LAr) calorimeter cells.
A third technique, using Medipix radiation monitors, measures the average particle flux observed in these devices.

\subsubsection{Photomultiplier currents in the central hadronic calorimeter}

The 
TileCal~\cite{bib:TileTDR} is constructed from plastic-tile scintillators as the active medium and from steel absorber plates. It covers the pseudorapidity range $|\eta| < 1.7$ and consists of a long central cylindrical barrel and two smaller extended barrels, one on each side of the long barrel. Each of these three cylinders is divided azimuthally into 64 modules and segmented into three radial sampling layers. Cells are defined in each layer according to a projective geometry, and each cell is connected by optical fibres to two photomultiplier tubes. 
The current drawn by each PMT is proportional to the total number of particles interacting in a given TileCal cell, and provides a signal proportional to the luminosity summed over all the colliding bunches.
This current is monitored by an integrator system with a time constant of 10~ms and is sensitive to currents from 0.1~nA to 1.2~$\mu$A. 
The calibration and the monitoring of the linearity of the integrator electronics are ensured by a dedicated high-precision current-injection system.

\par
The collision-induced PMT current depends on the pseudorapidity of the cell considered and on the radial sampling in which it is located. The cells most sensitive to luminosity variations are located near $|\eta| \approx 1.25$; at a given pseudorapidity, the current is largest in the innermost sampling layer, because the hadronic showers are progressively absorbed as they expand in the middle and outer radial layers.  
Long-term variations of the TileCal response are monitored, and corrected if appropriate~\cite{Aad:2013ucp}, by injecting a laser pulse directly into the PMT, as well as by integrating the counting rate from a $^{137}\mathrm{Cs}$ radioactive source that circulates between the calorimeter cells during calibration runs.

\par
The TileCal luminosity measurement is not directly calibrated by the \vdM procedure, both because its slow and asynchronous readout is not optimized to keep in step with the scan protocol, and because the luminosity is too low during the scan for many of its cells to provide accurate measurements. Instead, the TileCal luminosity calibration is performed in two steps. The PMT currents, corrected for electronics pedestals and for non-collision backgrounds\footnote{For each LHC fill, the currents are baseline-corrected using data recorded shortly before the LHC beams are brought into collision.} and averaged over the most sensitive cells, are first cross-calibrated to the absolute luminosity reported by the BCM during the April 2012 \vdM scan session
(Sect.~\ref{sec:calibration}). 
Since these high-sensitivity cells would incur radiation damage at the highest luminosities encountered during 2012, thereby requiring large calibration corrections, their luminosity scale is transferred, during an early intermediate-luminosity run and on a cell-by-cell basis, to the currents measured in the remaining cells (the sensitivities of which are insufficient under the low-luminosity conditions of \vdM scans).
The luminosity reported in any other physics run is then computed as the average, over the usable cells, of the individual cell luminosities, determined by multiplying the baseline-subtracted PMT current from that cell by the corresponding calibration constant.

\subsubsection{LAr-gap currents}

The electromagnetic endcap (EMEC) and forward (FCal) calorimeters are sampling devices that cover the pseudorapidity ranges of, respectively, $1.5 < |\eta| < 3.2$ and $3.2 < |\eta| < 4.9$. They are housed in the two endcap cryostats along with the hadronic endcap calorimeters.

\par
The EMECs consist of accordion-shaped lead/stainless-steel absorbers interspersed with honeycomb-insulated electrodes that distribute the high voltage (HV) to the LAr-filled gaps where the ionization electrons drift, and that collect the associated electrical signal by capacitive coupling. In order to keep the electric field across each LAr gap constant over time, the HV supplies are regulated such that any voltage drop induced by the particle flux through a given HV sector is counterbalanced by a continuous injection of electrical current. The value of this current is proportional to the particle flux and thereby provides a relative-luminosity measurement using the EMEC HV line considered.

\par
Both forward calorimeters are divided longitudinally into three modules. Each of these consists of a metallic absorber matrix (copper in the first module, tungsten elsewhere) containing cylindrical electrodes arranged parallel to the beam axis. The electrodes are formed by a copper (or tungsten) tube, into which a rod of slightly smaller diameter is inserted. This rod, in turn, is positioned concentrically using a helically wound radiation-hard plastic fibre, which also serves to electrically isolate the anode rod from the cathode tube. The remaining small annular gap is filled with LAr as the active medium. Only the first sampling is used for luminosity measurements. It is divided into 16 azimuthal sectors, each fed by 4 independent HV lines. As in the EMEC, the HV system provides a stable electric field across the LAr gaps and the current drawn from each line is directly proportional to the average particle flux through the corresponding FCal cells. 

\par
After correction for electronic pedestals and single-beam backgrounds, the observed currents are assumed to be proportional to the luminosity summed over all bunches; the validity of this assumption is assessed in Sect.~\ref{sec:stability}. The EMEC and FCal gap currents cannot be calibrated during a {\em vdM} scan, because the instantaneous luminosity during these scans remains below the sensitivity of the current-measurement circuitry. Instead, the calibration constant associated with an individual HV line is evaluated as the ratio of the absolute luminosity reported by the baseline bunch-by-bunch luminosity algorithm (BCMH\_EventOR) and integrated over one high-luminosity reference physics run, to the HV current drawn through that line, pedestal-subtracted and integrated over exactly the same time interval. This is done for each usable HV line independently. The luminosity reported in any other physics run by either the EMEC or the FCal, separately for the A and C detector arms, is then computed as the average, over the usable cells, of the individual HV-line luminosities.

\subsubsection{Hit counting in the Medipix system}
\label{subsubsec:MPXhits}

The Medipix (MPX) detectors are hybrid silicon pixel devices, which are distributed around the ATLAS detector~\cite{bib:MPX} and are primarily used to monitor radiation conditions in the experimental hall. Each of these 12 devices consists of a 2~cm$^2$ silicon sensor matrix, segmented in $256\times256$ cells and bump-bonded to a readout chip. Each pixel in the matrix counts hits from individual particle interactions observed during a software-triggered ``frame'', which integrates over 5 to 120 seconds, depending upon the typical particle flux at the location of the detector considered.  
In order to provide calibrated luminosity measurements, the total number of pixel clusters observed in each sensor is counted and scaled to the TileCal luminosity in the same reference run as the EMEC and  FCal. The six MPX detectors with the highest counting rate are analysed in this fashion for the 2012 running period; their mutual consistency is discussed in Sect.~\ref{sec:stability}.

\par
The  hit-counting algorithm described above is primarily sensitive to charged particles. The MPX detectors offer the additional capability to detect thermal neutrons via 
$^6{\mathrm{Li}}(n,\alpha)^3{\mathrm{H}}$ 
reactions in a $\mathrm{{}^{6}LiF}$ converter layer. This neutron-counting rate provides a further measure of the luminosity, which is consistent with, but statistically inferior to, the MPX hit counting measurement~\cite{bib:MPX} .

\section{Absolute luminosity calibration by the van der Meer method}
\label{sec:calibration}

In order to use the measured interaction rate $\mu_\mathrm{vis}$ as a luminosity monitor, 
each detector and algorithm must be calibrated by determining its visible cross-section 
$\sigma_\mathrm{vis}$.
The primary calibration technique to determine the absolute luminosity scale of each bunch-by-bunch
luminosity detector and algorithm employs dedicated {\em vdM} scans to infer the delivered 
luminosity at one point in time from the measurable parameters of the colliding bunches.
By comparing the known luminosity delivered in the {\em vdM} scan to the visible interaction rate
$\mu_\mathrm{vis}$, the visible cross-section can be determined from Eq.~(\ref{eqn:defmu}).

\par
This section is organized as follows. The formalism of the van der Meer method is recalled in Sect.~\ref{subsec:vdMform}, followed in Sect.~\ref{subsec:lumiScanData} by a description of the \vdM-calibration datasets collected during the 2012 running period. The step-by-step determination of the visible cross-section is outlined in Sect.~\ref{subsec:scanAna}, and each ingredient is discussed in detail in Sects.~\ref{subsec:scanBgds} to \ref{subsec:currents}. The resulting absolute calibrations of the bunch-by-bunch luminometers, as applicable to the low-luminosity conditions of \vdM\ scans, are summarized in Sect.~\ref{subsec:vdMResults}.
\subsection{Absolute luminosity from measured beam parameters}
\label{subsec:vdMform}

In terms of colliding-beam parameters, the bunch luminosity ${\mathcal L}_{\mathrm b}$ is given by
\begin{equation}
{\mathcal L}_{\mathrm b} =  f_{\mathrm r}\, n_1 n_2\, \int {\hat{\rho} _1 (x,y)}\, \hat{\rho} _2(x,y)\,\mathrm{d}x\,\mathrm{d}y~,
\label{eqn:lumi}
\end{equation}
where the beams are assumed to collide with zero crossing angle, $n_1 n_2$ is the bunch-population product and $\hat{\rho}_{1(2)}(x,y)$ is the normalized particle density in the transverse ($x$--$y$) plane of beam 1 (2) at the IP. With the standard assumption that the particle densities can be factorized
into independent horizontal and vertical component distributions,
$\hat{\rho}(x,y)=\rho_x(x)\,\rho_y(y)$, Eq.~(\ref{eqn:lumi}) can be rewritten as
\begin{equation}
{\mathcal L}_{\mathrm b} =  f_{\mathrm r}\, n_1 n_2 \,\Omega _x (\rho _{x1},\rho _{x2}) \,\Omega _y (\rho _{y1},\rho _{y2})~,
\label{eqn:lumi1}
\end{equation}
where \[ \Omega _x (\rho _{x1},\rho _{x2} ) = \int {\rho _{x1} (x)\,\rho _{x2} (x)\,\mathrm{d}x}\] 
is the beam-overlap integral in the $x$ direction (with an analogous definition in the $y$ direction). 
In the method proposed by van der Meer~\cite{bib:vdm}, the overlap integral (for example in the $x$ direction) can be calculated as
\begin{equation}
\Omega _x (\rho _{x1},\rho _{x2}) = \frac{{R_x (0)}}{{\int {R_x (\delta)\,\mathrm{d}\delta} }}~,
\label{eqn:vdm}
\end{equation}
where $R_x(\delta)$ is the luminosity (at this stage in arbitrary units) measured during a horizontal scan at the time the two beams are 
separated horizontally by the distance $\delta$, and $\delta=0$ represents the 
case of zero beam separation.  
Because the luminosity $R_x(\delta)$ is normalized to that at zero separation $R_x(0)$, any quantity
proportional to the luminosity (such as $\mu_\mathrm{vis}$) can be substituted in Eq.~(\ref{eqn:vdm}) in place of $R$.

Defining the horizontal convolved beam size $\Sigma_x$ \cite{bib:LumR, bib:bbDeflCalc} as
\begin{equation}
\Sigma _x  = \frac{1}{{\sqrt {2\pi } }}\frac{{\int {R_x (\delta)\,\mathrm{d}\delta} }}{{R_x (0)}}~,
\label{eqn:caps}
\end{equation}
and similarly for $\Sigma_y$, the bunch luminosity in Eq.~(\ref{eqn:lumi1}) can be rewritten as
\begin{equation}
{\mathcal L}_{\mathrm b} = \frac{ f_{\mathrm r} n_1 n_2 }{{2\pi \Sigma _x \Sigma _y}}~,
\label{eqn:lumifin}
\end{equation}
which allows the absolute bunch luminosity to be determined from the revolution frequency
$f_{\mathrm r}$, the bunch-population product $n_1 n_2$,  and the product
$\Sigma_x \Sigma_y$ which is measured directly during a pair of orthogonal \vdM (beam-separation) scans. 
In the case where the luminosity curve $R_x(\delta)$ is Gaussian, 
$\Sigma _x $ coincides with the standard deviation of that distribution. It is important to note that the \vdM method does not rely on any particular functional form of $R_x(\delta)$:
the quantities $\Sigma_x$ and $\Sigma_y$ can be determined for any observed luminosity curve
from Eq.~(\ref{eqn:caps}) and used with Eq.~(\ref{eqn:lumifin}) to determine the absolute luminosity
at $\delta=0$.

\par
In the more general case where the factorization assumption breaks down, \ie when the particle densities (or more precisely the dependence of the luminosity on the beam separation ($\delta_x,\delta_y$)) cannot be factorized into a product of uncorrelated $x$ and $y$ components, the formalism can be extended to yield~\cite{bib:Rubbia}
\begin{equation}
\Sigma _x  \Sigma_y= \frac{1}{{2\pi}}\frac{{\int {R_{x,y} (\delta_x,\delta_y)\,\mathrm{d}\delta_x\,\mathrm{d}\delta_y}
}}{{R_{x,y} (0,0)}}~,
\label{eq:CapSNonFact}
\end{equation}
with Eq.~(\ref{eqn:lumifin}) remaining formally unaffected. Luminosity calibration in the presence of non-factorizable bunch-density distributions is discussed extensively in Sect.~\ref{subsec:nonFctrztnCrctn}.

\par
The measured product of the transverse convolved beam sizes $\Sigma_x  \Sigma_y$ is directly related to the reference specific luminosity:\footnote{The specific luminosity is defined as the luminosity per bunch and per unit bunch-population product~\cite{bib:LumR}.}
\begin{equation*}
{\cal L}_{\mathrm{spec}} \equiv \frac{{\cal L}_{\mathrm{b}}}{ n_1 n_2} = \frac{f_{\mathrm{r}}}{2\pi \Sigma_x \Sigma_y}
\end{equation*}
which, together with the bunch currents, determines the absolute luminosity scale.
To calibrate a given luminosity algorithm, one can equate the absolute luminosity computed from beam parameters
using Eq.~(\ref{eqn:lumifin}) to that measured according to Eq.~(\ref{eqn:defmu}) to get
\begin{equation}
\sigmavis =\muvismax \frac{2\pi\, \Sigma_x \Sigma_y}{n_1 n_2}~,
\label{eqn:sigmaVis}
\end{equation}
where \muvismax\ is the visible interaction rate per bunch crossing reported at the peak of the scan curve by that particular algorithm.
Equation~(\ref{eqn:sigmaVis}) provides a direct calibration of the visible cross-section 
\sigmavis\ for each algorithm in terms of the peak visible interaction rate \muvismax, the product of the convolved beam widths $\Sigma_x \Sigma_y$, and the bunch-population product $n_1 n_2$.

In the presence of a significant crossing angle in one of the scan planes, the formalism becomes considerably more involved~\cite{bib:Lconcept}, but the conclusions remain unaltered and Eqs.~(\ref{eqn:caps})--(\ref{eqn:sigmaVis}) remain valid. The non-zero vertical crossing angle in some scan sessions widens the luminosity curve by a factor that depends on the bunch length, the transverse beam size and the crossing angle, but reduces the peak luminosity by the same factor. 
The corresponding increase in the measured value of $\Sigma_y$ is exactly compensated by the decrease in \muvismax, so that no correction for the crossing angle is needed in the determination of \sigmavis.
\subsection{Luminosity-scan datasets}
\label{subsec:lumiScanData}

The beam conditions during {\em vdM}  scans are  different from those in normal physics operation, with lower bunch intensities and only a few tens of widely spaced bunches circulating. These conditions are optimized to reduce various systematic uncertainties in the calibration procedure~\cite{bib:LumR}.
Three scan sessions were performed during 2012: in April, July, and November (Table~\ref{tab:vdmScan}). The April scans were performed with nominal collision optics $(\beta^\star = 0.6\,\mathrm{m}$), which minimizes the accelerator set-up time but yields conditions which are inadequate for achieving the best possible calibration accuracy.\footnote{The $\beta$ function describes the single-particle motion and determines the variation of the beam envelope along the beam trajectory. It is calculated from the focusing properties of the magnetic lattice (see for example Ref.~\cite{bib:betaDef}). The symbol $\beta^\star$ denotes the value of the $\beta$ function at the IP.}
 The July and November scans were performed using dedicated  {\em vdM}-scan optics with $\beta^\star = 11\,\mathrm{m}$, in order to increase the transverse beam sizes while retaining a sufficiently high collision rate even in the tails of the scans. This strategy limits the impact of the vertex-position resolution on the non-factorization analysis, which is detailed in Sect.~\ref{subsec:nonFctrztnCrctn}, and also reduces potential $\mu$-dependent calibration biases. 
 In addition, the observation of large non-factorization effects in the April and July scan data motivated, for the November scan, a dedicated set-up of the LHC injector chain~\cite{bib:injBunchPrep} to produce more Gaussian and less correlated transverse beam profiles.

\par
Since the luminosity can be different for each colliding-bunch pair, both because the beam sizes differ from bunch to bunch and because the bunch populations $n_1$ and $n_2$ can each vary by up to $\pm10$\%, the determination of  $\Sigma_x$ and $\Sigma_y$ and the measurement of \muvismax\ are performed independently for each colliding-bunch pair. As a result, and taking the November session as an example, each scan set provides 29 independent measurements of \sigmavis, allowing detailed consistency checks.  

\par
To further test the reproducibility of the calibration procedure, multiple centred-scan\footnote{A {\em centred} (or {\em on-axis}) beam-separation scan is one where the beams are kept centred on each other in the transverse direction orthogonal to the scan axis. An {\em offset} (or {\em off-axis}) scan is one where the beams are partially separated in the non-scanning direction.} sets, each consisting of one horizontal scan and one vertical scan, are executed in the same scan session.  In November for instance, two sets of centred scans (X and XI) were performed in quick succession, followed by two sets of off-axis scans (XII and XIII), where the beams were separated by 340~$\muup$m and 200~$\muup$m respectively in the non-scanning direction. A third set of centred scans (XIV) was then performed as a reproducibility check. A fourth centred scan set (XV) was carried out approximately one day later in a different LHC fill.

\par
The variation of the calibration results between individual scan sets in a given scan session is used to quantify the reproducibility of the optimal relative beam position, the convolved beam sizes, and the visible cross-sections.
The reproducibility and consistency of the visible cross-section results across the April, July and November scan sessions provide a measure of the long-term stability of the response of each detector, and are used to assess potential systematic biases in the \vdM-calibration technique under different accelerator conditions.

\begin{sidewaystable}
  \centering
\begin{tabular}[h]{lccc}
\hline
Scan labels 			&				I--III 						& IV--IX 								& X--XV 						\\
\hline
Date                        	& 				16 April 2012				& 19 July 2012 						& 22, 24 November 2012		\\
LHC fill number 		& 				2520					& 2855, 2856							& 3311, 3316					\\
\hline
Total number of bunches per beam          	& 48 					& 48 								& 39 						\\
Number of bunches colliding in ATLAS 		& 35 					& 35 								& 29							\\ 					
Typical number of protons per bunch $n_{1,2}$ & $0.6\cdot10^{11}$ 	& $0.9\cdot10^{11}$						& $0.9\cdot10^{11}$			\\
Nominal $\beta$-function at the IP ($\beta^\star$) [m]   & $ 0.6$ 		& $11$ 								& $11$ 						\\
Nominal transverse single-beam size $\sigma_{\mathrm{b}}^{\mathrm{nom}}$ [$\muup$m]
									& 23 					& 98									& 98							\\
Actual transverse emittance $\epsilon_N$ [$\muup$m-radians] 	
									& $ 2.3$ 					& 3.2								& 3.1						\\
Actual transverse single-beam size $\sigma_{\mathrm{b}}$ [$\muup$m] 
									& 18 					& 91									& 89							\\
Actual transverse luminous size $\sigma_{\mathcal L}$ ($\approx \sigma_{\mathrm{b}}/\sqrt{2}$) [$\muup$m] 
									& 13 					& 65									& 63							\\
Nominal vertical half crossing-angle [$\muup$rad]    	&  $\pm 145$ 		& 0 									& 0							\\
 \hline
Typical luminosity/bunch [$\muup{\mathrm b}^{-1}\,\mathrm{s}^{-1}$] & 0.8 	& 0.09							& 0.09 						\\
Pile-up parameter $\mu$ [interactions/crossing] 	
									& 5.2 					& 0.6 								& 0.6 						\\
\hline
Scan sequence 		& 				3 sets				 	& 4 sets  								& 4 sets 						\\
					&				of centred $x + y$ scans	& of centred $x + y$ scans				& of centred $x + y$ scans		\\
					&				(I-III)						& (IV--VI, VIII)							& (X, XI, XIV, XV)				\\
		                     	& 			 							& plus 2 sets  							& plus 2  sets 					\\
		                     	&										& of $x + y$ off-axis scans				& of $x + y$ off-axis scans		\\
					&										& (VII, IX)								& (XII, XIII)					\\
Total scan steps per plane 				& 25 					& 25 (sets IV--VII)						& 25 						\\
									&						& 17 (sets VIII--IX)						&							\\
Maximum beam separation  & 	$\pm 6\sigma_{\mathrm{b}}^{\mathrm{nom}}$ 	
															& $\pm 6\sigma_{\mathrm{b}}^{\mathrm{nom}}$ 			
															& $\pm 6\sigma_{\mathrm{b}}^{\mathrm{nom}}$							\\
Scan duration per step [seconds]			& 20 					& 30 								& 30						 	\\
 \hline
\end{tabular}
\caption{Summary of the main characteristics of the 2012 \vdM scans performed at the ATLAS interaction point. 
The nominal tranverse beam size is computed using the nominal LHC emittance ($\epsilon_N = 3.75~\muup$m-radians).
The actual transverse emittance and single-beam size are estimated by combining the convolved transverse widths measured in the first scan of each session with the nominal IP $\beta$-function. 
The values of the luminosity/bunch and of $\mu$ are given for zero beam separation during the first scan.  
The specific luminosity decreases by 6--17\% over the duration of a given scan session.}
\label{tab:vdmScan}
\end{sidewaystable}
\subsection{\vdM-scan analysis methodology}
\label{subsec:scanAna}

The 2012 \vdM scans were used to derive calibrations for the LUCID\_EventOR, BCM\_EventOR and track-counting algorithms.
Since there are two distinct BCM readouts, calibrations are determined separately for the horizontal (BCMH) and vertical (BCMV) detector pairs. Similarly, the fully inclusive (EventOR) and single-arm inclusive (EventA, EventC) algorithms are calibrated independently. For the April scan session, the dedicated track-counting event stream (Sect.~\ref{subsec:IDalgos}) used the same random trigger as during physics operation. For the July and November sessions, where the typical event rate was lower by an order of magnitude, track counting was performed on events triggered by the ATLAS Minimum Bias Trigger Scintillator (MBTS)~\cite{bib:ATLASDetectorPaper}. Corrections for MBTS trigger inefficiency and for CTP-induced deadtime are applied, at each scan step separately, when calculating the average number of tracks per event.

\par
For each individual algorithm, the {\em vdM} data are analysed in the same manner. The specific visible interaction rate $\muvis/(n_1 n_2)$ is measured, for each colliding-bunch pair, as a function of the nominal beam separation (\ie the separation specified by the LHC control system) in two orthogonal scan directions ($x$ and $y$). The value of \muvis is determined from the raw counting rate using the formalism described in Sect.~\ref{subsec:bunchCapDets} or \ref{subsec:IDalgos}. The specific interaction rate is used so that the calculation of $\Sigma_x$ and $\Sigma_y$ properly takes into account the bunch-current variation during the scan; the measurement of the bunch-population product $n_1 n_2$ is detailed in Sect.~\ref{subsec:currents}.

\par
\begin{figure}
\centering
\subfigure[]{
\includegraphics[width=0.80\textwidth]{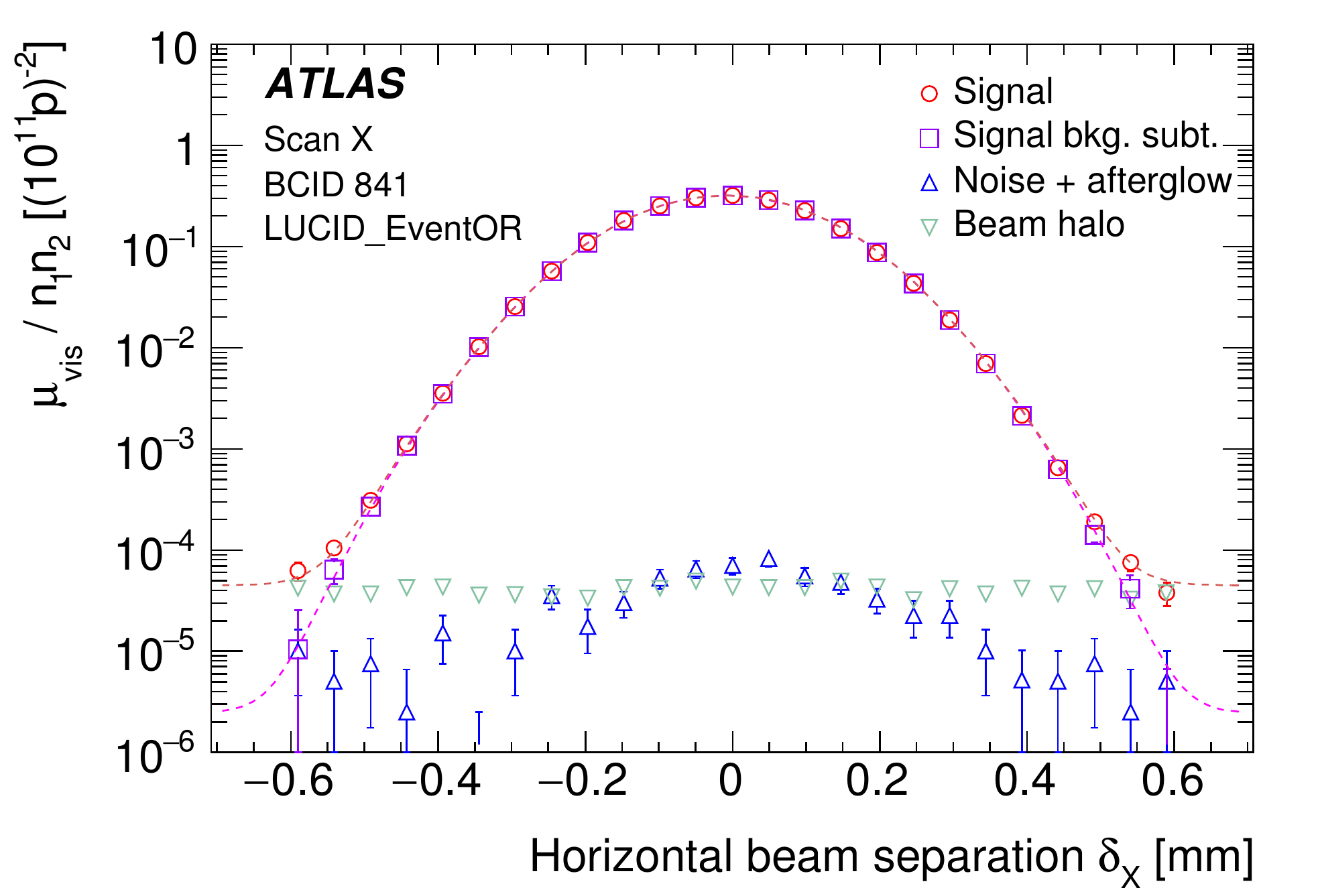}
}
\subfigure[]{
\includegraphics[width=0.80\textwidth]{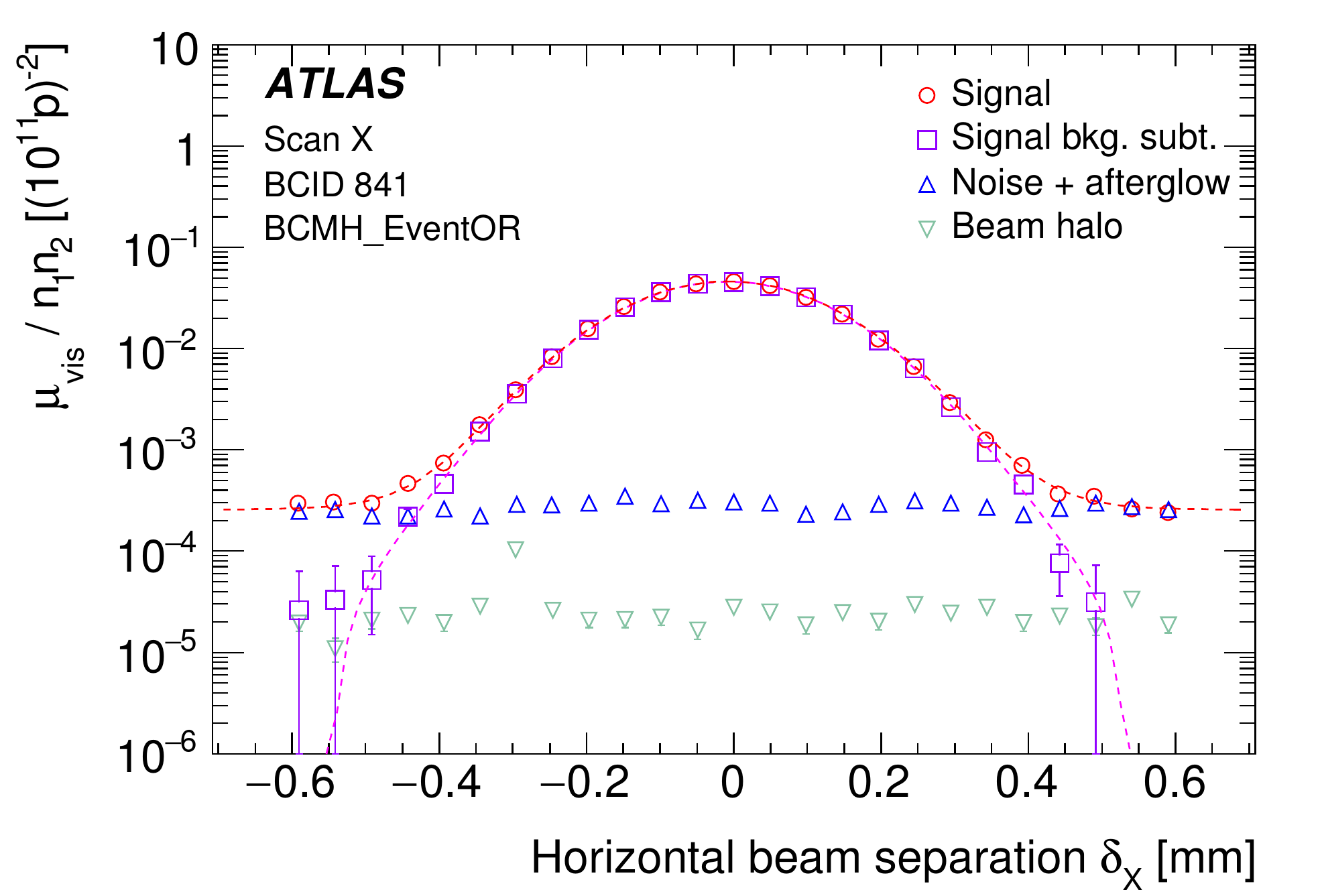}
}
\caption{Beam-separation dependence of the specific visible interaction rate measured using the (a) LUCID\_EventOR and (b) BCMH\_EventOR algorithms during horizontal scan X, before (red circles) and after (purple squares) afterglow, noise and single-beam background subtraction. The subtracted contributions are shown as triangles. The scan curves are fitted to a Gaussian function multiplied by a sixth-order polynomial, plus a constant.
}
\label{fig:scanCurves}
\end{figure}

Figure~\ref{fig:scanCurves} shows examples of horizontal-scan curves measured for a single BCID using two different algorithms. At each scan step, the visible interaction rate \muvis is first corrected for afterglow, instrumental noise and beam-halo backgrounds as described in Sect.~\ref{subsec:scanBgds}, and the nominal beam separation is rescaled using the calibrated beam-separation scale (Sect.~\ref{subsec:LengthScale}). The impact of orbit drifts is addressed in Sect.~\ref{subsec:orbit}, and that of beam--beam deflections and of the dynamic-$\beta$ effect is discussed in Sect.~\ref{subsec:beambeam}. 
For each BCID and each scan independently, a characteristic function is fitted to the corrected data; the peak of the fitted function provides a measurement of \muvismax , while the convolved width $\Sigma$ is computed from the integral of the function using Eq.~(\ref{eqn:caps}).
Depending on the beam conditions, this function can be a single-Gaussian function plus a constant term, a double-Gaussian function plus a constant term, a Gaussian function times a polynomial (plus a constant term), or other variations. As described in Sect.~\ref{sec:vdMerrors}, the differences between the results extracted using different characteristic functions are taken into account as a systematic uncertainty in the calibration result. 

\par
The combination of one horizontal ($x$) scan and one vertical ($y$) scan is the minimum needed to perform a measurement of \sigmavis. In principle, while the \muvismax parameter is detector- and algorithm-specific, the convolved widths $\Sigma_x$ and $\Sigma_y$, which together specify the head-on reference luminosity, do not need to be determined using that same detector and algorithm. 
In practice, it is convenient to  extract all the parameters associated with a given algorithm consistently from a single set of scan curves, and the average value of  \muvismax\ between the two scan planes is used. The correlations between the fitted values of \muvismax, $\Sigma_x$ and $\Sigma_y$  are taken into account when evaluating the statistical uncertainty affecting \sigmavis.

\par
Each BCID should yield the same measured \sigmavis\ value, and so the average over all BCIDs is taken as the \sigmavis\ measurement for the scan set under consideration.  The bunch-to-bunch consistency of the visible cross-section for a given luminosity algorithm, as well as the level of agreement between $\Sigma$ values measured by different detectors and algorithms in a given scan set, are discussed in Sect.~\ref{sec:vdMerrors} as part of the systematic uncertainty.

\par
Once visible cross-sections have been determined from each scan set as described above, two beam-dynamical effects must be considered (and if appropriate corrected for), both associated with the shape of the colliding bunches in transverse phase space: non-factorization and emittance growth. These are discussed in Sects.~\ref{subsec:nonFctrztnCrctn} and \ref{subsec:epsGrowth} respectively.

\subsection{Background subtraction}
\label{subsec:scanBgds}

The \vdM calibration procedure is affected by three distinct background contributions to the luminosity signal: afterglow, instrumental noise, and single-beam backgrounds.

\par
As detailed in Refs.~\cite{Aad:2013ucp, bib:ATLASLumiPaper}, both the LUCID and BCM detectors observe some small activity in the BCIDs immediately following a collision, which in later BCIDs decays  to a baseline value with several different time constants. This afterglow is most likely caused by photons from nuclear de-excitation, which in turn is induced by the hadronic cascades initiated by \pp\ collision products. For a given bunch pattern, the afterglow level is observed to be proportional to the luminosity in the colliding-bunch slots. During \vdM scans, it lies three to four orders of magnitude below the luminosity signal, but reaches a few tenths of a percent during physics running because of the much denser bunch pattern.

\par 
Instrumental noise is, under normal circumstances, a few times smaller than the single-beam backgrounds, and remains negligible except  at the largest beam separations. However, during a one-month period in late 2012 that includes the November \vdM scans, the A arm of both BCM detectors was affected by high-rate electronic noise corresponding to about 0.5\% (1\%) of the visible interaction rate, at the peak of the scan, in the BCMH (BCMV) diamond sensors (Fig.~\ref{fig:scanCurves}(b)). This temporary perturbation, the cause of which could not be identified, disappeared a few days after the scan session. Nonetheless, it was large enough that a careful subtraction procedure had to be implemented in order for this noise not to bias the fit of the BCM luminosity-scan curves.

\par
Since afterglow and instrumental noise both induce random hits at a rate that varies slowly from one BCID to the next, they are subtracted together from the raw visible interaction rate \muvis\ in each colliding-bunch slot. Their combined magnitude is estimated using the rate measured in the immediately preceding bunch slot, assuming that the variation of the afterglow level from one bunch slot to the next can be neglected. 

\par
A third background contribution arises from activity correlated with the passage of a single beam through the detector. This activity is attributed to a combination of shower debris from beam--gas interactions and from beam-tail particles that populate the beam halo and impinge on the luminosity detectors in time with the circulating bunch. It is observed to be proportional to the bunch population, can differ slightly between beams 1 and 2, but is otherwise uniform for all bunches in a given beam. The total single-beam background in a colliding-bunch slot is estimated by measuring the single-beam rates in unpaired bunches (after subtracting the afterglow and noise as done for colliding-bunch slots), separately for beam 1 and beam 2, rescaling them by the ratio of the bunch populations in the unpaired and colliding bunches, and summing the contributions from the two beams. This background typically amounts to $2\times 10^{-4}$ ($8\times 10^{-4}$) of the luminosity at the peak of the scan for the LUCID (BCM) EventOR algorithms. Because it depends neither on the luminosity nor on the beam separation, it can become comparable to the actual luminosity in the tails of the scans.

\subsection{Determination of the absolute beam-separation scale}
\label{subsec:LengthScale}

\begin{figure}[htbp] 
   \centering
   \includegraphics[width=0.75\textwidth]{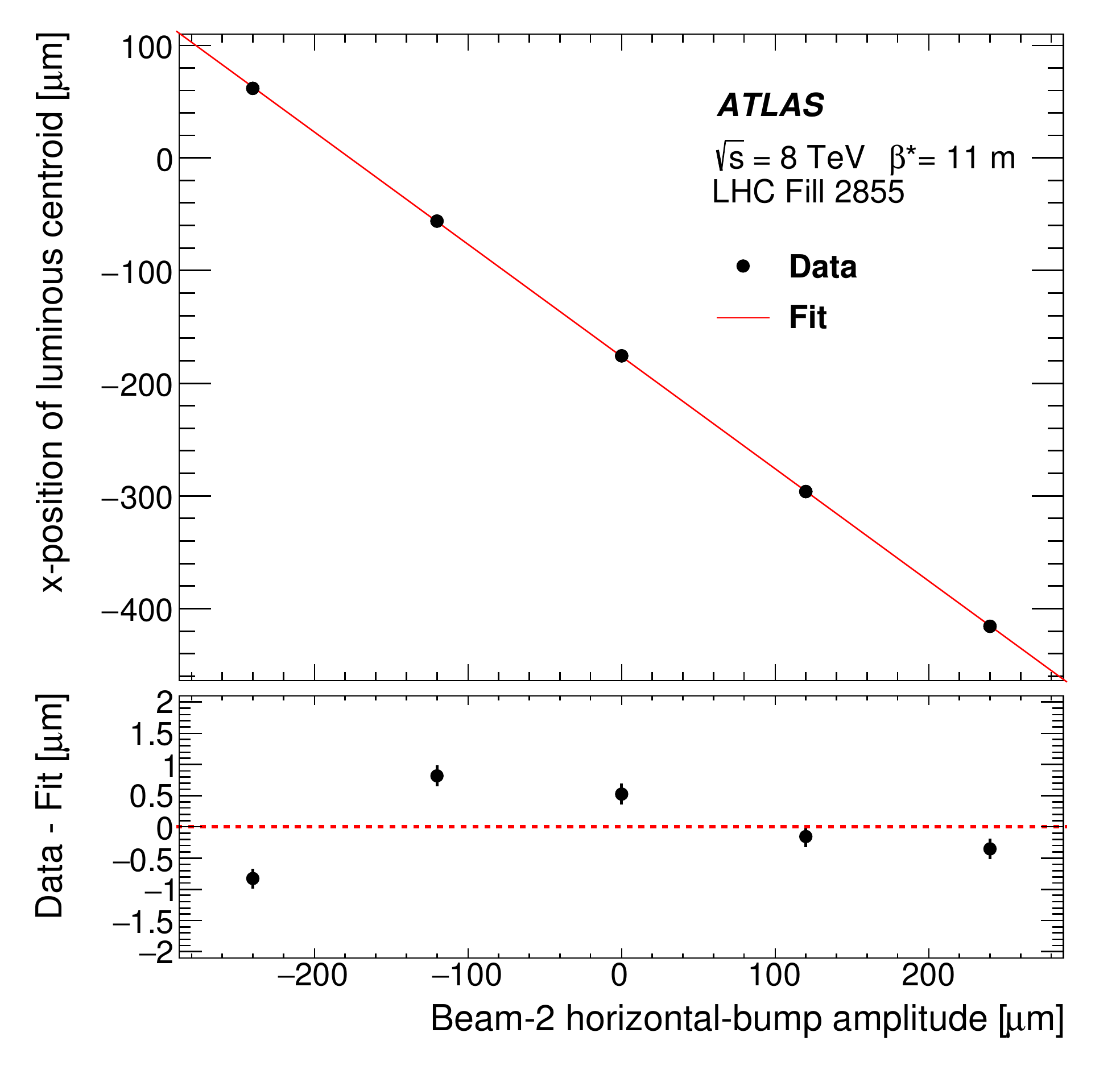} 
   \caption{Length-scale calibration scan for the $x$ direction of beam 2.  Shown is the measured displacement of the luminous centroid as a function of the expected displacement based on the corrector bump amplitude.  The line is a linear fit to the data, and the residual is shown in the bottom panel.  Error bars are statistical only.
}
   \label{fig:length}
\end{figure}

Another key input to the \vdM scan technique is the knowledge of the beam separation at 
each scan step.
The ability to measure \Capsig depends upon knowing the 
absolute distance by which the beams are separated during the \vdM scan, which is controlled by a set of 
closed orbit bumps\footnote{A closed orbit bump is a local distortion of the beam orbit
that is implemented using pairs of steering dipoles located on either
side of the affected region. In this particular case, these bumps are tuned to offset the trajectory of
 either beam parallel to itself at the IP, in either the horizontal or the vertical direction.} 
applied locally near the ATLAS IP. 
To determine this beam-separation scale, dedicated calibration measurements were performed close in time to the April and July scan sessions using the same optical configuration at the interaction point.
Such length-scale scans are performed by displacing both beams transversely by five steps over a range of up to $\pm 3 \sigma_\mathrm{b}^{\mathrm{nom}}$, at each step keeping the beams well centred on each other in the scanning plane. The actual displacement of the luminous region can then be measured with high accuracy using the primary-vertex position reconstructed by the ATLAS tracking detectors. 
Since each of the four bump amplitudes (two beams in two transverse directions) depends 
on different magnet and lattice functions, the length-scale calibration scans are 
performed so that each of these four calibration constants can be extracted independently.
The July 2012 calibration data for the horizontal bump of beam 2 are presented in Fig.~\ref{fig:length}.
The scale factor which relates the nominal beam displacement to the measured displacement of the luminous centroid is given by the slope of the fitted straight line; the intercept is irrelevant.

\par
Since the coefficients relating magnet currents to beam displacements depend on the interaction-region optics, the absolute length scale depends on the $\beta^\star$ setting and must be recalibrated when the latter changes. The results of the 2012 length-scale calibrations are summarized in Table~\ref{tab:length}.
Because the beam-separation scans discussed in Sect.~\ref{subsec:lumiScanData} are performed by displacing the two beams symmetrically in opposite directions, the relevant scale factor in the determination of \Capsig is the average of the scale factors for beam 1 and beam 2 in each plane.
A total correction of $-2.57$\%  ($-0.77$\%) is  
applied to the convolved-width product $\Sigma_x \Sigma_y$ and to the visible cross-sections measured during the April (July and November) 2012 {\em vdM} scans.

\begin{table*}[htbp]
   \centering
\begin{tabular}{c|cc|cc}
  \hline
Calibration session(s)			&  \multicolumn{2}{c|} {April 2012} 		  				& \multicolumn{2}{c}  {July 2012 (applicable to November)}							\\ 
 $\beta^\star$						&  \multicolumn{2}{c|} {0.6\,m} 		  					& \multicolumn{2}{c}  {11\,m} 						\\ 
\hline
							& Horizontal					& Vertical 			& Horizontal			&  Vertical				\\
\hline
Displacement scale				& 							&   					&  					&   					\\
Beam 1 			 			& $0.9882\pm0.0008$ 			&  $0.9881\pm 0.0008$ 	& $0.9970 \pm 0.0004$	& $0.9961 \pm 0.0006$		\\
Beam 2 			 			& $0.9822\pm0.0008$ 			&  $0.9897\pm 0.0009$	& $0.9964 \pm 0.0004$	& $0.9951 \pm 0.0004$		 \\
\hline
Separation scale				& $0.9852\pm0.0006$ 			& $0.9889\pm0.0006$	& $0.9967 \pm 0.0003$	& $0.9956 \pm 0.0004$		\\ \hline
\end{tabular}
   \caption{Length-scale calibrations at the ATLAS interaction point at $\sqrt{s} = 8$\,TeV.  Values shown are the ratio of the beam displacement measured by ATLAS using the average primary-vertex position, to the nominal displacement entered into the accelerator control system.  Ratios are shown  for each individual beam in both planes, as well as for the beam-separation scale that determines that of the convolved beam sizes in the \vdM scan. The uncertainties are statistical only.}
\label{tab:length}
\end{table*}
\subsection{Orbit-drift corrections}
\label{subsec:orbit}

Transverse drifts of the individual beam orbits at the IP during a scan session can distort the luminosity-scan curves and, if large enough, bias the determination of the overlap integrals and/or of the peak interaction rate. Such effects are monitored by extrapolating to the IP beam-orbit segments measured using beam-position monitors (BPMs) located in the LHC arcs~\cite{bib:bbDefl}, where the beam trajectories should remain unaffected by the {\em vdM} closed-orbit bumps across the IP. This procedure is applied to each beam separately and provides measurements of the relative drift of the two beams during the scan session, which are used to correct the beam separation at each scan step as well as between the $x$ and $y$ scans. The resulting impact on the visible cross-section varies from one scan set to the next; it does not exceed $\pm 0.6$\% in any 2012 scan set, except for scan set X where the orbits drifted rapidly enough for the correction to reach +1.1\%. 
\subsection{Beam--beam corrections}
\label{subsec:beambeam}

When charged-particle bunches collide, the electromagnetic field generated by a bunch in beam 1 distorts the individual particle trajectories in the corresponding bunch of beam 2 (and vice-versa). This so-called {\em beam--beam interaction} affects the scan data in two ways.

\par
First, when the bunches are not exactly centred on each other in the $x$--$y$ plane, their electromagnetic repulsion induces a mutual angular kick~\cite{bib:SLCdefl} of a fraction of a microradian and modulates the actual transverse separation at the IP in a manner that depends on the separation itself. The phenomenon is well known from $e^+e^-$ colliders and has been observed at the LHC at a level consistent with predictions~\cite{bib:bbDefl}. If left unaccounted for, these {\em beam--beam deflections} would bias the measurement of the overlap integrals in a manner that depends on the bunch parameters.

\par
The second phenomenon, called {\em dynamic $\beta$}~\cite{bib:dynBeta}, arises from the mutual defocusing of the two colliding bunches: this effect is conceptually analogous to inserting a small quadrupole at the collision point. The resulting fractional change in $\beta^\star$, or equivalently the optical demagnification between the LHC arcs and the collision point, varies with the transverse beam separation, slightly modifying, at each scan step, the effective beam separation in both planes (and thereby also the collision rate), and resulting in a distortion of the shape of the \vdM scan curves. 

\par
The amplitude and the beam-separation dependence of both effects depend similarly on the beam energy, the tunes\footnote{The tune of a storage ring is defined as the betatron phase advance per turn, or equivalently as the number of betatron oscillations over one full ring circumference.} and the unperturbed $\beta$-functions, as well as on the bunch intensities and transverse beam sizes. The beam--beam deflections and associated orbit distortions are calculated analytically~\cite{bib:bbDeflCalc} assuming elliptical Gaussian beams that collide in ATLAS only. For a typical bunch, the peak angular kick during the November 2012 scans is about $\pm 0.25\,\muup$rad, and the corresponding peak increase in relative beam separation amounts to $\pm 1.7\,\muup$m.
The MAD-X optics code~\cite{bib:MADX} is used to validate this analytical calculation, and to verify that higher-order dynamical effects (such as the orbit shifts induced at other collision points by beam--beam deflections at the ATLAS IP) result in negligible corrections to the analytical prediction.

\par
The dynamic evolution of $\beta^\star$ during the scan is modelled using the MAD-X simulation assuming bunch parameters representative of the May 2011 {\em vdM} scan~\cite{Aad:2013ucp}, and then scaled using the beam energies, the $\beta^\star$ settings, as well as the measured intensities and convolved beam sizes of each colliding-bunch pair. The correction function is intrinsically independent of whether the bunches collide in ATLAS only, or also at other LHC interaction points~\cite{bib:dynBeta}. For the November session, the peak-to-peak $\beta^\star$ variation during a scan is about $1.1\%$.

\par
At each scan step, the predicted deflection-induced change in beam separation is added to the nominal beam separation, and the dynamic-$\beta$ effect is accounted for by rescaling both the effective beam separation  and the measured visible interaction rate to reflect the beam-separation dependence of the IP $\beta$-functions.
Comparing the results of the 2012 scan analysis without and with beam--beam corrections, it is found that the visible cross-sections are increased by 1.2--1.8\% by the deflection correction, and reduced by 0.2--0.3\% by the dynamic-$\beta$ correction. The net combined effect of these beam--beam corrections is a 0.9--1.5\% increase of the visible cross-sections, depending on the scan set considered.

\subsection{Non-factorization effects}
\label{subsec:nonFctrztnCrctn}
 The original \vdM formalism~\cite{bib:vdm} explicitly assumes that 
the particle densities in each bunch can be factorized into independent horizontal and vertical 
components, such that the term $1/2 \pi \Sigma_x \Sigma_y$ in Eq.~(\ref{eqn:lumifin})
fully describes the overlap integral of the two beams.
If this factorization assumption is violated, the horizontal (vertical) convolved beam width $\Sigma_x$ ($\Sigma_y$) is no longer independent of the vertical (horizontal) beam separation $\delta_y$ ($\delta_x$); similarly, the transverse luminous size~\cite{bib:LumR} in one plane (\sigxL or \sigyL), as extracted from the spatial distribution of reconstructed collision vertices, depends on the separation in the other plane. The generalized \vdM formalism summarized by Eq.~(\ref{eq:CapSNonFact}) correctly handles such two-dimensional luminosity distributions, provided the dependence of these distributions on the beam separation in the transverse plane is known with sufficient accuracy.

\par
Non-factorization effects are unambiguously observed in some of the 2012 scan sessions, both from significant differences in $\Sigma_x$ ($\Sigma_y$) between a standard scan and an off-axis scan, during which the beams are partially separated in the non-scanning plane (Sect.~\ref{subsubsec:offAxis}), and from the $\delta_x$ ($\delta_y$) dependence of \sigyL (\sigxL) during a standard horizontal (vertical) scan (Sect.~\ref{subsubsec:lumRegAna}). Non-factorization effects can also be quantified, albeit with more restrictive assumptions, by performing a simultaneous fit to horizontal and vertical \vdM scan curves using a non-factorizable function to describe the simultaneous dependence of the luminosity on the $x$ and $y$ beam separation (Sect.~\ref{subsubsec:cpldFits}).

\par
A large part of the scan-to-scan irreproducibility observed during the April and July scan sessions can be attributed to non-factorization effects, as discussed for ATLAS in Sect.~\ref{subsubsec:nonFactCrctns} below and as independently reported by the LHCb Collaboration~\cite{bib:LHCbLumPap2}. The strength of the effect varies widely across \vdM scan sessions, differs somewhat from one bunch to the next and evolves with time within one LHC fill.
Overall, the body of available observations can be explained neither by residual linear $x$--$y$ coupling in the LHC optics~\cite{Aad:2013ucp, bib:emittanceSimon}, nor by crossing-angle or beam--beam effects; instead, it points to non-linear transverse correlations in the phase space of the individual bunches. This phenomenon was never envisaged at previous colliders, and was considered for the first time at the LHC~\cite{Aad:2013ucp} as a possible source of systematic uncertainty in the absolute luminosity scale. More recently, the non-factorizability of individual bunch density distributions was demonstrated directly by an LHCb beam--gas imaging analysis~\cite{bib:LHCbLumPap2}.

\subsubsection{Off-axis \vdM scans}
\label{subsubsec:offAxis}

\begin{figure}
\centering
\subfigure[]{
\includegraphics[width=0.77\textwidth]{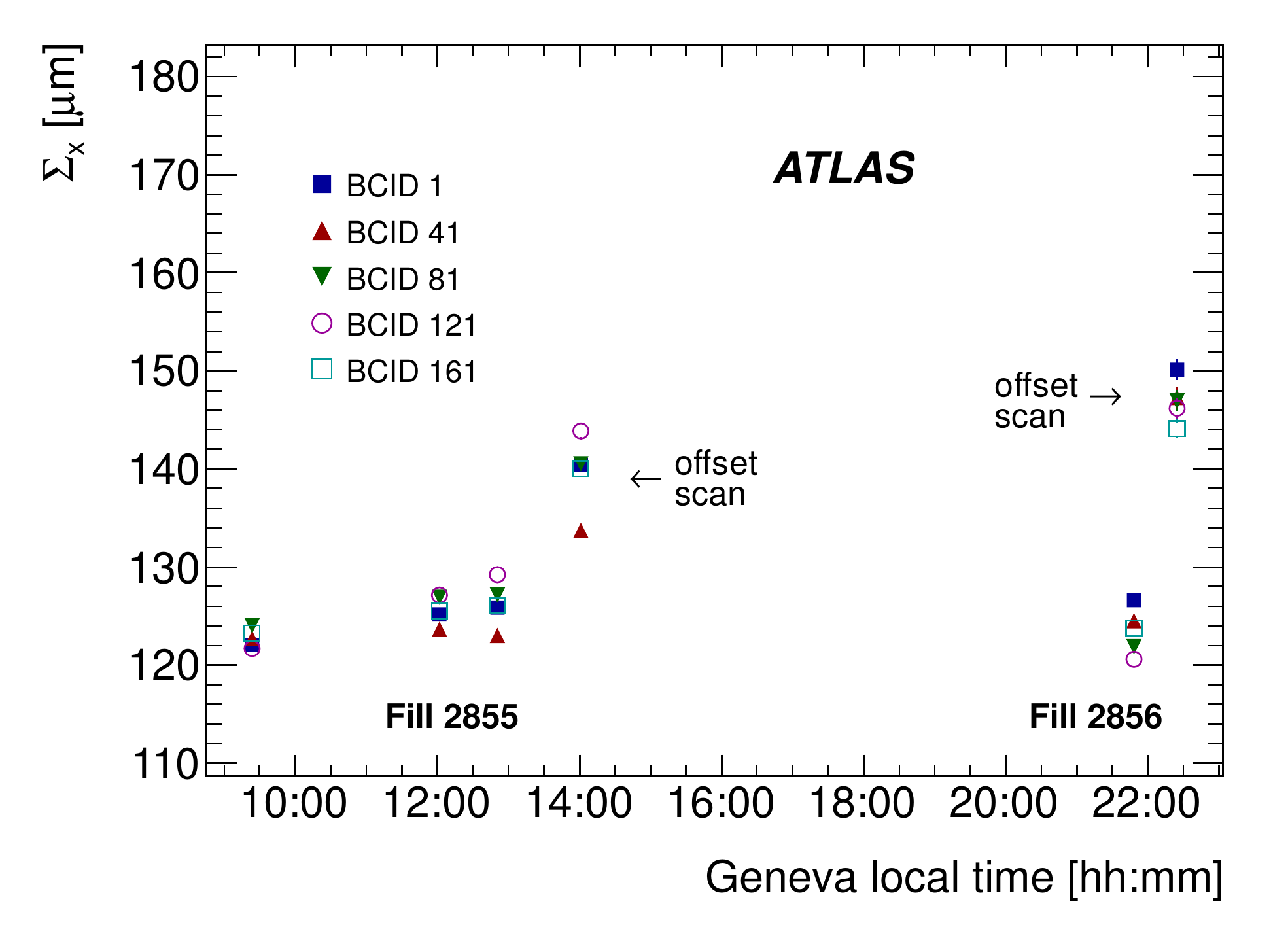}
}
\subfigure[]{
\includegraphics[width=0.77\textwidth]{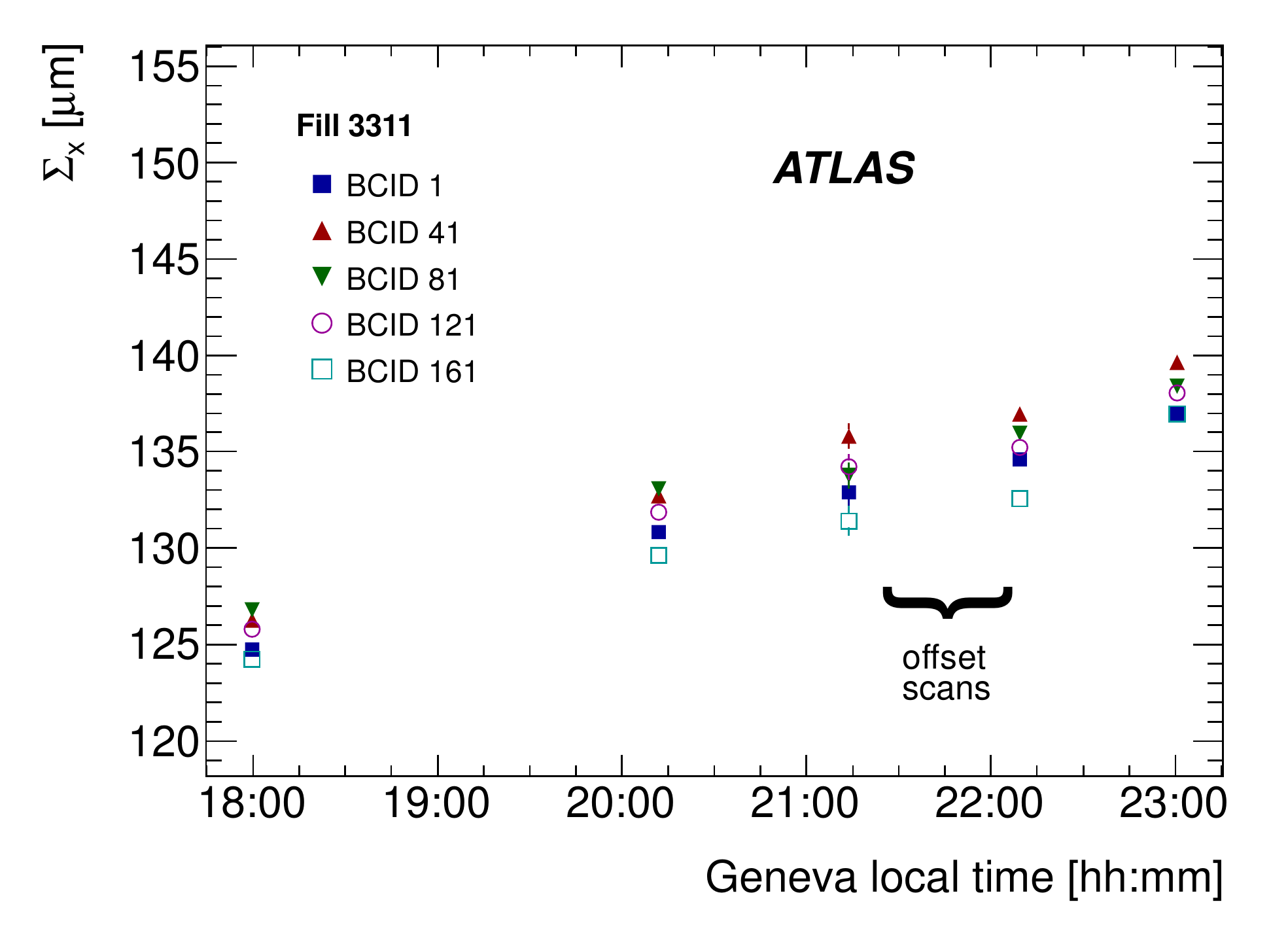}
}
\caption{Time evolution of the horizontal convolved beam size $\Sigma_x$ for five different colliding-bunch pairs (BCIDs), measured using the LUCID\_EventOR luminosity algorithm during the (a) July and (b)  November 2012 \vdM-scan sessions.}
\label{fig:nonFactProof}
\end{figure}

An unambiguous signature of non-factorization can be provided by comparing the transverse convolved width measured during centred (or on-axis) \vdM scans with the same quantity extracted from an offset (or off-axis) scan, \ie one where the two beams are significantly separated in the direction orthogonal to that of the scan. This is illustrated in Fig.~\ref{fig:nonFactProof}(a). The beams remained vertically centred on each other during the first three horizontal scans (the first horizontal scan) of LHC fill 2855 (fill 2856), and were separated vertically by approximately 340\,$\muup$m (roughly $4 \sigma_{\mathrm{b}}$) during the last horizontal scan in each fill. In both fills, the horizontal convolved beam size is significantly larger when the beams are vertically separated, demonstrating that the horizontal luminosity distribution depends on the vertical beam separation, \ie that the horizontal and vertical luminosity distributions do not factorize. 

\par
The same measurement was carried out during the November scan session: the beams remained vertically centred on each other during the first, second and last scans (Fig.~\ref{fig:nonFactProof}(b)), and were separated vertically by about 340 (200)\,$\muup$m during the third (fourth) scan. The horizontal convolved beam size increases with time at an approximately constant rate, reflecting transverse-emittance growth. No significant deviation from this trend is observed when the beams are separated vertically, suggesting that the horizontal luminosity distribution is independent of the vertical beam separation, \ie that during the November scan session the horizontal and vertical luminosity distributions approximately factorize.

\subsubsection{Determination of single-beam parameters from luminous-region and luminosity-scan data}
\label{subsubsec:lumRegAna}

While a single off-axis scan can provide convincing evidence for non-factorization, it samples only one thin slice in the ($\delta_x$, $\delta_y$) beam-separation space  and is therefore insufficient to fully determine the two-dimensional luminosity distribution. Characterizing the latter by performing an $x$--$y$ grid scan (rather than two one-dimensional $x$ and $y$ scans) would be prohibitively expensive in terms of beam time, as well as limited by potential emittance-growth biases. The strategy, therefore, is to retain the standard \vdM technique (which assumes factorization) as the baseline calibration method, and to use the data to constrain possible non-factorization biases. 
In the absence of input from beam--gas imaging (which requires a vertex-position resolution within the reach of LHCb only), the most powerful approach so far has been the modelling of the simultaneous beam-separation-dependence of the luminosity and of the luminous-region geometry. In this procedure,
the parameters describing the transverse proton-density distribution of individual bunches are determined by fitting the evolution, during \vdM\ scans, not only of the luminosity itself but also of the position, orientation and shape of its spatial distribution, as reflected by that of reconstructed \pp-collision vertices~\cite{bib:CONF027}. 
Luminosity profiles are then generated for simulated \vdM scans using these fitted single-beam parameters, and analysed in the same fashion as real \vdM scan data. 
The impact of non-factorization on the absolute luminosity scale is quantified by the ratio $R_{\mathrm{NF}}$ of the ``measured'' luminosity extracted from the one-dimensional simulated luminosity profiles using the standard \vdM method, to the ``true'' luminosity from the computed four-dimensional ($x$, $y$, $z$, $t$) overlap integral~\cite{bib:LumR} of the single-bunch distributions at zero beam separation.
This technique is closely related to beam--beam imaging~\cite{bib:LumR, bib:BalagNIM, bib:LHCbLumPap1}, with the notable difference that it is much less sensitive to the vertex-position resolution because it is used only to estimate a small fractional correction to the overlap integral, rather than its full value.

\par
The luminous region is modelled by a three-dimensional (3D) ellipsoid~\cite{bib:LumR}. Its parameters are extracted, at each scan step, from an unbinned maximum-likelihood fit of a 3D Gaussian function to the spatial distribution of the reconstructed primary vertices that were collected, at the corresponding beam separation, from the limited subset of colliding-bunch pairs monitored by the high-rate, dedicated ID-only data stream (Sect.~\ref{subsec:IDalgos}). The vertex-position resolution, which is somewhat larger (smaller) than the transverse luminous size during scan sets I--III (scan sets IV--XV), is determined from the data as part of the fitting procedure~\cite{bib:CONF027}. It potentially impacts the reported horizontal and vertical luminous sizes, but not the measured position, orientation nor length of the luminous ellipsoid.

\par
\begin{figure}
\centering
\subfigure[]{\label{fig:lum:scanx:scan4:bcid1}\includegraphics[scale=0.39]{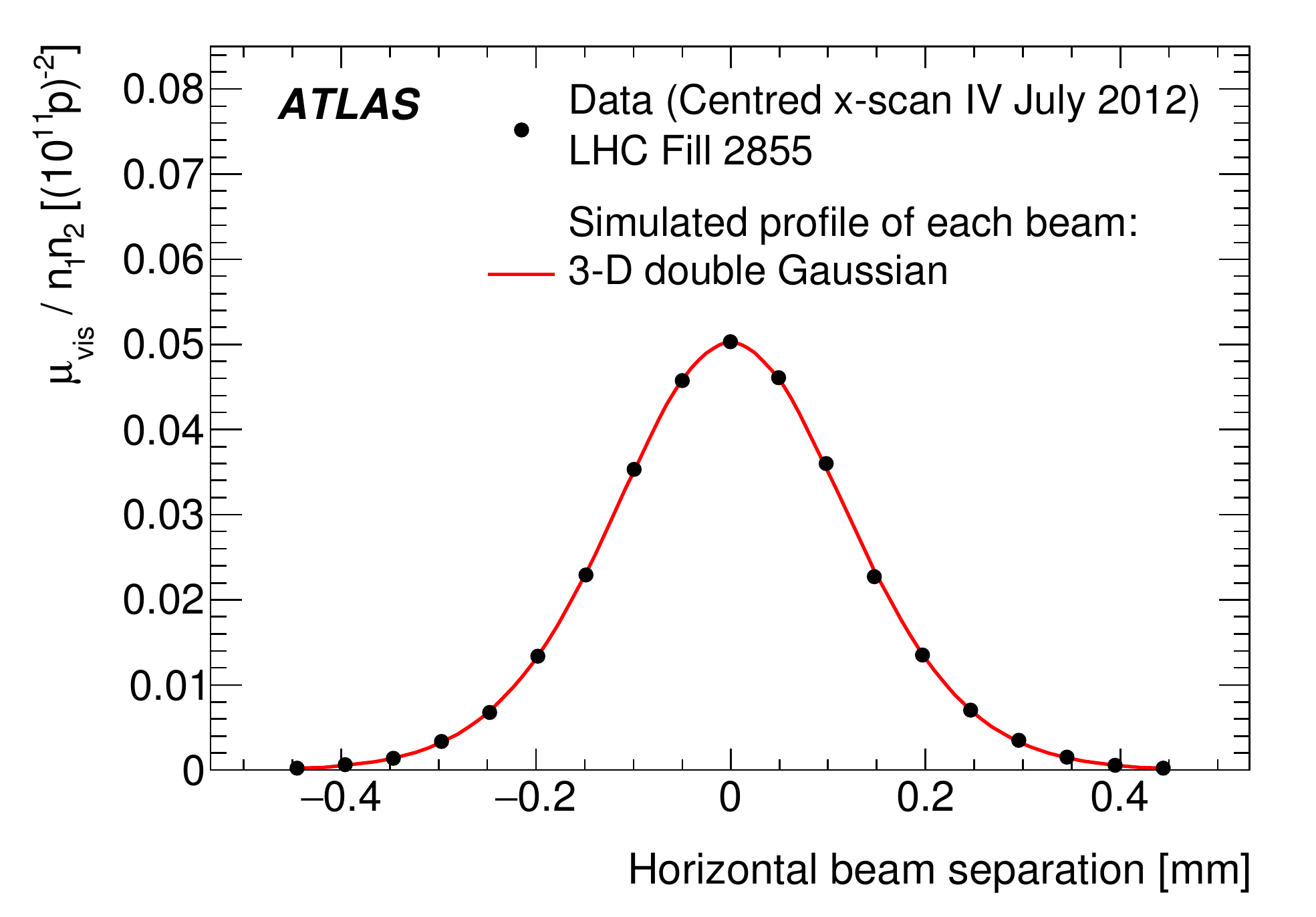}}\quad
\subfigure[]{\label{fig:posx:scanx:scan4:bcid1}\includegraphics[scale=0.39]{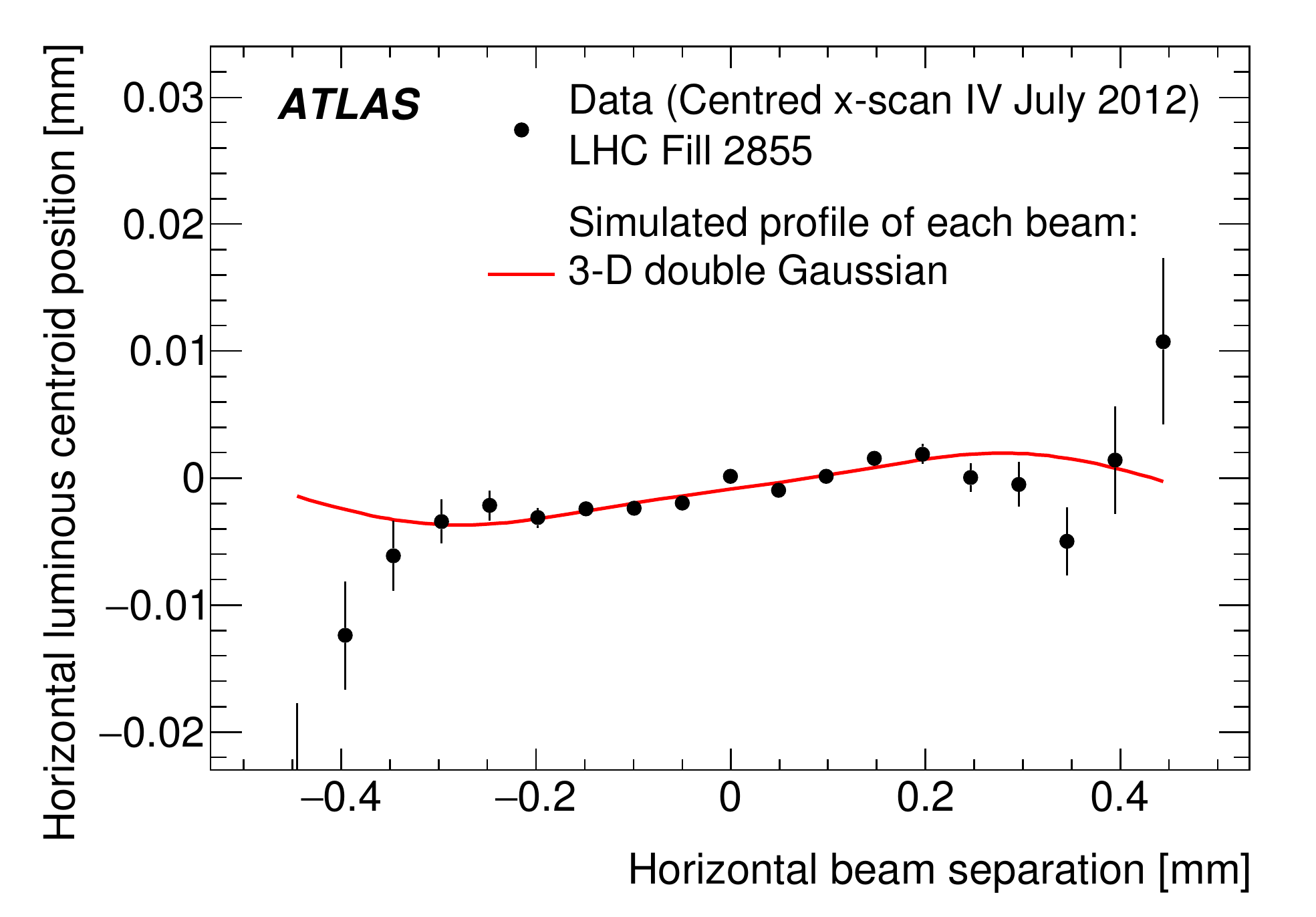}}\quad
\subfigure[]{\label{fig:widthx:scanx:scan4:bcid1}\includegraphics[scale=0.39]{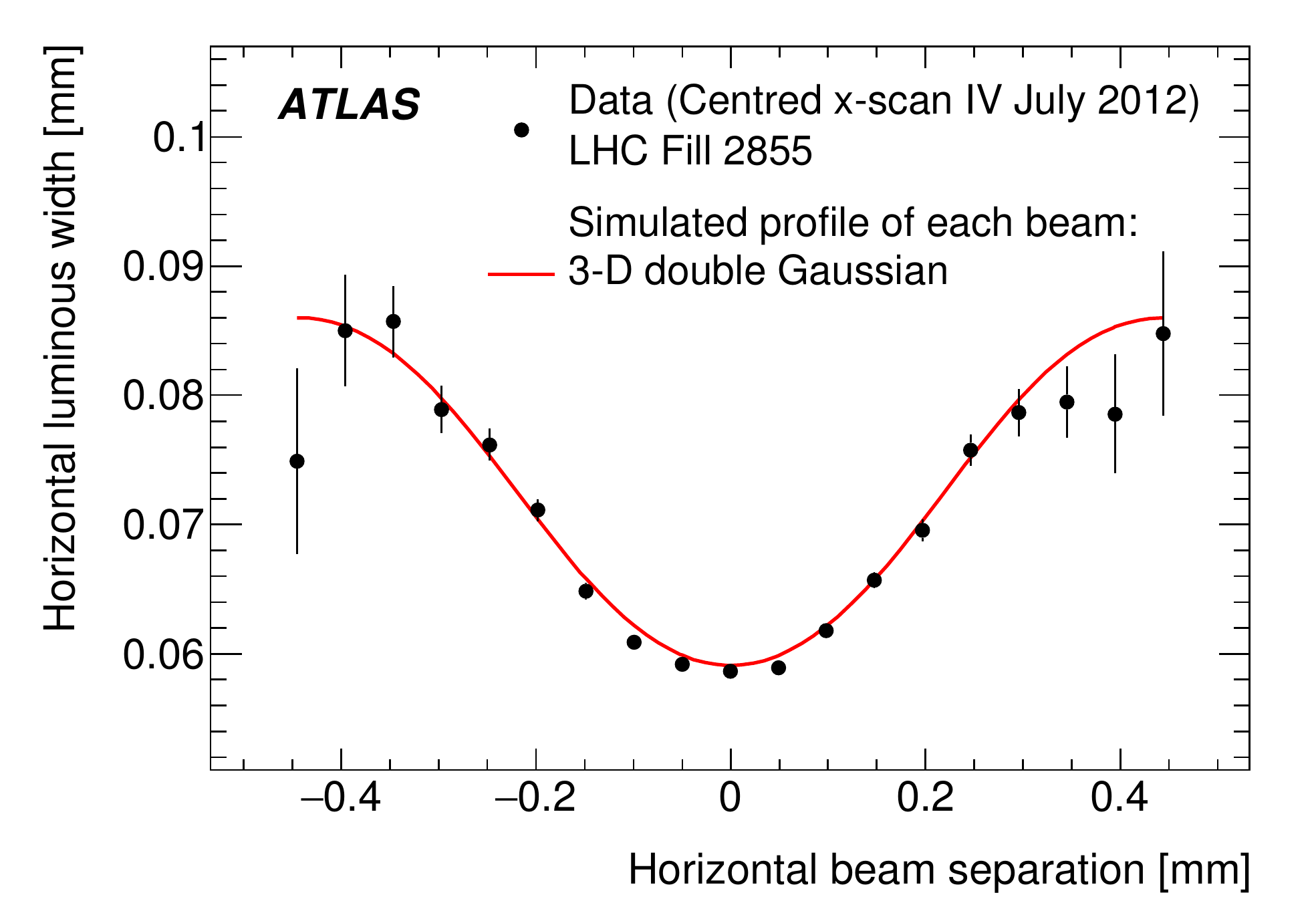}}\quad
\subfigure[]{\label{fig:widthy:scanx:scan4:bcid1}\includegraphics[scale=0.39]{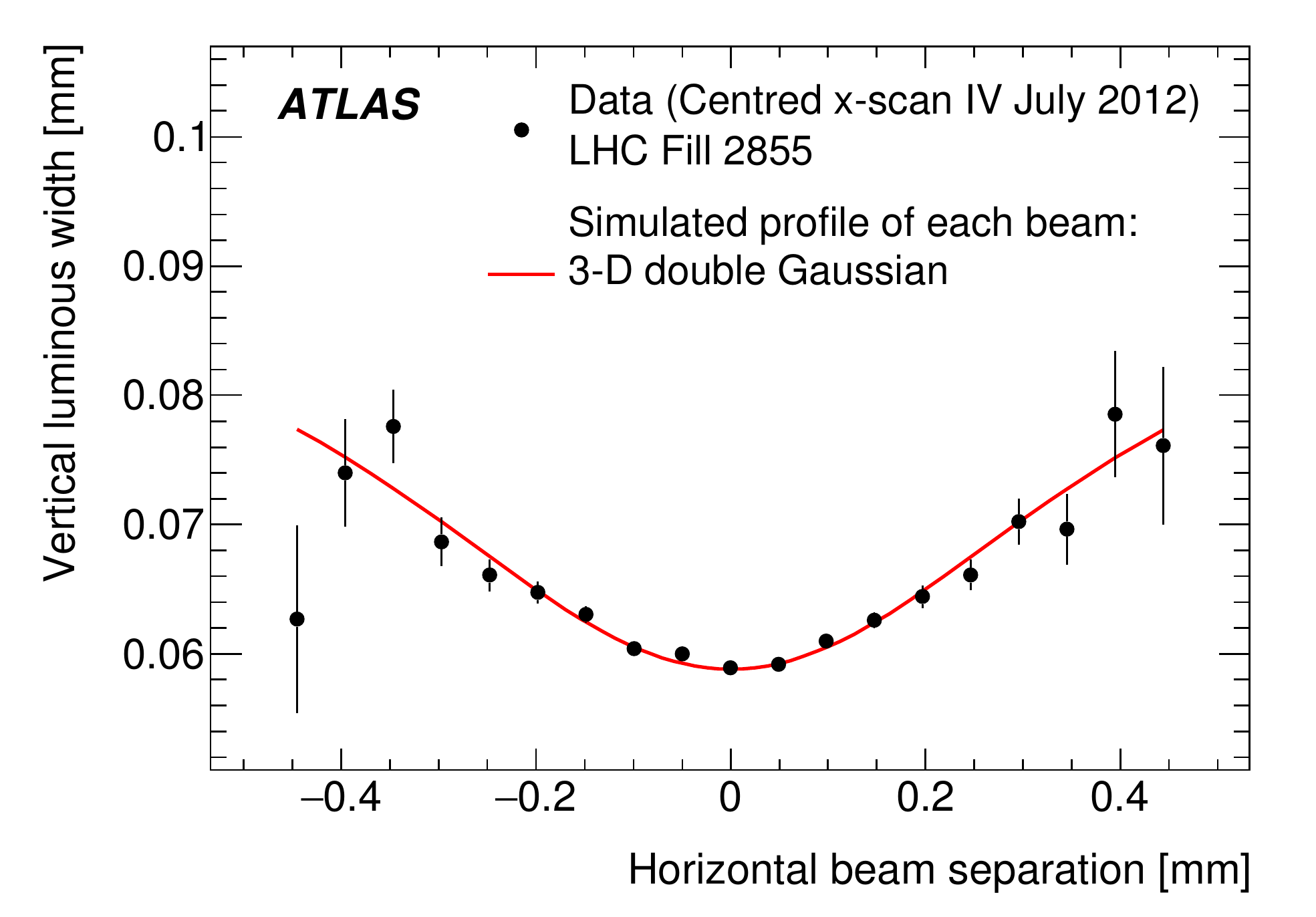}}\quad
\caption{Beam-separation dependence of the luminosity and of a subset of luminous-region parameters during horizontal \vdM scan IV. The points represent (a) the specific visible interaction rate (or equivalently the specific luminosity), (b) the horizontal position of the luminous centroid, (c) and (d) the horizontal and vertical luminous widths \sigxL and \sigyL. The red line is the result of the fit described in the text. 
}
\label{fig:lumRegAna}
\end{figure}

The single-bunch proton-density distributions $\rho_B(x, y,z)$ are parameterized, independently for each beam $B$ ($B$ = 1, 2), as the non-factorizable sum of up to three 3D Gaussian or super-Gaussian~\cite{decker_1994} distributions ($G_{\mathrm a},  G_{\mathrm b}, G_{\mathrm c}$) with arbitrary widths and orientations~\cite{bib:CBthesis, bib:SWthesis}:
\begin{equation*}
\rho_B = w_{\mathrm{a}B} \times G_{\mathrm{a}B} 
                         + (1-w_{\mathrm{a}B})[w_{\mathrm{b}B} \times G_{\mathrm{b}B} + (1-w_{\mathrm{b}B}) \times G_{\mathrm{c}B}]\, ,
\end{equation*}
where the weights $w_{\mathrm{a(b)}B}$, $(1-w_{\mathrm{a(b)}B})$ add up to one by construction.
The overlap integral of these density distributions, which allows for a crossing angle in both planes, is evaluated at each scan step to predict the produced luminosity and the geometry of the luminous region for a given set of bunch parameters. This calculation takes into account the impact, on the relevant observables, of the luminosity backgrounds, orbit drifts and beam--beam corrections. The bunch parameters are then adjusted, by means of a $\chi^2$-minimization procedure, to provide the best possible description of the centroid position, the orientation and the resolution-corrected  widths of the luminous region measured at each step of a given set of on-axis $x$ and $y$  scans.
Such a fit is illustrated in Fig.~\ref{fig:lumRegAna} for one of the horizontal scans in the July 2012 session. The goodness of fit is satisfactory 
($\chi^2 = 1.3$ per degree of freedom), 
even if some systematic deviations are apparent in the tails of the scan. The strong horizontal-separation dependence of the vertical luminous size (Fig.~\ref{fig:lumRegAna}(d)) confirms the presence of significant non-factorization effects, as already established from the off-axis luminosity data for that scan session (Fig.~\ref{fig:nonFactProof}(a)).

\begin{figure}
  \centering
  \includegraphics[width=1.00\textwidth]{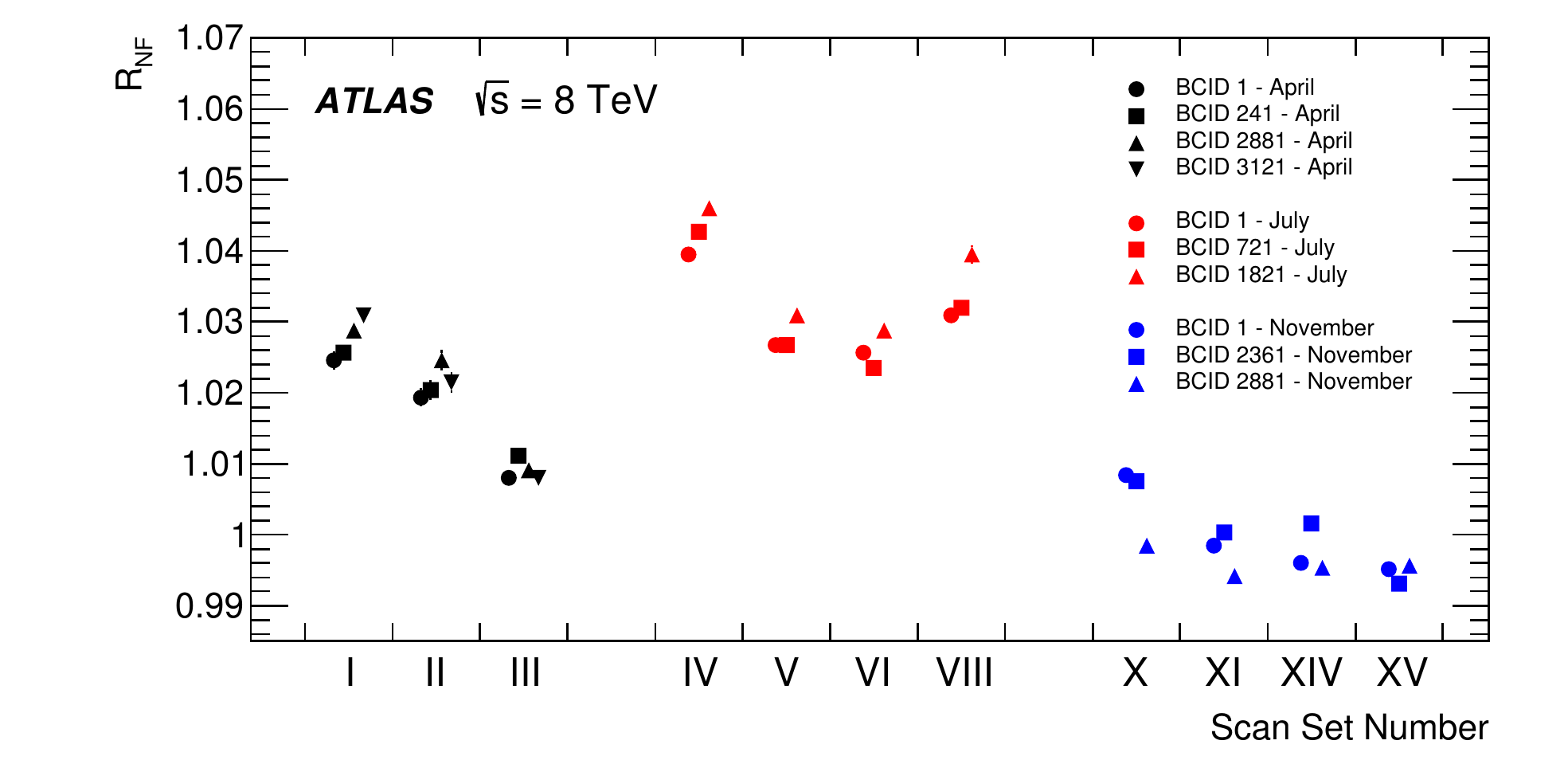} 
  \caption{Ratio $R_{\mathrm{NF}}$ of the luminosity determined by the \vdM method assuming factorization, to that evaluated from the overlap integral of the reconstructed single-bunch profiles at the peak of each  scan set.
The results are colour-coded by scan session. Each point corresponds to one colliding-bunch pair in the dedicated ID-only stream. The statistical errors are smaller than the symbols.
}
 \label{fig:rNF}
\end{figure}

\par
This procedure is applied to all 2012 \vdM scan sets, and the results are summarized in Fig.~\ref{fig:rNF}. 
The luminosity extracted from the standard \vdM analysis with the assumption that factorization is valid, is larger than that computed from the reconstructed single-bunch parameters. 
This implies that neglecting non-factorization effects in the \vdM calibration leads to overestimating the absolute luminosity scale (or equivalently underestimating the visible cross-section) by up to 3\% (4.5\%) in the April (July) scan session. Non-factorization biases remain below 0.8\% in the November scans, thanks to bunch-tailoring in the LHC injector chain~\cite{bib:injBunchPrep}. These observations are consistent, in terms both of absolute magnitude and of time evolution within a scan session, with those reported by LHCb~\cite{bib:LHCbLumPap2} and CMS~\cite{bib:CMS_LumP2012_2, bib:CMS_LumP2012_3} in the same fills. 

\subsubsection{Non-factorizable \vdM fits to luminosity-scan data}
\label{subsubsec:cpldFits}

A second approach, which does not use luminous-region data, performs a combined fit of the measured beam-separation dependence of the specific visible interaction rate to horizontal- and vertical-scan data simultaneously, in order to determine the overlap integral(s) defined by either Eq.~(\ref{eqn:caps}) or Eq.~(\ref{eq:CapSNonFact}). Considered fit functions include factorizable or non-factorizable combinations of two-dimensional Gaussian or other functions (super-Gaussian, Gaussian times polynomial) where the (non-)factorizability between the two scan directions is imposed by construction.

\par
The fractional difference between \sigmavis\ values extracted from such factorizable and non-factorizable fits, \ie the multiplicative correction factor to be applied to visible cross-sections extracted from a standard \vdM analysis, is consistent with the equivalent ratio $R _{\mathrm{NF}}$ extracted from the analysis of Sect.~\ref{subsubsec:lumRegAna} within 0.5\% or less for all scan sets.
Combined with the results of the off-axis scans, this confirms that while the April and July \vdM analyses require substantial non-factorization corrections, non-factorization biases during the November scan session remain small.

\subsubsection{Non-factorization corrections and scan-to-scan consistency}
\label{subsubsec:nonFactCrctns}

\begin{figure} 
   \centering
   \includegraphics[width=0.77 \columnwidth]{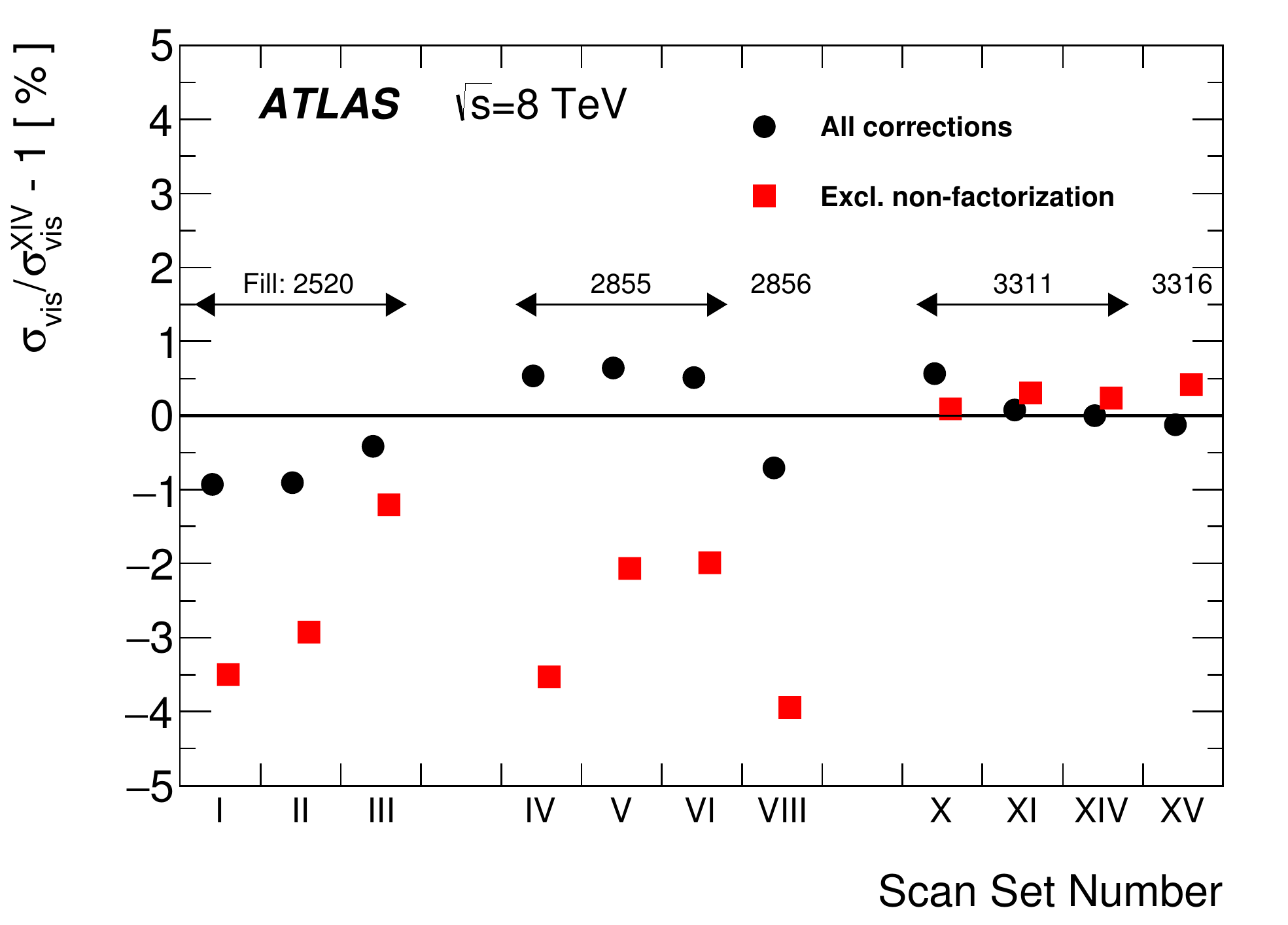} 
   \caption{Comparison of {\em vdM}-calibrated visible cross-sections for the default track-counting algorithm, with all corrections applied (black circles) and with all corrections except for non-factorization (red squares). Shown is the fractional difference between the visible cross-section from a given scan set, and the fully corrected visible cross-section from scan set XIV. The LHC fill numbers corresponding to each scan set are indicated.
 }
 \label{fig:nonFactImpact}
\end{figure}

Non-factorization corrections significantly improve the reproducibility of the calibration results (Fig.~\ref{fig:nonFactImpact}). Within a given LHC fill and in the absence of non-factorization corrections, the visible cross-section increases with time, as also observed at other IPs in the same fills~\cite{bib:LHCbLumPap2, bib:CMS_LumP2012_2}, suggesting that the underlying non-linear correlations evolve over time. Applying the non-factorization corrections extracted from the luminous-region analysis dramatically improves the scan-to-scan consistency within the April and July scan sessions,  as well as from one session to the next. 
The 1.0--1.4\% inconsistency between the fully corrected cross-sections (black circles) in scan sets I--III and in later scans, as well as the difference between fills 2855 and 2856 in the July session, are discussed in Sect.~\ref{subsec:vdMResults}.

\subsection{Emittance-growth correction}
\label{subsec:epsGrowth}

The \vdM scan formalism assumes that both convolved beam sizes $\Sigma_x$, $\Sigma_y$ (and therefore the transverse emittances of each beam) remain constant, both during a single $x$ or $y$ scan and in the interval between the horizontal scan and the associated vertical scan.

\par
Emittance growth within a scan would manifest itself by a slight distortion of the scan curve. The associated systematic uncertainty, determined from pseudo-scans simulated with the observed level of emittance growth, was found to be negligible.

\par
\begin{figure}
\centering
\subfigure[]{
\includegraphics[width=0.485\textwidth]{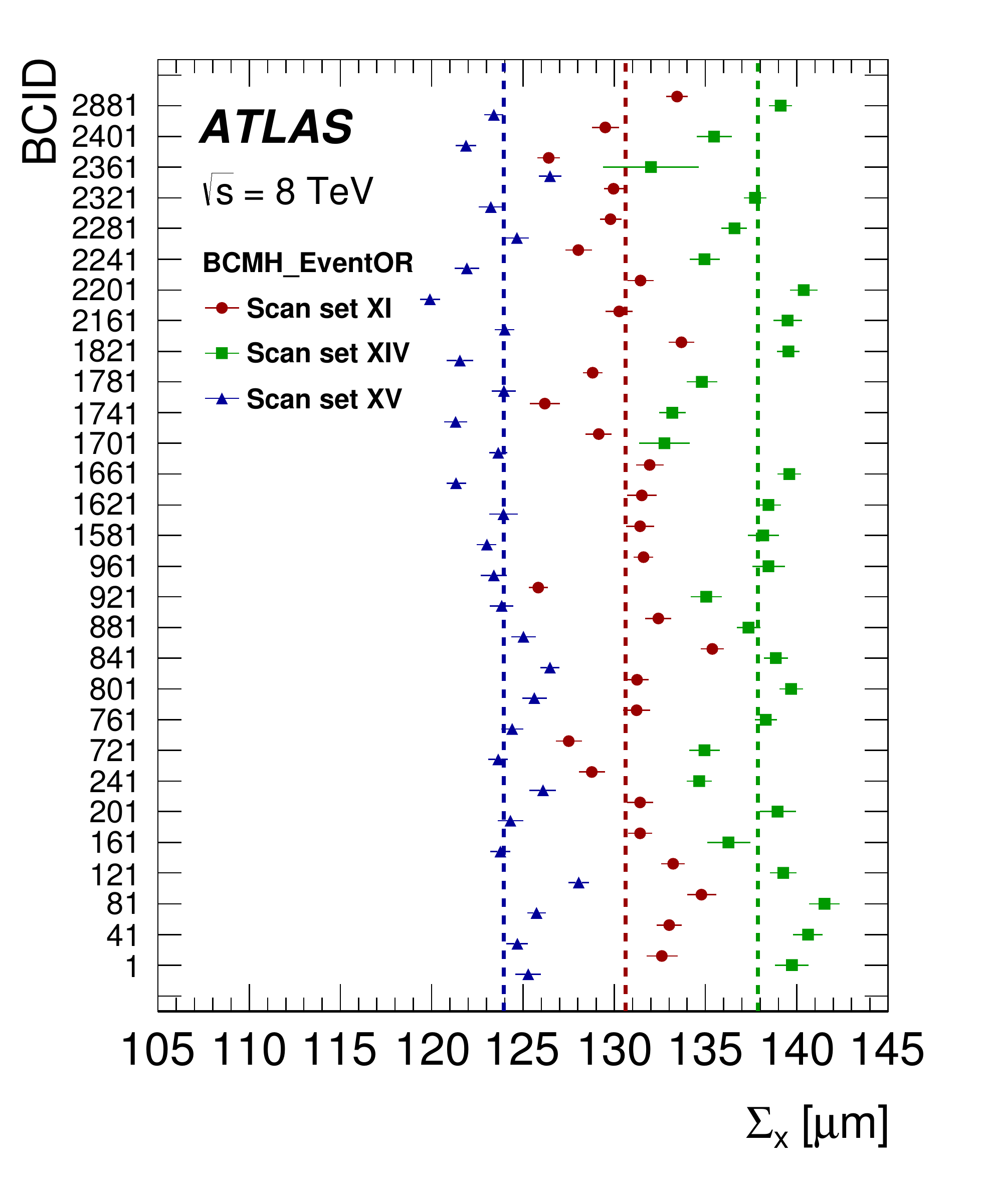}
}
\subfigure[]{
\includegraphics[width=0.485\textwidth]{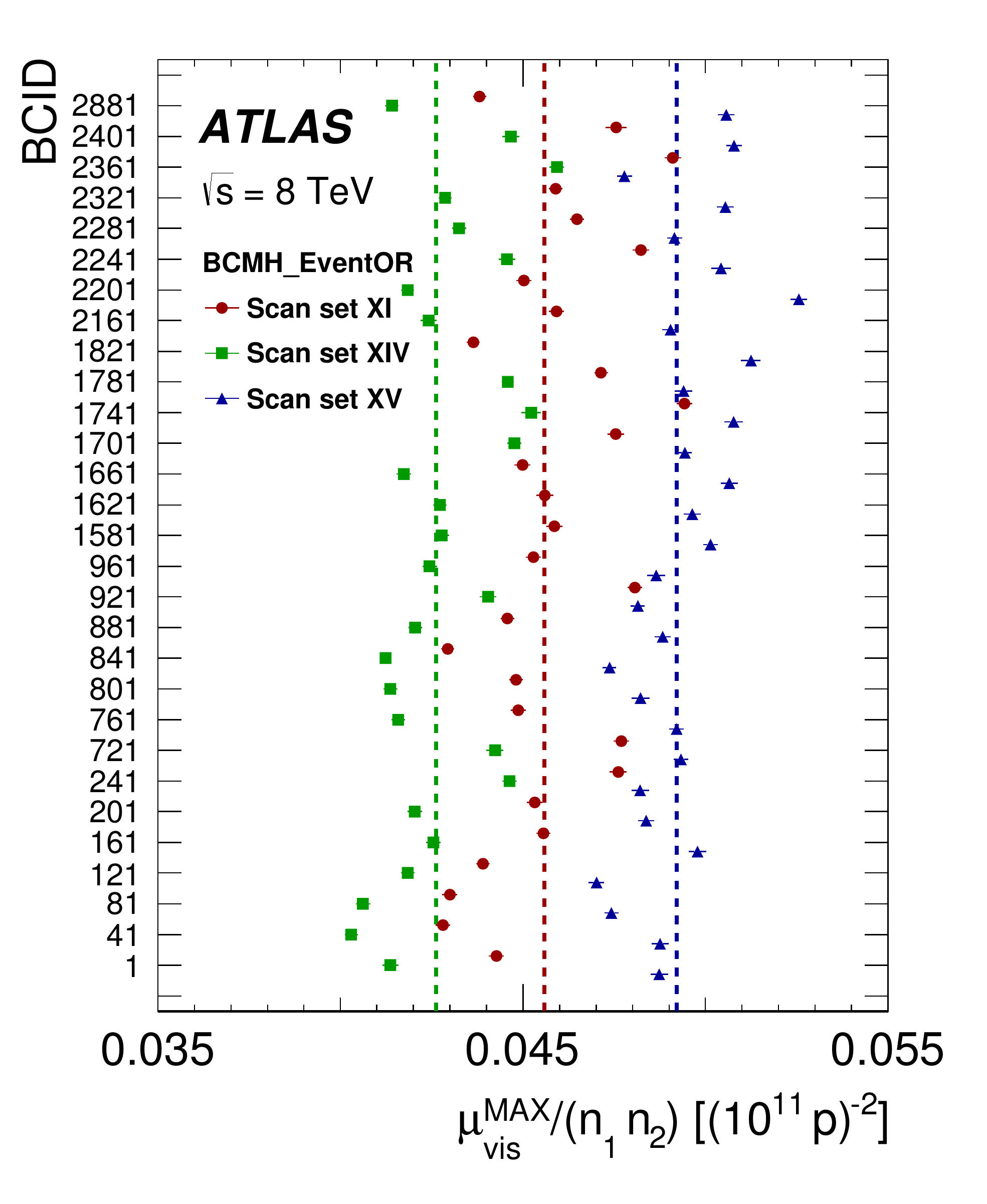}
}
\caption{Bunch-by-bunch (a) horizontal convolved beam size and (b) peak specific interaction rate measured in scan sets XI, XIV, and XV for the BCMH\_EventOR algorithm. The vertical lines represent the weighted average over colliding-bunch pairs for each scan set separately. The error bars are statistical only, and are approximately the size of the marker.
 }  
\label{fig:epsGrowthEx}
\end{figure}

Emittance growth between scans manifests itself by a slight increase of the measured value of $\Capsig$ from one scan to the next, and by a simultaneous decrease in specific luminosity. Each scan set requires 40 to 60 minutes, during which time the convolved beam sizes each grow by 1--2\%, and the peak specific interaction rate decreases accordingly as $1/(\Sigma_x \Sigma_y)$. 
This is illustrated in Fig~\ref{fig:epsGrowthEx}, which displays the $\Sigma_x$ and $\mu_{\mathrm{vis}}^{\mathrm{MAX}}/(n_1 n_2)$ values measured by the BCMH\_EventOR algorithm during scan sets XI, XIV and XV. For each BCID, the convolved beam sizes increase, and the peak specific interaction rate decreases, from scan XI to scan XIV; since scan XV took place very early in the following fill, the corresponding transverse beam sizes (specific rates) are smaller (larger)  than for the previous scan sets.

\par
If the horizontal  and  vertical emittances grow at identical rates, the procedure described in Sect.~\ref{subsec:scanAna} remains valid without any need for correction, provided the decrease in peak rate is fully accounted for by the increase in ($\Sigma_x \Sigma_y$), and that the peak specific interaction rate in Eq.~(\ref{eqn:sigmaVis}) is computed as the average of the specific rates at the peak of the horizontal and the vertical scan:
\begin{equation*}
\muvismax / n_1 n_2 = \frac { (\muvismax / n_1 n_2)_x \,+\, (\muvismax / n_1 n_2)_y  } {2}~.
\end{equation*}

\par
The horizontal-emittance growth rate is measured from the bunch-by-bunch difference in fitted convolved width between two consecutive horizontal scans in the same LHC fill, and similarly for the vertical emittance. For LHC fill 3311 (scan sets X--XIV), these measurements reveal that the horizontal convolved width grew 1.5--2 times faster than the vertical width. The potential bias associated with unequal horizontal and vertical growth rates can be corrected for by interpolating  the measured values of $\Sigma_x$, $\Sigma_y$ and \muvismax\ to a common reference time, assuming that all three observables evolve linearly with time. This reference time is in principle arbitrary: it can be, for instance, the peak of the $x$ scan (in which case only $\Sigma_y$ needs to be interpolated), or the peak of the $y$ scan, or any other value. The visible cross-section, computed from Eq.~(\ref{eqn:sigmaVis})  using measured values projected to a common reference time, should be independent of the reference time chosen.

\par
Applying this procedure to the November scan session results in fractional corrections to \sigmavis\ of 1.38\%, 0.22\% and 0.04\% for scan sets X, XI and XIV, respectively. The correction for scan set X is exceptionally large because operational difficulties forced an abnormally long delay (almost two hours) between the horizontal  scan and the vertical scan, exacerbating the impact of the unequal horizontal and vertical growth rates; its magnitude is validated by the noticeable improvement it brings to the scan-to-scan reproducibility of \sigmavis. 

\par
No correction is available for scan set XV, as no other scans were performed in LHC fill 3316. However, in that case the delay between the $x$ and $y$ scans was short enough, and the consistency of the resulting \sigmavis\ values with those in scan sets XI and XIV sufficiently good (Fig.~\ref{fig:nonFactImpact}), that this missing correction is small enough to be covered by the systematic uncertainties discussed in Sects.~\ref{subsubsec:epsGrowthSyst} and \ref{subsubsec:sTsReprod}.

\par
Applying the same procedure to the July scan session yields emittance-growth corrections below 0.3\% in all cases. However, the above-described correction procedure is, strictly speaking, applicable only when non-factorization effects are small enough to be neglected. When the factorization hypothesis no longer holds, the very concept of separating horizontal and vertical emittance growth is ill-defined. In addition, the time evolution of the fitted one-dimensional convolved widths and of the associated peak specific rates is presumably more influenced by the progressive dilution, over time, of the non-factorization effects discussed in Sect.~\ref{subsec:nonFctrztnCrctn} above. Therefore, and given that the non-factorization corrections applied to scan sets I-VIII (Fig.~\ref{fig:rNF}) are up to ten times larger than a typical emittance-growth correction, no such correction is applied to the April and July scan results; an appropriately conservative systematic uncertainty must be assigned instead.
\subsection{Bunch-population determination}
\label{subsec:currents}

The bunch-population measurements are performed by the LHC Bunch-Current Normalization Working Group and have been described in detail in 
Refs.~\cite{bib:LHCbLumPap2, bib:BCNWG3, bib:BCNWG4, bib:CBthesis, bib:LDM}. 
A brief summary of the analysis is presented here. The fractional uncertainties affecting the bunch-population product ($n_1 n_2$) are summarized in Table~\ref{tab:BCNWG}.

\begin{table}
   \centering
   \begin{tabular*}{\columnwidth}{@{\extracolsep{\fill}}lccccc@{}}
      	\hline
      	Scan Set Number 					& I--III 		& IV--VII 		& VIII--IX 		&  X--XIV		&	XV			\\
      	LHC Fill Number 					& 2520 		& 2855		& 2856 		& 3311		&	3316			 \\
      	\hline
	      							 	&  \multicolumn{5}{c}{Fractional systematic uncertainty [\%]}   					\\
	Total intensity scale (DCCT)			& 0.26 		& 0.21 		& 0.21 		& 0.22	 	&	0.23			\\
	Bunch-by-bunch fraction (FBCT)		& 0.03 		& 0.04 		& 0.04 		& 0.04	  	&	0.04			\\
 	Ghost charge (LHCb beam--gas)		& 0.04 		& 0.03  		& 0.04 		& 0.04	  	&	0.02			\\
	Satellites (longitudinal density monitor)	& 0.07 		& 0.02 		& 0.03 		& 0.01	  	&	$< 0.01$		\\

	\hline
	Total 							& 0.27	 	& 0.22	 	& 0.22 		& 0.24	 	& 	0.23		\\
\hline  
   \end{tabular*}
  \caption{Systematic uncertainties affecting the bunch-population product $n_1 n_2$ during the 2012 \vdM scans.}
   \label{tab:BCNWG}
\end{table}

\par
The LHC bunch currents are determined in a multi-step process due to the different capabilities of the available instrumentation. First, the total intensity of each beam is monitored by two identical and redundant DC current transformers (DCCT), which are high-accuracy devices but have no ability to distinguish individual bunch populations. Each beam is also monitored by two fast beam-current transformers (FBCT), which measure relative bunch currents individually for each of the 3564 nominal 25\,ns slots in each beam; these fractional bunch populations are converted into absolute bunch currents using the overall current scale provided by the DCCT. Finally, corrections are applied to account for out-of-time charge present in a given BCID but not colliding at the interaction point.

\par
A precision current source with a relative accuracy of 0.05\% is used to calibrate the DCCT at regular intervals. An exhaustive analysis of the various sources of systematic uncertainty in the absolute scale of the DCCT, including in particular residual non-linearities, long-term stability and dependence on beam conditions, is documented in Ref.~\cite{bib:BCNWG3}. In practice, the uncertainty depends on the beam intensity and the acquisition conditions, and must be evaluated on a fill-by-fill basis; it typically translates into a 0.2--0.3\% uncertainty in the absolute luminosity scale.

\par
Because of the highly demanding bandwidth specifications dictated by single-bunch current measurements, the FBCT response is potentially sensitive to the frequency spectrum radiated by the circulating bunches, timing adjustments with respect to the RF phase, and bunch-to-bunch intensity or length variations. Dedicated laboratory measurements and beam experiments, comparisons with the response of other bunch-aware beam instrumentation (such as the ATLAS beam pick-up timing system), as well as the imposition of constraints on the bunch-to-bunch consistency of the measured visible cross-sections, resulted in a $<$\,0.04\% systematic luminosity-calibration uncertainty in the luminosity scale arising from the relative-intensity measurements~\cite{bib:BCNWG4, bib:CBthesis}.

\par
Additional corrections to the bunch-by-bunch population are made to correct for  {\em ghost charge} and {\em satellite bunches}. Ghost charge refers to protons that are present in nominally empty bunch slots at a level below the FBCT threshold (and hence invisible), but which still contribute to the current measured by the more accurate DCCT.
Highly precise measurements of these tiny currents (normally at most a few per mille of the total intensity) have been achieved~\cite{bib:CBthesis} by comparing the number of beam--gas vertices reconstructed by LHCb in nominally empty bunch slots, to that in non-colliding bunches whose current is easily measurable. For the 2012 luminosity-calibration fills, the ghost-charge correction to the bunch-population product ranges from $-0.21$\% to $-0.65$\%; its systematic uncertainty is dominated by that affecting the LHCb trigger efficiency for beam--gas events.

\par
Satellite bunches describe out-of-time protons present in collision bunch slots that are measured by the FBCT, but that remain captured in an RF bucket at least one period (2.5\,ns) away from the nominally filled LHC bucket. As such, they experience at most long-range encounters with the nominally filled bunches in the other beam. The best measurements are obtained using the longitudinal density monitor. This instrument uses avalanche photodiodes with 90\,ps timing resolution to compare the number of infrared synchrotron-radiation photons originating from satellite RF buckets, to that from the nominally filled buckets. The corrections to the bunch-population product range from $-0.03$\% to $-0.65$\%, with the lowest satellite fraction achieved in scans X--XV. The measurement techniques, as well as the associated corrections and systematic uncertainties, are detailed in Ref.~\cite{ bib:LDM}.

\subsection{Calibration Results}
\label{subsec:vdMResults}
\subsubsection{Summary of calibration corrections}
\label{subsubsec:calibCrctnSmry}
With the exception of the noise and single-beam background subtractions (which depend on the location, geometry and instrumental response of individual subdetectors), all the above corrections to the \vdM-calibrated visible cross-sections are intrinsically independent of the luminometer and luminosity algorithm considered. The beam-separation scale, as well as the orbit-drift and beam--beam corrections, impact the effective beam separation at each scan step; the non-factorization and emittance-growth corrections depend on the properties of each colliding bunch-pair and on their time evolution over the course of a fill; and corrections to the bunch-population product translate into an overall scale factor that is common to all scan sets within a given LHC fill. The mutual consistency of these corrections was explicitly verified for the LUCID\_EventOR  and BCM\_EventOR visible cross-sections, for which independently determined corrections are in excellent agreement. As the other algorithms (in particular track counting) are statistically less precise  during \vdM\ scans, their visible cross-sections are corrected using scale factors extracted from the LUCID\_EventOR scan analysis.

\par
\begin{figure}[htbp] 
   \centering
   \includegraphics[width=0.77 \columnwidth]{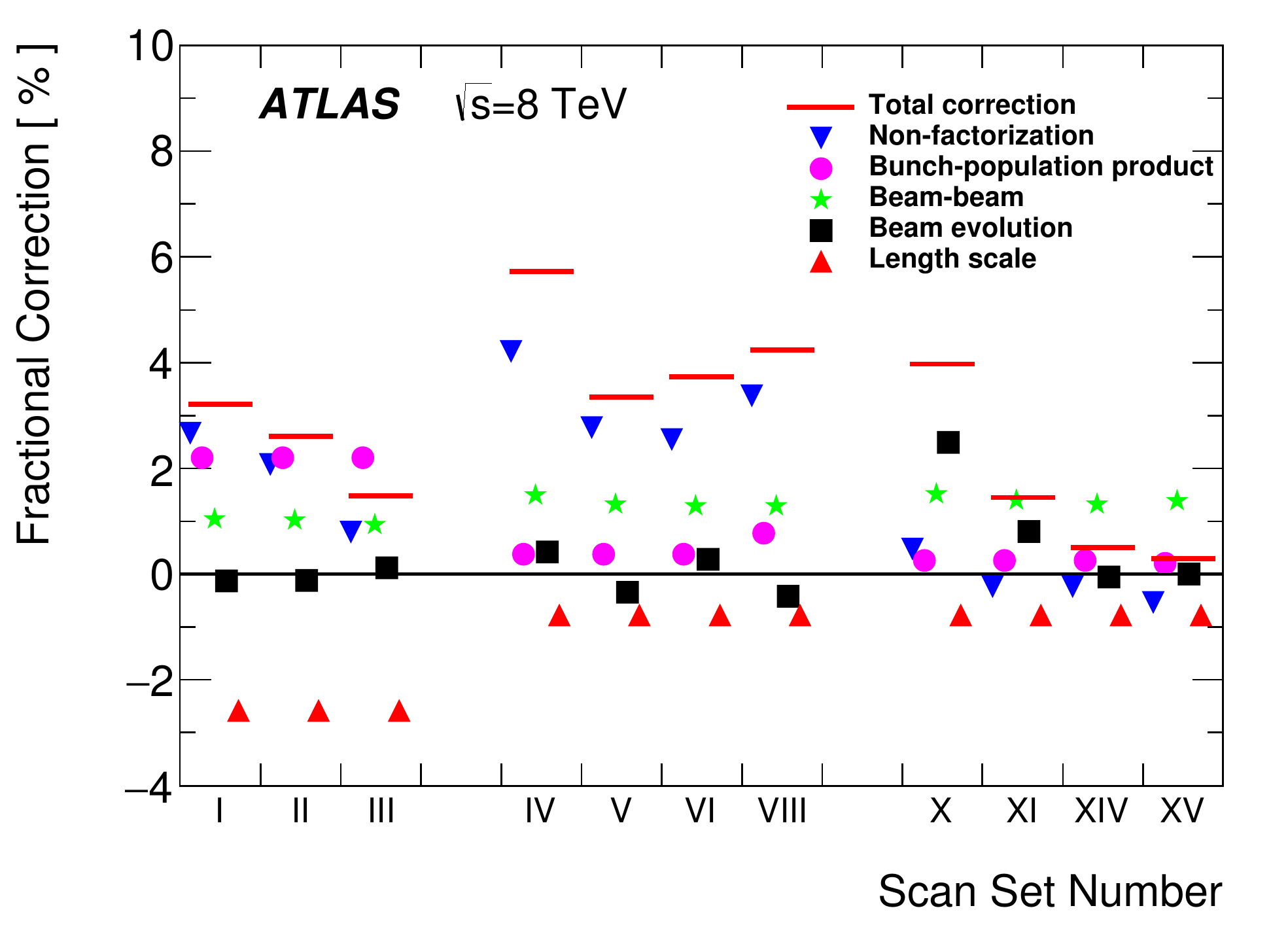} 
   \caption{Luminometer-independent corrections to the visible cross-sections calibrated by the van der Meer method, averaged over all colliding bunches and displayed separately for each scan set. The length--scale, beam--beam, non-factorization and bunch--population corrections are discussed in Sects.~\ref{subsec:LengthScale}, \ref{subsec:beambeam},~\ref{subsec:nonFctrztnCrctn} and \ref{subsec:currents}, respectively. The orbit--drift (Sect.~\ref{subsec:orbit}) and emittance--growth (Sect.~\ref{subsec:epsGrowth}) corrections are combined for clarity, and their cumulative effect is displayed as ``beam evolution''. The sum of all corrections is shown, for each scan set, by the red line.
}
   \label{fig:vdMCrctnsSmry}
\end{figure}

The dominant correction in scan sets I--VIII (Fig.~\ref{fig:vdMCrctnsSmry}) is associated with non-factorization; it is also the most uncertain, because it is sensitive to the vertex-position resolution, especially in scan sets I--III where the transverse luminous size is significantly smaller than the resolution. In contrast, non-factorization corrections are moderate in scan sets X--XV, suggesting a correspondingly minor contribution to the systematic uncertainty for the November scan session.

\par
The next largest correction in scan sets I--III is that of the beam-separation scale, which, because of different \bst settings, is uncorrelated between the April session and the other two sessions, and fully correlated across scan sets IV--XV (Sect.~\ref{subsubsec:LSCsyst}). The correction to the bunch-population product is equally shared among FBCT, ghost-charge and satellite corrections in scan sets I--III, and dominated by the ghost-charge subtraction in scans IV--XV. This correction is uncorrelated between scan sessions, but fully correlated between scan sets in the same fill.

\par
Of comparable magnitude across all scan sets, and partially correlated between them, is the beam--beam correction; its systematic uncertainty is moderate and can be calculated reliably (Sect.~\ref{subsubsec:bbsysts}). The uncertainties associated with orbit drifts (Sect.~\ref{subsubsec:orbitSyst}) and emittance growth (Sect.~\ref{subsubsec:epsGrowthSyst}) are small, except for scan set X where these corrections are largest.

\subsubsection{Consistency of \vdM\ calibrations across 2012 scan sessions}

\begin{figure}
\centering
\subfigure[]{
\includegraphics[width=0.77\textwidth]{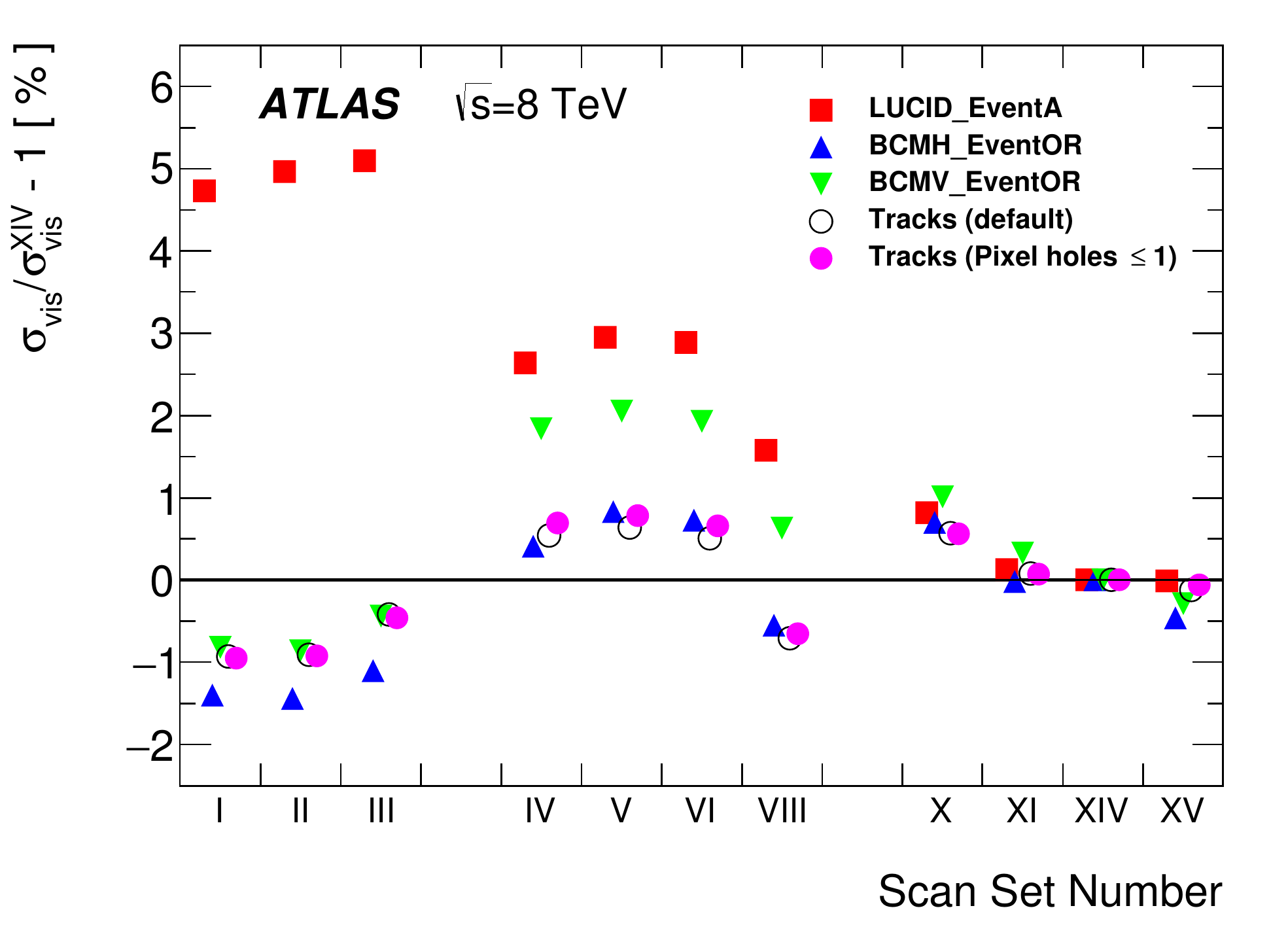}
}
\subfigure[]{
\includegraphics[width=0.77\textwidth]{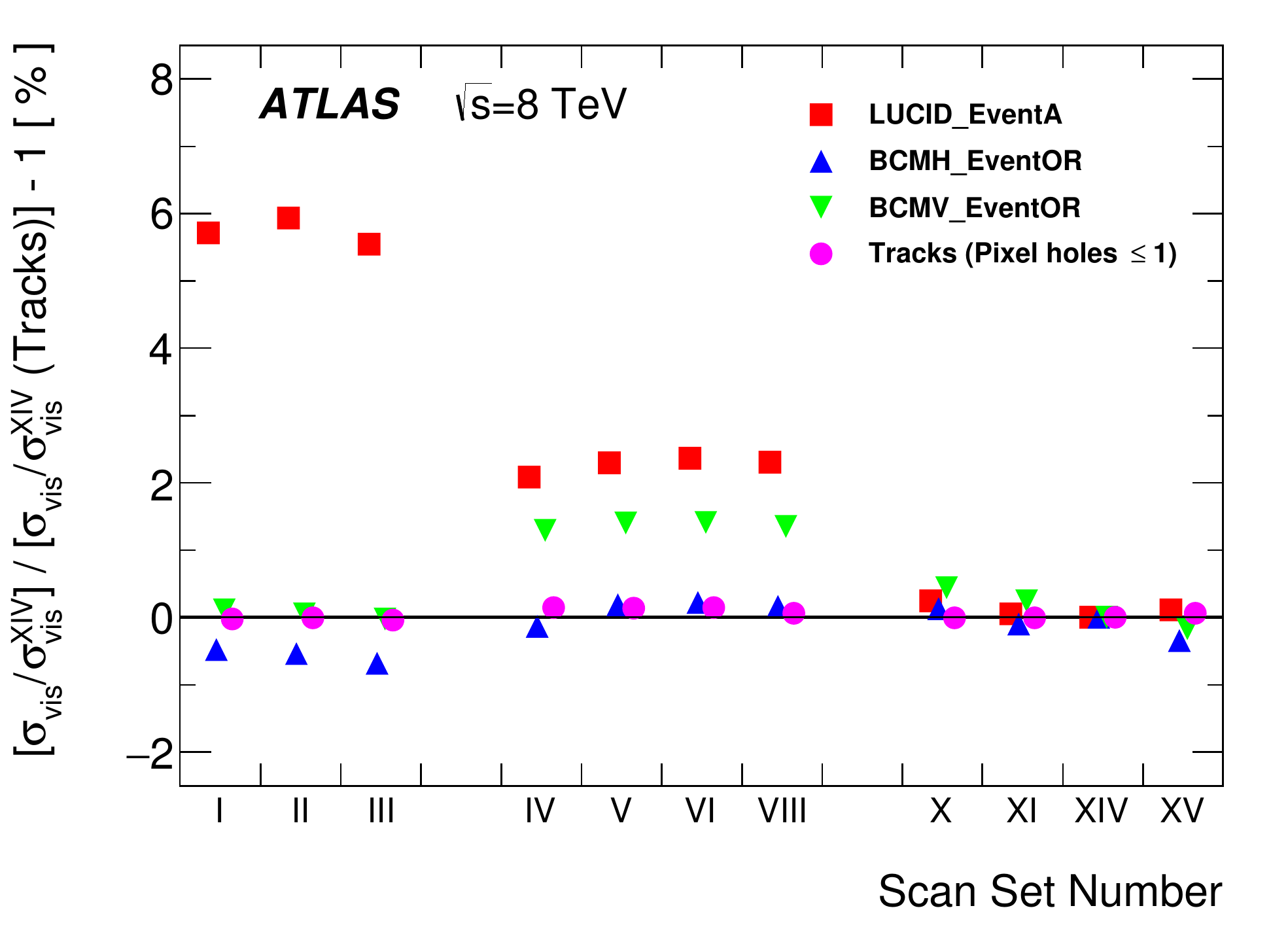}
}
\caption{(a) Stability of absolutely calibrated visible cross-sections across scan sets, as quantified by the ratio of the visible cross-section in a given scan set to that of  the same luminosity algorithm in scan set XIV.
(b) Relative instrumental stability of different luminosity algorithms across scan sets, as quantified by the ratio shown in (a) for a given algorithm, divided by the same ratio for the default track-counting algorithm.
}
\label{figvdMCalStblty}
\end{figure}

The relative stability of \vdM\ calibrations, across scan sets within a scan session and from one scan session to the next, can be quantified by the ratio $S^k_{\mathrm{calib},j}$ of the visible cross-section for luminosity algorithm $k$ ($k =$ BCMH\_EventOR, BCMV\_EventOR, LUCID\_EventA, ...) in a given scan set $j$ to that in a reference scan set, arbitrarily chosen as scan set XIV:
\begin{equation*}
S^k_{\mathrm{calib}, j} = \sigma^k_{\mathrm{vis}, j} / \sigma^k_{\mathrm{vis,\,XIV}}\, .
\end{equation*}
The ratio $S^k_{\mathrm{calib}, j}$ is presented in Fig.~\ref{figvdMCalStblty}(a) for a subset of BCM, LUCID and track-counting algorithms. Several features are apparent.
\begin{itemize}
\item
The visible cross-section associated with the LUCID\_EventA algorithm drops significantly between the April and July scan sessions, and then again between July and November. 
\item
For each algorithm separately, the \sigmavis variation across scan sets within a given LHC fill (scan sets I--III, IV--VI and X--XIV) remains below 0.5\%, except for scan set X which stands out by 1\%.
\item
The absolute calibrations of the BCMH\_EventOR and track-counting algorithms are stable to better than $\pm 0.8$\% across scan sets IV--VI and X--XV, with the inconsistency being again dominated by scan set  X. 
\item
Between scan sets IV--VI and X--XV, the calibrations of the track counting, BCMH\_EventOR and BCMV\_EventOR algorithms drop on the average by 0.5\%, 0.6\%  and 1.7\% respectively.
\item
The calibrations of the BCM\_EventOR (track-counting) algorithm in scan sets I--III and VIII are lower by up to 1.4\% (2\%) compared to the other scan sets. This structure, which is best visible in Fig.~\ref{fig:nonFactImpact}, is highly correlated across all algorithms. Since the corresponding luminosity detectors use very different technologies, this particular feature cannot be caused by luminometer instrumental effects.
\end{itemize}

\par
In order to separate purely instrumental drifts in the ATLAS luminometers from \vdM-calibration inconsistencies linked to other sources (such as accelerator parameters or beam conditions), Fig.~\ref{figvdMCalStblty}(b) shows the variation, across scan sets $j$, of the double ratio
\begin{equation*}
S^k_{\mathrm{instr}, j} = S^k_{\mathrm{calib}, j} /  S^{\mathrm{track~counting}}_{\mathrm{calib}, j} = 
          \frac{\sigma^k_{\mathrm{vis}, j} / \sigma^k_{\mathrm{vis,\, XIV}}}
                 {\sigma^{\mathrm{track~counting}}_{\mathrm{vis}, j} / \sigma^{\mathrm{track~counting}}_{\mathrm{vis,\, XIV}}}~~,
\end{equation*}
which quantifies the stability of algorithm $k$ relative to that of the default track-counting algorithm. Track counting is chosen as the reference here because it is the bunch-by-bunch algorithm whose absolute calibration is the most stable over time (Figs.~\ref{fig:nonFactImpact} and \ref{figvdMCalStblty}(a)), and that displays the best stability relative to all bunch-integrating luminosity algorithms during physics running across the entire 2012 running period (this is demonstrated in Sect.~\ref{subsec:relStbltyAll}). By construction, the instrumental-stability parameter $S^k_{\mathrm{instr}, j}$ is  sensitive only to instrumental effects, because the corrections described in Sects.~\ref{subsec:LengthScale}--\ref{subsec:currents} are intrinsically independent of the luminosity algorithm considered. The following features emerge.
\begin{itemize}
\item
For each algorithm individually, the instrumental stability is typically better than 0.5\% within each scan session. 
\item
The instrumental stability of both the ``Pixel holes $\le 1$'' selection and the vertex-associated track selection (not shown) is better than 0.2\% across all scan sets.
\item
Relative to track counting, the LUCID efficiency drops by 3.5\% between the April and July scan sessions, and by an additional 2.2\% between July and November. This degradation is understood to be caused by PMT aging. 
\item
The BCMH\_EventOR efficiency increases by about 0.7\% with respect to that of track counting between the April and July sessions, and then remains stable to within 0.2--0.4\% across the July and November sessions.
In contrast, the efficiency of the BCMV\_EventOR algorithm compared to that of track counting increases by about 1.3\% from April to July, and drops back to its original level by the November session. These long-term variations in the response of various subsets of diamond sensors in the low-luminosity regime of \vdM scans are possibly related to subtle solid-state physics effects arising from the combination of radiation damage during physics running~\cite{Aad:2013ucp, bib:CMS_BCM} and of partial annealing during beam-off and low-luminosity periods. Aging effects of comparable magnitude are observed at high luminosity (Sect.~\ref{sec:stability}).
\item
Given the 0.7\% relative stability, between scan sets I--III and IV--VI,  of the track-counting and BCMH\_EventOR calibrations (Fig.~\ref{figvdMCalStblty}(b)), the 1.4--2.0\% discrepancy, between the April and July \vdM-scan sessions, that affects the absolute calibrations of both the BCMH\_EventOR and the track-counting algorithms (Fig.~\ref{figvdMCalStblty}(a)) cannot be primarily instrumental in nature. 
The actual cause could not be identified with certainty. Since the transverse luminous size $\sigma_{\mathcal L}\xspace$ in the April session (Table~\ref{tab:vdmScan}) is approximately three times smaller than the vertex-position resolution, a plausible scenario is that a small error in the estimated resolution biases the reconstructed luminous size in such a way as to underestimate the non-factorization corrections $R_{\mathrm{NF}}$, and thereby the visible cross-sections, in scan sets I--III. 
\item
Similarly, the 1.3\% discrepancy, between scan sets IV--VI and scan set VIII, of the absolute calibrations of all algorithms (Fig.~\ref{figvdMCalStblty}(a)) cannot be instrumental either. Here however, the luminous size is 1.5 times larger than the resolution: resolution biases (if any) should be noticeably smaller than in the April scan session. But as scan sets IV--VIII were carried out in two consecutive LHC fills under very similar beam conditions, such biases should impact scan sets IV--VI and VIII in the same manner. 
\end{itemize}
\subsubsection{Final visible cross-sections for bunch-by-bunch luminosity algorithms}
\label{subsubsec:finSigvis}

The percent-level inconsistencies of the absolute calibrations between April and July and within the July session itself, as well as the excellent internal consistency of the November results for all algorithms (Fig.~\ref{figvdMCalStblty}(a)), suggest that the November calibrations are the most reliable. In addition, the calibrations extracted from scan sets I--VIII are affected by  several large adjustments that in some cases partially cancel (Fig.~\ref{fig:vdMCrctnsSmry}); of these the most uncertain are the non-factorization corrections, which affect the November scans much less. The cumulated magnitude of the corrections is also smallest for scan sets XI--XV (scan set X suffers from larger orbit-drift and emittance-growth corrections because of the long delay between the $x$ and $y$ scans).

\par
The combination of these arguments suggests that the visible cross-sections, averaged over all colliding bunches in each scan set and then averaged over scan sets XI--XV, should be adopted as the best estimate $\overline{\sigma}_{\mathrm{vis}}$ of the absolute luminosity scale at the time of, and applicable to the beam conditions during, the November 2012 \vdM session. Table~\ref{tab:sigvisresult} lists the $\overline{\sigma}_{\mathrm{vis}}$ values for the main luminosity algorithms considered in this paper; the associated systematic uncertainties are detailed in Sect.~\ref{sec:vdMerrors}. Transferring the BCM and LUCID calibrations to the high-luminosity regime of routine physics operation, and accounting for time-dependent variations in luminometer response over the course of the 2012 running period, is addressed in Sect.~\ref{subsec:highLCrctns}.
\begin{table}[htbp]
   	\centering
   	\begin{tabular}{p{6 cm}ccc} 
      		\hline
      		 Luminosity algorithm 				& $\overline{\sigma}_{\mathrm{vis}}$ [mb]  & Statistical uncertainty [\%] \\	
		\hline	
		BCMH\_EventOR					& $  5.0541 $				& 0.05		\\
		BCMV\_EventOR					& $  5.0202 $				& 0.06		\\
		LUCID\_EventOR					& $ 35.316  $				& 0.02		\\
		LUCID\_EventA					& $ 23.073  $				& 0.02		\\	
		LUCID\_EventC					& $ 20.422  $				& 0.02		\\	
		Track counting (Pixel holes $\le 1$)		& $243.19   $				& 0.14		\\
		Track counting (default) 				& $241.27   $				& 0.14		\\
		Track counting (vertex-associated)		& $226.24   $				& 0.14		\\
		\hline      
  	 \end{tabular}
	 \caption{Visible cross-sections averaged over scan sets XI--XV.}
   	\label{tab:sigvisresult}
\end{table}

\section{van der Meer calibration uncertainties}
\label{sec:vdMerrors}

 This section details the systematic uncertainties affecting the visible cross-sections reported in Table~\ref{tab:sigvisresult}. The contributions from instrumental effects (Sect.~\ref{subsec:instrumSysts}) are comparable in magnitude to those associated with beam conditions (Sect.~\ref{subsec:beamCondSysts}), while those from the bunch-population product (Sect.~\ref{subsec:bunchPopSysts}) are about three times smaller.  A summary is presented in Table~\ref{tab:vdMsysts}.

\subsection{Instrumental effects}
\label{subsec:instrumSysts}

\subsubsection{Reference specific luminosity}

For simplicity, the visible cross-section extracted from \vdM scans for a given luminometer utilizes the specific luminosity measured by that same luminometer. Since this quantity depends only on the convolved beam sizes, consistent results should be reported by all detectors and algorithms for a given scan set. 

\par
\begin{figure} 
   \centering
   \subfigure[]{\includegraphics[width=0.485\textwidth]{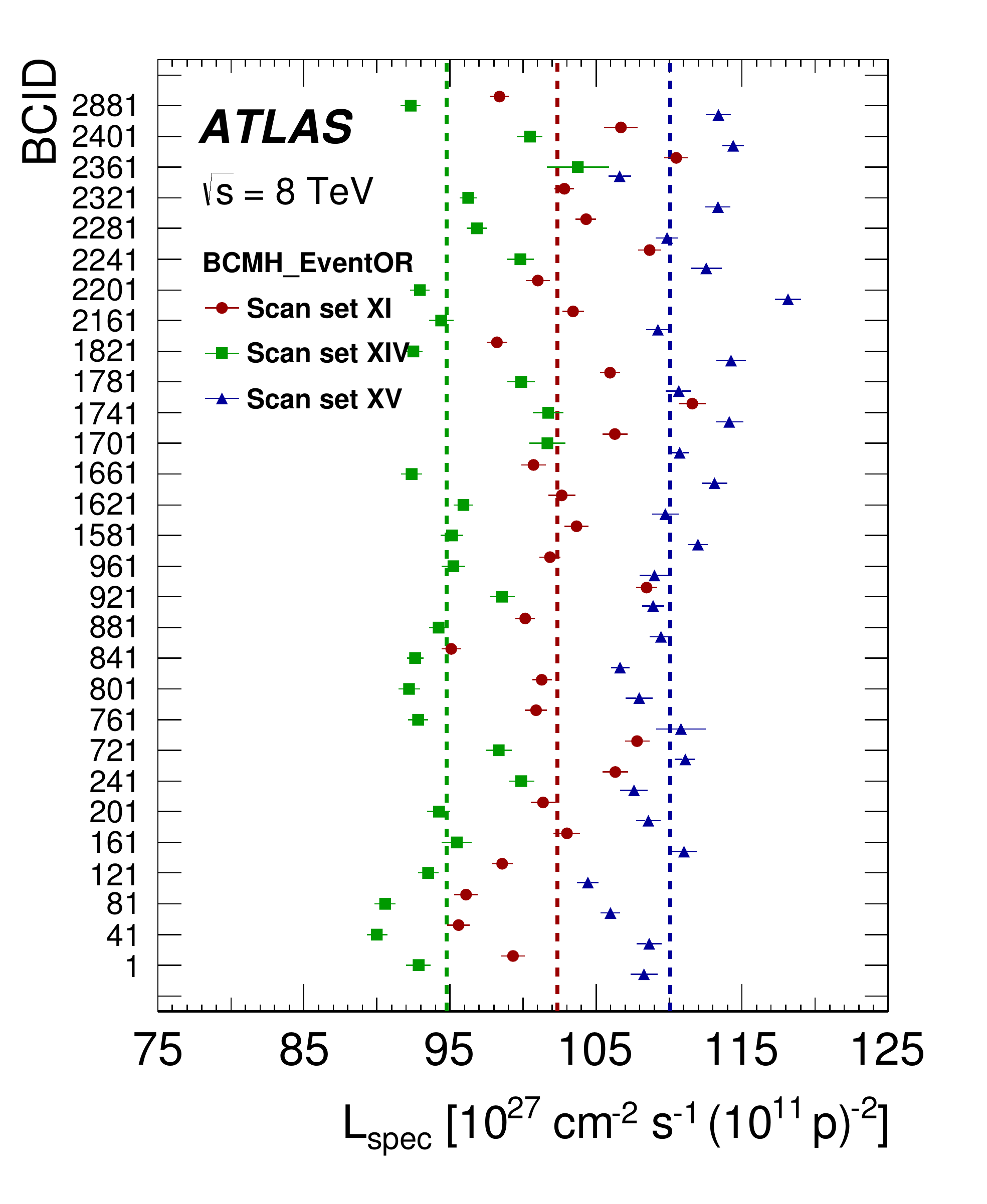}}
   \subfigure[]{\includegraphics[width=0.485\textwidth]{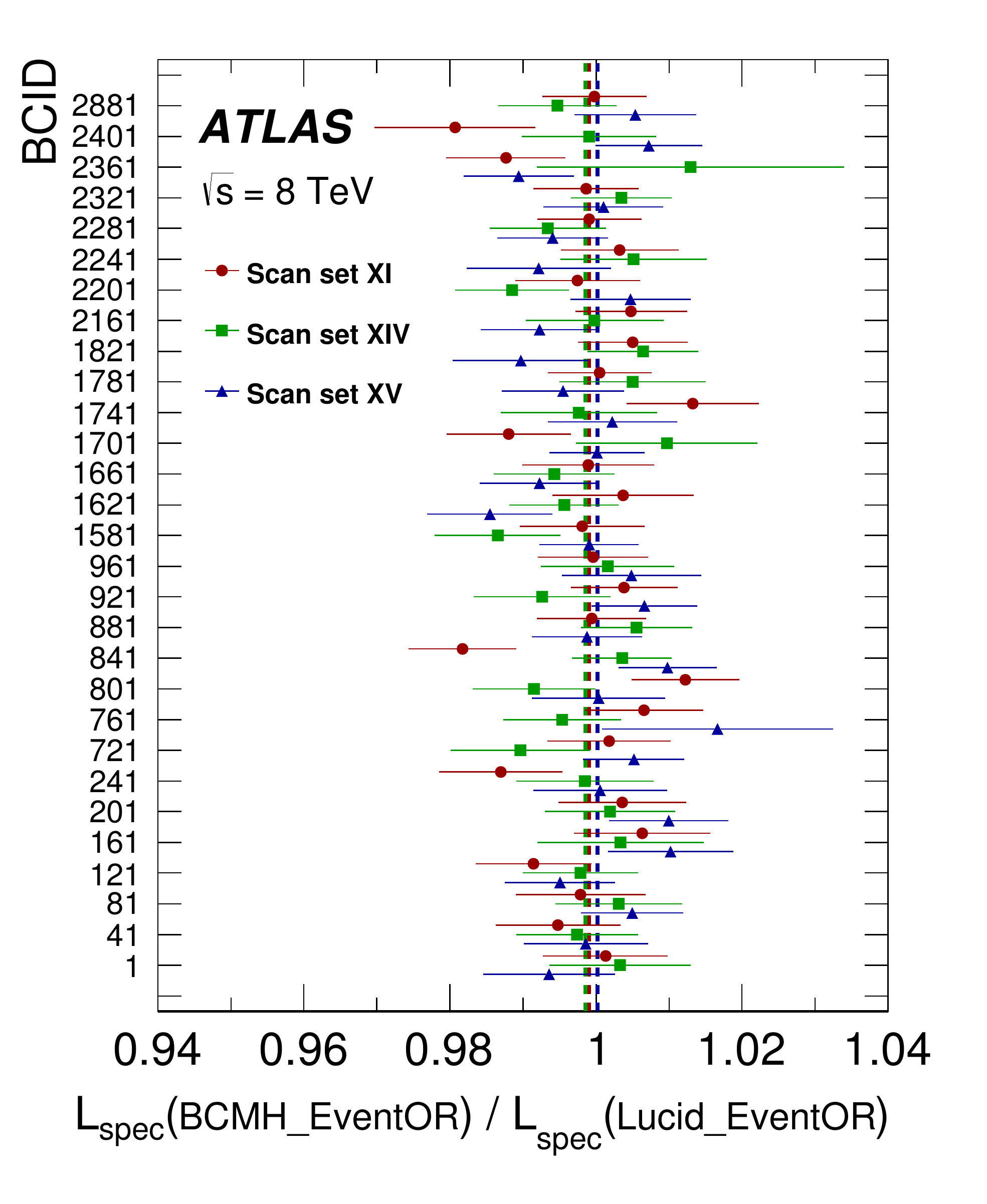}}
   \caption{ (a) Bunch-by-bunch specific luminosity for scan sets XI, XIV and XV determined using the BCMH\_EventOR algorithm. 
(b)  Bunch-by-bunch ratio of the  ${\cal L}_{\mathrm{spec}}$ values reported by the BCMH\_EventOR and LUCID\_EventOR algorithms. The vertical lines indicate the weighted average over BCIDs for the three scan sets separately. The error bars represent statistical uncertainties only.
}
   \label{fig:LSpec}
\end{figure}

Figure~\ref{fig:LSpec} compares the  ${\cal L}_{\mathrm{\mathrm{spec}}}$ values measured by two independent luminosity algorithms in three consecutive scan sets. Bunch-to-bunch variations of the specific luminosity are typically 5--10\% (Fig.~\ref{fig:LSpec}(a)), reflecting bunch-to-bunch differences in transverse emittance also seen during normal physics fills. A systematic reduction in ${\cal L}_{\mathrm{spec}}$ can be observed from scan XI to scan XIV, caused by emittance growth over the duration of the fill. Although the two algorithms appear statistically consistent for each bunch pair separately (Fig.~\ref{fig:LSpec}(b)), their bunch-averaged ratio systematically differs from unity by a small amount. The largest such discrepancy in scan sets XI--XV among the BCM, LUCID and track-counting algorithms amounts to 0.5\%  and is adopted as the systematic uncertainty associated with the choice of reference specific-luminosity value.

\subsubsection{Noise and background subtraction}

To assess possible uncertainties in the default subtraction scheme, an alternative fit is performed to data without applying the background-correction procedure of Sect.~\ref{subsec:scanBgds}, but interpreting the constant (\ie separation-independent) term in the fitting function as the sum of instrumental noise and single-beam backgrounds. The maximum difference observed between these two background treatments, averaged over scan sets XI--XV, amounts to less than 0.3\% (0.02\%) for the BCMH\_EventOR (LUCID\_EventOR) algorithm. A systematic uncertainty of $\pm 0.3$\% is thus assigned to the background-subtraction procedure during \vdM scans.

\subsubsection{Length-scale calibration}
\label{subsubsec:LSCsyst}

The length scale of each scan step enters the extraction of $\Sigma_{x,y}$ and hence
directly affects the absolute luminosity scale. The corresponding calibration procedure is described in Sect.~\ref{subsec:LengthScale}.  Combining in quadrature the statistical errors in the horizontal and vertical beam-separation scales (Table~\ref{tab:length}) yields a statistical uncertainty of $\pm0.08$\% in the length-scale product.

\par
The residual non-linearity visible in Fig.~\ref{fig:length}, and also observed in length-scale calibration scans performed in 2011, could be caused either by the power converters that drive the steering correctors forming the closed-orbit bumps, by the response of the steering correctors themselves, or by magnetic imperfections (higher multipole components) at large betatron amplitudes in the quadrupoles located within those orbit bumps. The potential impact of such a non-linearity on the luminosity calibration is estimated to be less than 0.05\%.

\par
Another potential source of bias is associated with orbit drifts. These were monitored during each of the four length-scale scans using the method outlined in Sect.~\ref{subsec:orbit}, revealing no significant drift. Small inconsistencies in the transverse beam positions extrapolated to the IP from the BPMs in the left and right arcs are used to set an upper limit on the potential orbit drift, during each scan, of the beam being calibrated, resulting in an overall $\pm 0.4$\% uncertainty in the length-scale product and therefore in the visible cross-section.

\subsubsection{Absolute length scale of the inner detector}
\label{subsubsec:LSCsyst2}

The determination of the beam-separation scale is based on comparing the scan step requested by the LHC control system with the actual transverse displacement of the luminous centroid measured by ATLAS.
This measurement relies on the length scale of the ATLAS inner detector tracking system 
(primarily the Pixel detector) being correct in measuring displacements of vertex positions
away from the centre of the detector.
The determination of the uncertainty in this absolute length scale is described in Ref.~\cite{Aad:2013ucp}; its impact amounts to a systematic uncertainty of $\pm0.3$\% in the visible cross-section.

\subsection{Beam conditions}
\label{subsec:beamCondSysts}

\subsubsection{Orbit drifts during \vdM scans}
\label{subsubsec:orbitSyst}

The systematic uncertainty associated with orbit drifts is taken as half of the correction described in Sect.~\ref{subsec:orbit}, averaged over scan sets XI--XV. It translates into a $\pm 0.1$\% systematic uncertainty in $\overline{\sigma}_{\mathrm{vis}}$. Because the sign and amplitude of the orbit drifts vary over time, this uncertainty is uncorrelated with that affecting the length-scale calibration.

\subsubsection{Beam-position jitter}

At each step of a scan, the actual beam separation may be affected by random deviations of the beam positions from their nominal settings, which in turn induce fluctuations in the luminosity measured at each scan point. The magnitude of this potential jitter was evaluated from the variation between consecutive measurements, a few seconds apart, of the relative beam separation at the IP extracted from single-beam orbits measured by BPMs in the nearby LHC arcs and extrapolated to the IP (Sect.~\ref{subsec:orbit}). The typical jitter in transverse beam separation observed during the November scan session amounts to 0.75\,$\muup$m RMS. The resulting systematic uncertainty in \sigmavis\ is obtained by random Gaussian smearing of the nominal separation by this amount, independently at each scan step, in a series of simulated scans. The RMS of the resulting fluctuations in fitted visible cross-section yields a $\pm 0.2$\% systematic uncertainty associated with beam-position jitter.

\subsubsection{Beam--beam corrections}
\label{subsubsec:bbsysts}

For given values of the bunch intensity and transverse convolved beam sizes, which are precisely measured, the deflection-induced orbit distortion and the relative variation of $\beta^\star$ are both proportional to $\beta^\star$ itself; they also depend on the fractional tune. Assigning a $\pm 20$\% uncertainty to each $\beta$-function value at the IP and a $\pm 0.01$ upper limit to each tune variation results in a $\pm 0.28$\% uncertainty in $\sigma_{\mathrm{vis}}$.
This uncertainty is computed with the conservative assumption that $\beta$-function and tune uncertainties are correlated between the horizontal and vertical planes, but uncorrelated between the two LHC rings.

\subsubsection{Fit model}

The choice of the fit function is arbitrary, but guided by the requirement that the fit provides faithful measurements of the integral under the luminosity-scan curve and of the rate at zero beam separation. The choice of functional form therefore depends on the underlying shapes of the colliding bunches, as manifested in the beam-separation dependence of the luminosity. Scan sets I--VIII are best modelled using a double Gaussian function plus a constant. The beam shapes are different in scan sets X--XV~\cite{bib:injBunchPrep}: here the best fit is obtained using a Gaussian function multiplied by a sixth-order polynomial. Additional fits are performed with different model assumptions: a super-Gaussian function, and a Gaussian function multiplied by a fourth-order polynomial (plus a constant term in all cases).
 The maximum fractional difference between the results of these different fits, across scan sets XI--XV and across the BCM, LUCID and track-counting algorithms,  amounts to 0.5\%. This value is assigned as the uncertainty associated with the fit model.

\subsubsection{Non-factorization correction}
\label{subsubsec:nonFactSyst}

The non-factorization corrections extracted from the luminous-region analysis (Sect. \ref{subsubsec:lumRegAna}) and the non-factorizable \vdM fits (Sect.~\ref{subsubsec:cpldFits}), are consistent to within 0.5\% or less in all scan sets. This value is chosen as the systematic uncertainty associated with non-factorization biases in the November scans.

\subsubsection{Emittance-growth correction}
\label{subsubsec:epsGrowthSyst}

The uncertainty in the correction described in Sect.~\ref{subsec:epsGrowth} is estimated as the largest difference in the scan-averaged correction for extreme choices of reference times, and amounts to $\pm 0.1$\% in $\overline{\sigma}_{\mathrm{vis}}$.

\subsubsection{Consistency of bunch-by-bunch visible cross-sections}

The calibrated \sigmavis\ value associated with a given luminometer and algorithm should be a universal scale factor independent of beam conditions or BCID. The variation in $\sigma_{\mathrm{vis}}$ across colliding-bunch pairs in a given scan set, as well as between scan sets, is used to quantify the reproducibility and stability of the calibration procedure during a scan session. 

\par
\begin{figure}
\centering
\subfigure[]{
\includegraphics[width=0.485\textwidth]{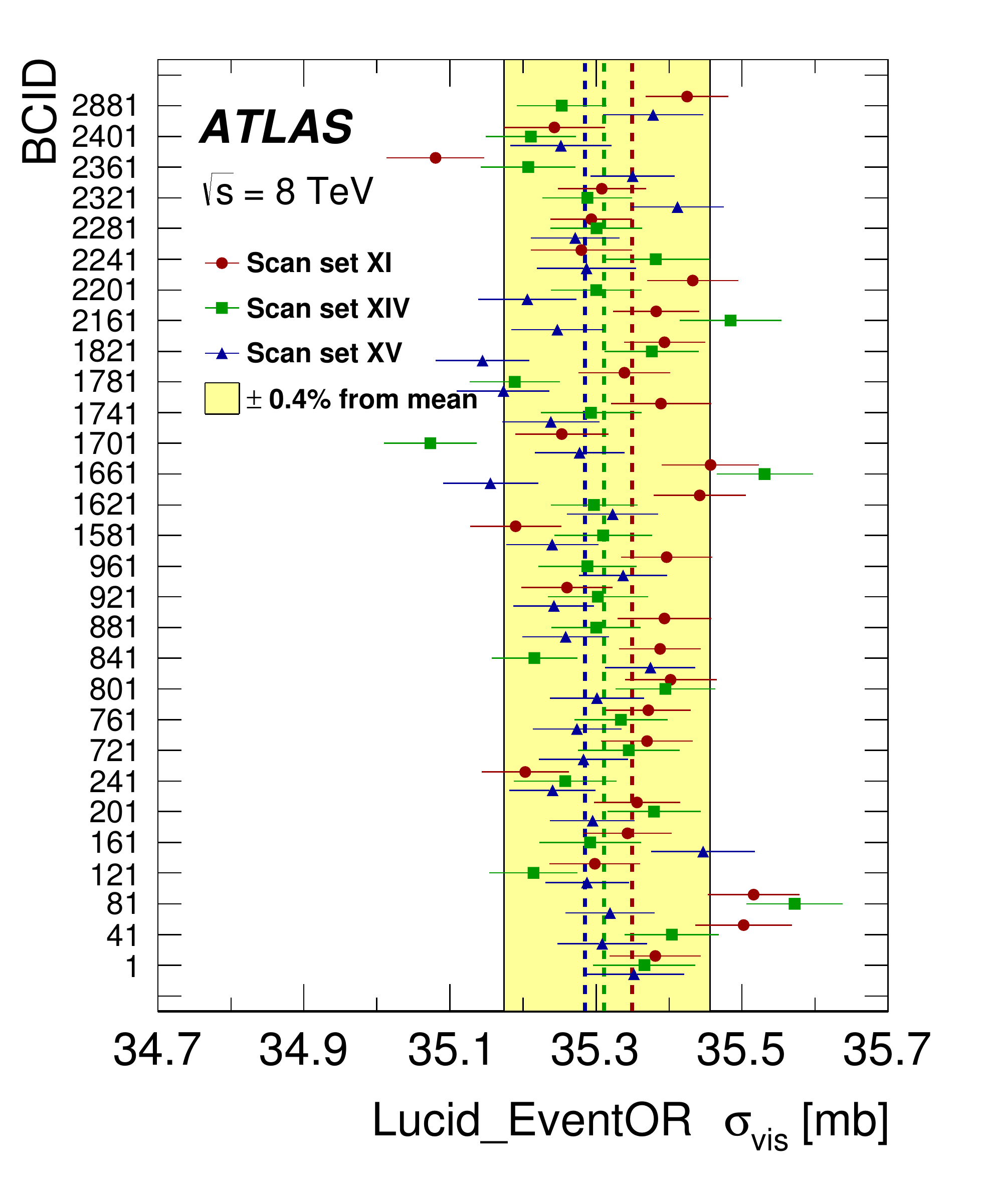}
}
\subfigure[]{
\includegraphics[width=0.485\textwidth]{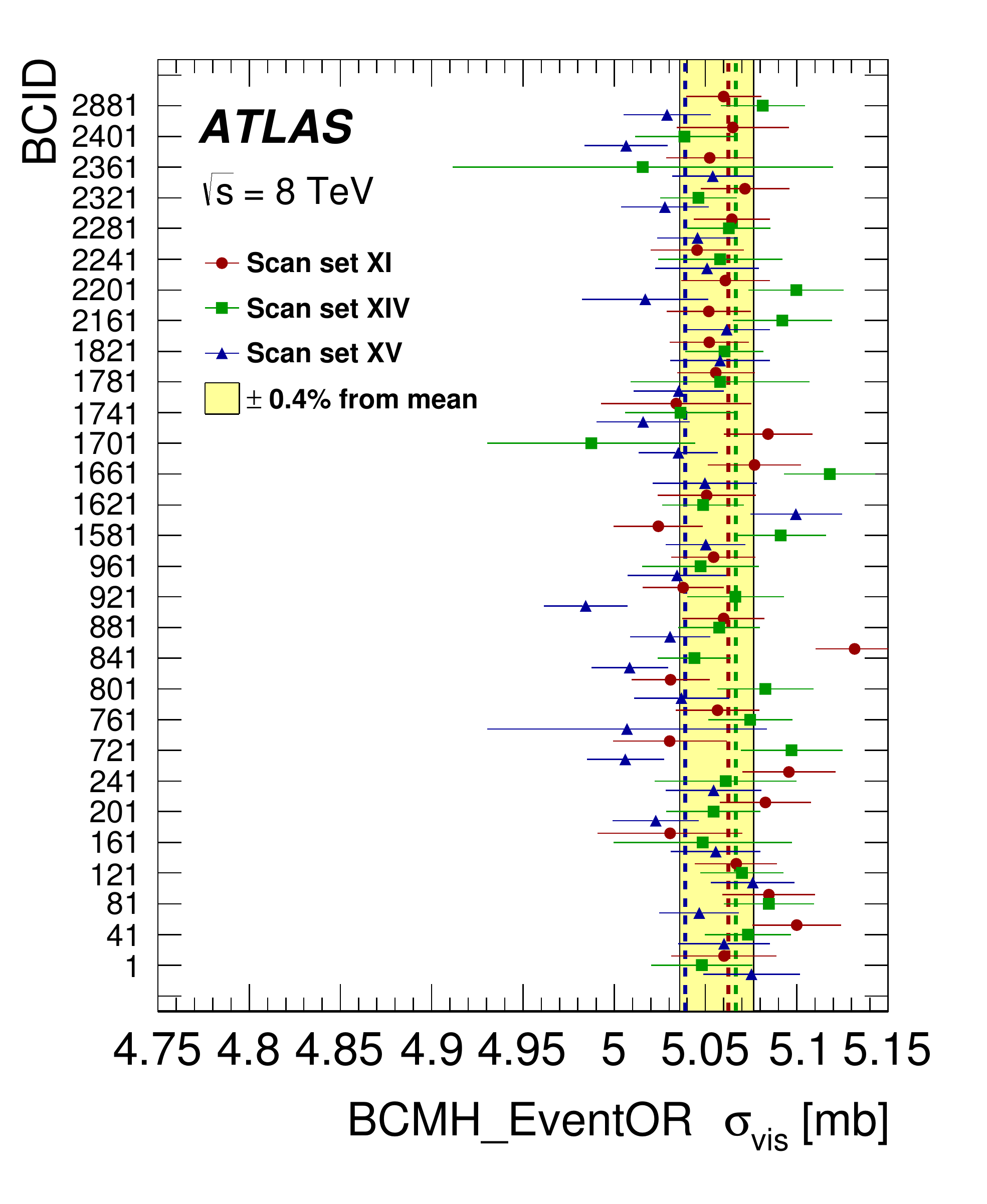}
}
\caption{Bunch-by-bunch \sigmavis\ values measured in scan sets XI, XIV, and XV for the (a) LUCID\_EventOR and (b) BCMH\_EventOR algorithm. The error bars are statistical only. The vertical lines represent the weighted average over colliding-bunch pairs, separately for each scan set. The shaded band indicates a  $\pm 0.4\%$ variation from the average, which is the sum in quadrature of the systematic uncertainties associated with bunch-by-bunch and scan-to-scan \sigmavis\ consistency.
 }  
\label{fig:bbbSigvis}
\end{figure}

The comparison of Figs.~\ref{fig:bbbSigvis}(a) and \ref{fig:bbbSigvis}(b) for scan sets XI, XIV and XV suggests that some of the \sigmavis\ variation from one bunch pair to the next is not statistical in nature, but rather correlated across bunch slots. The non-statistical component of this variation, \ie the difference in quadrature between the RMS bunch-by-bunch variation of \sigmavis\ within a given scan set and the average statistical uncertainty affecting a single-BCID \sigmavis\ measurement, is taken as a systematic uncertainty in the calibration technique. 
The largest such difference across scan sets XI--XV, evaluated using the measured LUCID\_EventOR visible cross-section, amounts to 0.23\%. The RMS bunch-by-bunch fluctuation of the BCM cross-sections is, in all cases but one, slightly smaller than the corresponding bunch-averaged statistical uncertainty, indicating that the statistical sensitivity of the BCM algorithms is insufficient to provide a reliable estimate of this uncertainty; the LUCID result is therefore adopted as a measure of the \sigmavis\ bunch-by-bunch consistency.

\subsubsection{Scan-to-scan reproducibility}
\label{subsubsec:sTsReprod}

The reproducibility of the visible cross-sections across the selected November scan sets, as illustrated in
Fig.~\ref{figvdMCalStblty}(a), is used as a measure of the residual inconsistencies potentially associated with imperfect correction procedures and unidentified sources of non-reproducibility. The largest such difference in visible cross-section between scan sets XI--XV, as reported by any of the BCM\_EventOR, LUCID\_EventOR or track-counting algorithms, amounts to $\pm 0.31$\%.

\subsection{Bunch-population product}
\label{subsec:bunchPopSysts}

The determination of this uncertainty ($\pm 0.24$\%) is discussed in Sect.~\ref{subsec:currents} and summarized in Table~\ref{tab:BCNWG}.

\subsection{Summary of van der Meer calibration uncertainties}
\label{subsec:vdMSystSmry}

The systematic uncertainties affecting the November 2012 \vdM calibration are summarized in Table~\ref{tab:vdMsysts}; they apply equally to all \vdM-calibrated luminosity algorithms. The statistical uncertainties, in contrast, are algorithm dependent (Table~\ref{tab:sigvisresult}), but small by comparison. 

\par
The uncertainties affecting the April and July 2012 calibrations have not been evaluated in detail. Most of them would be of comparable magnitude to their November counterparts, except for additional sizeable contributions from the non-factorization effects and scan-to-scan inconsistencies discussed in Sect.~\ref{subsec:vdMResults}.

\begin{table}[htbp]
   	\centering
      \vspace{3 mm}
   	\begin{tabular}{p{6 cm}cc} 
		\hline
      		Source 								& 				Uncertainty [\%]  		\\
		\hline	
		\hline
		Reference specific luminosity			& 					0.50				\\
		Noise and background subtraction		& 					0.30				\\
		Length-scale calibration					& 					0.40				\\
		Absolute ID length scale				& 					0.30				\\
		\hline
		Subtotal, instrumental effects			&					0.77				\\
		\hline
		\hline
		Orbit drifts							& 					0.10				\\
		Beam-position jitter					& 					0.20				\\
		Beam--beam corrections				& 					0.28				\\
		Fit model								& 					0.50				\\
		Non-factorization correction				& 					0.50				\\
		Emittance-growth correction				& 					0.10				\\
		Bunch-by-bunch \sigmavis\ consistency		
											& 					0.23				\\
		Scan-to-scan consistency				&					0.31				\\
		\hline
		Subtotal, beam conditions				&					0.89				\\
		\hline
		\hline
		Bunch-population product				& 					0.24				\\
		\hline
		\hline
		Total								& 					1.20				\\
		 \hline
  	 \end{tabular}
	 \caption{Fractional systematic uncertainties affecting the visible cross-section $\overline{\sigma}_{\mathrm{vis}}$ averaged over \vdM scan sets XI--XV (November 2012).}
   	\label{tab:vdMsysts}
\end{table}

\section{Consistency of relative-luminosity measurements during physics running}
\label{sec:stability}

\begin{figure}[htbp] 
   \centering
   \includegraphics[width=0.77\textwidth]{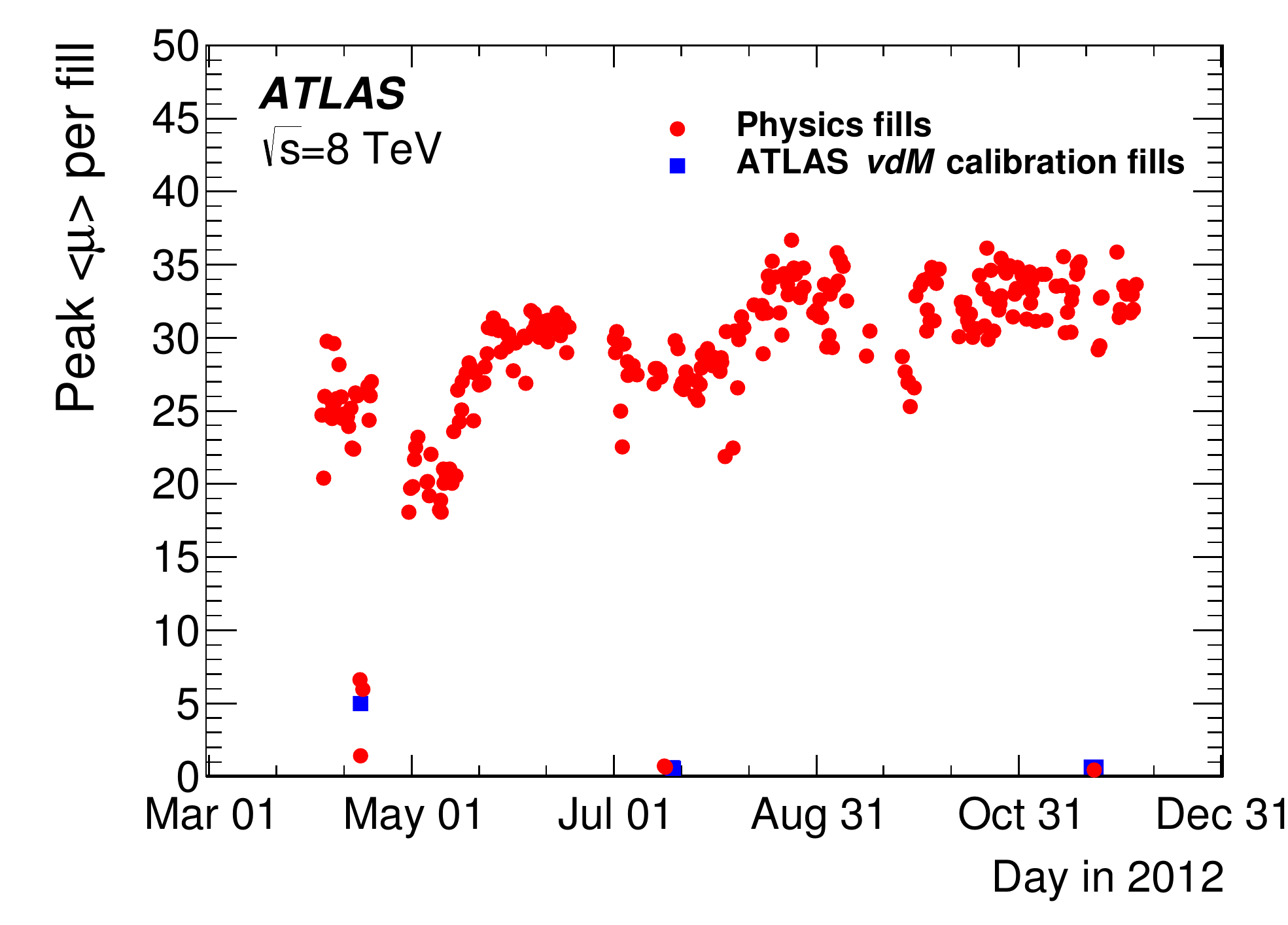} 
   \caption{History of the peak bunch-averaged pile-up parameter $\langle \mu \rangle$ during 2012, restricted to stable-beam periods. 
}
\label{fig:muvstime}
\end{figure}

The calibration of $\overline{\sigma}_{\mathrm{vis}}$ was performed at only a few points in time (Table~\ref{tab:vdmScan}), and at values of $\mu$ low compared to the pile-up levels routinely encountered during physics operation (Fig.~\ref{fig:muvstime}).
In this section, the stability of the luminosity measurement over the 2012 high-luminosity data sample is characterized from two distinct viewpoints: time stability of the relative response of various luminosity algorithms across the entire running period (Sect.~\ref{subsec:relStbltyAll}), and linearity of the calibrated luminosity values with respect to the actual pile-up parameter $\mu$ (Sect.~\ref{subsec:mudependence}). The relative consistency across all available luminosity detectors and algorithms is used to assess the robustness of the results and to quantify systematic variations in the response of the various luminometers.

\subsection{Relative stability of luminosity measurements over time}
\label{subsec:relStbltyAll}

\subsubsection{Consistency within individual luminometer subsystems}
\label{subsubsec:internDetStab}

\begin{figure}
\centering
\subfigure[]{
\includegraphics[width=0.485\textwidth]{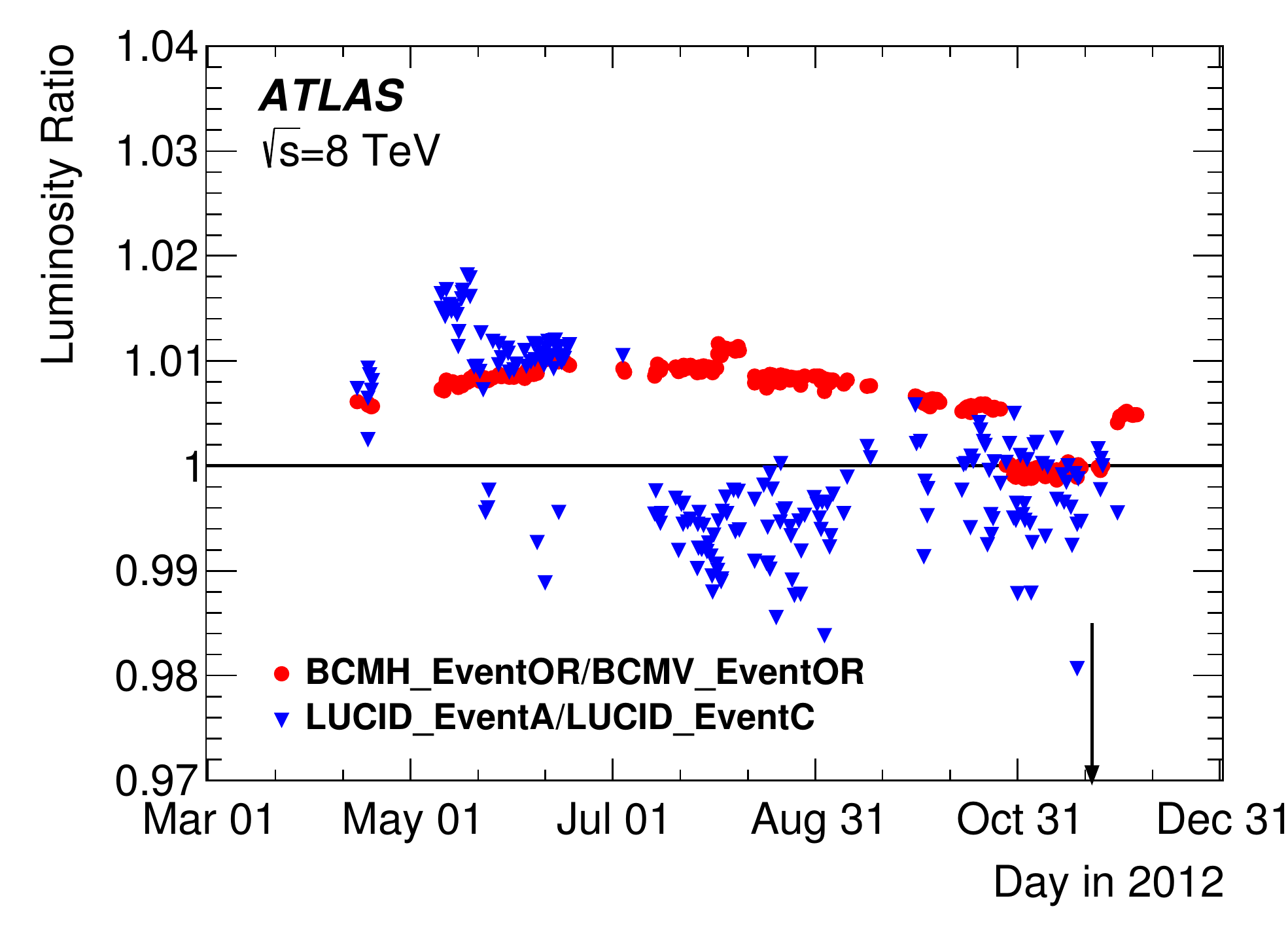}
}
\subfigure[]{
\includegraphics[width=0.485\textwidth]{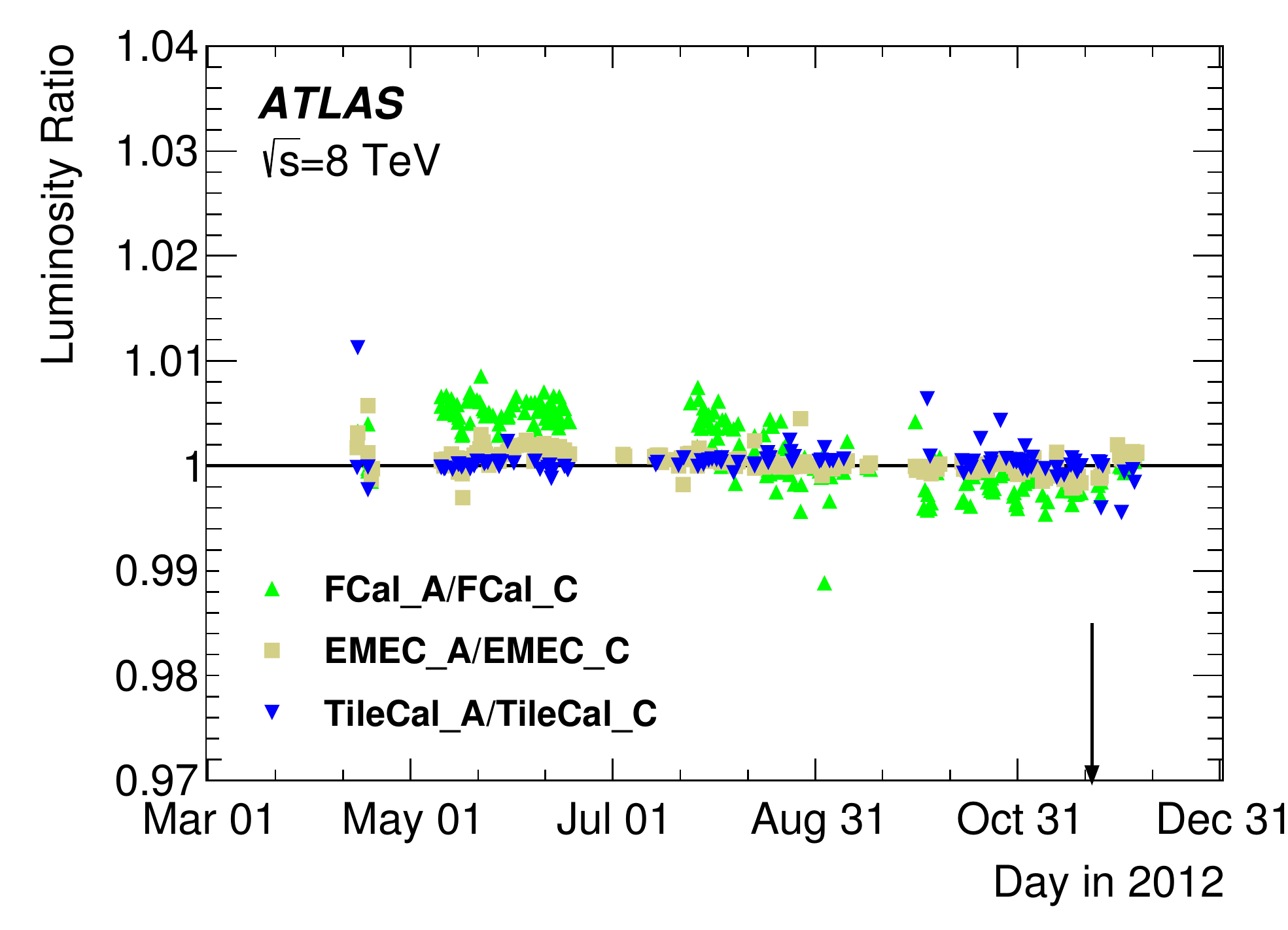}
}
\subfigure[]{
\includegraphics[width=0.485\textwidth]{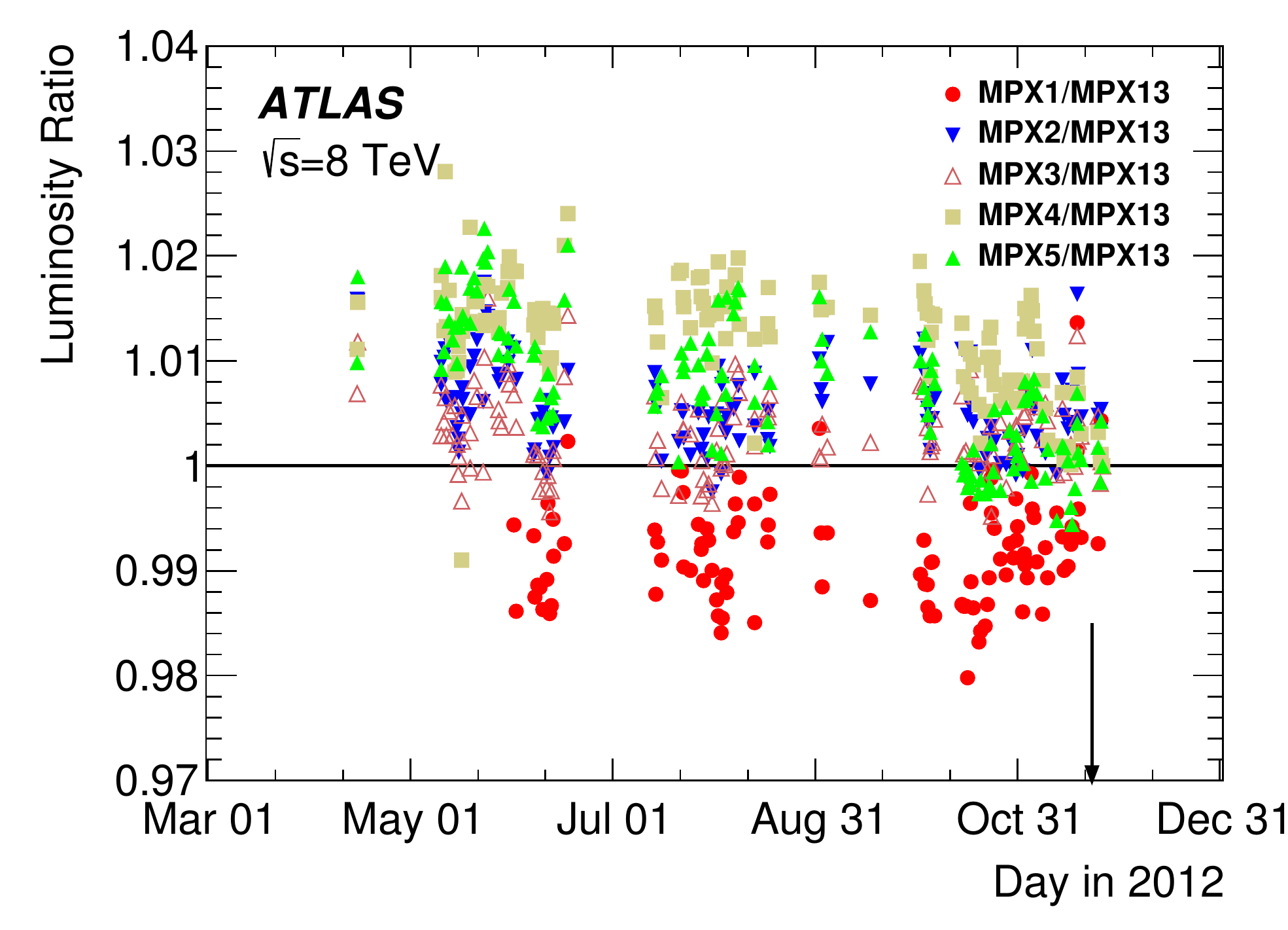}
}
\caption{(a) History of the ratio of the integrated luminosities per run reported by the BCM inclusive--OR algorithms (BCMV\_EventOR/BCMH\_EventOR)  and by the LUCID single-arm algorithms (LUCID\_EventA/LUCID\_EventC), during routine physics operation at high luminosity.
(b) History of the ratio of the integrated luminosities per run reported by the A and C arms of the electromagnetic endcap (EMEC), hadronic (TileCal) and forward (FCal) calorimeters. 
(c) History of the ratio of the integrated luminosities per run reported by five of the six individual MPX sensors,  to that reported by the sixth sensor in the same run. 
In all figures, each point shows the ratio for a single run relative to that in a reference run taken on November 25, 2012 (LHC fill 3323). Statistical uncertainties are negligible. The vertical arrows indicate the time of the November 2012 \vdM scan session.
}  
\label{fig:internStab}
\end{figure}

Figure~\ref{fig:internStab}(a) illustrates the internal consistency of the luminosity values reported by independent bunch-by-bunch algorithms during the 2012 running period, noise- and afterglow-subtracted as described in Sect.~\ref{subsec:scanBgds}, then summed over all colliding bunches and integrated over the stable-beam period in each ATLAS run. In order to better illustrate their relative time evolution, these run-integrated luminosity ratios are shown {\em anchored}, \ie normalized to the corresponding ratio in a high-luminosity run close in time to the November \vdM-scan session.

\par
During most of 2012, the ratio of the luminosity values reported by the horizontal and vertical pairs of BCM sensors is stable within a $\pm 0.4\%$ envelope, with the notable exception of a sharp $-0.6$\% step, lasting approximately one month, during which the BCM was affected by electronic noise (Sect.~\ref{subsec:scanBgds}). While during physics operation the noise itself has a negligible impact on the measured luminosity, its onset was accompanied by step changes in the response of individual diamond sensors; similar efficiency shifts in the opposite direction were observed when the noise disappeared, a few days after the November \vdM session. 

\par
The history of the luminosity ratio between the A and C arms of LUCID exhibits two distinct bands, each with a peak-to-peak scatter of up to $\pm 0.8$\% and separated by 1.5\% on the average. The step change in late June 2012 is associated with turning off two PMTs in the C arm, which were drawing excessive current. To mitigate the impact of this operational change on the LUCID performance, the LUCID luminosity before (after) this step change is determined using the April (November) 2012 \vdM calibrations.

\par
While relative efficiency variations among individual BCM sensors, or between the two LUCID arms, can be monitored using such internal luminosity ratios, quantifying 
the associated shifts in their absolute calibration requires an external reference.  
This can be provided, for instance, by the calorimeter- or MPX-based hit-counting luminosity algorithms presented in Sect.~\ref{subsec:bunchIntegDets}. Among these, the best internal performance is offered by the EMEC and the TileCal: in the high-luminosity regime, both achieve an arm-to-arm consistency better than $\pm 0.4$\% across the 2012 running period (Fig.~\ref{fig:internStab}(b)). 
The two FCal arms display a relative drift of about 1\% which is highly correlated among all channels in each arm. The run-to-run spread of the MPX luminosity ratios (Fig.~\ref{fig:internStab}(c)) lies in the 2\% range.

\par
While calorimeter algorithms lack sensitivity in the \vdM-calibration regime, the track-counting method  can be absolutely calibrated with a precision comparable to that of the BCM and LUCID algorithms (Table~\ref{tab:sigvisresult}). As demonstrated below, it also offers competitive precision for the run-integrated luminosity\footnote{Except for \vdM-scan sessions, track--counting-based luminosity measurements on shorter time scales (a few luminosity blocks), or on a bunch-by-bunch basis, are statistically limited by the available data-acquisition bandwidth.} during physics operation, thereby providing additional constraints on the performance of the other bunch-by-bunch algorithms. 

\par
\begin{figure}[htbp] 
   \centering
   \includegraphics[width=0.77\textwidth]{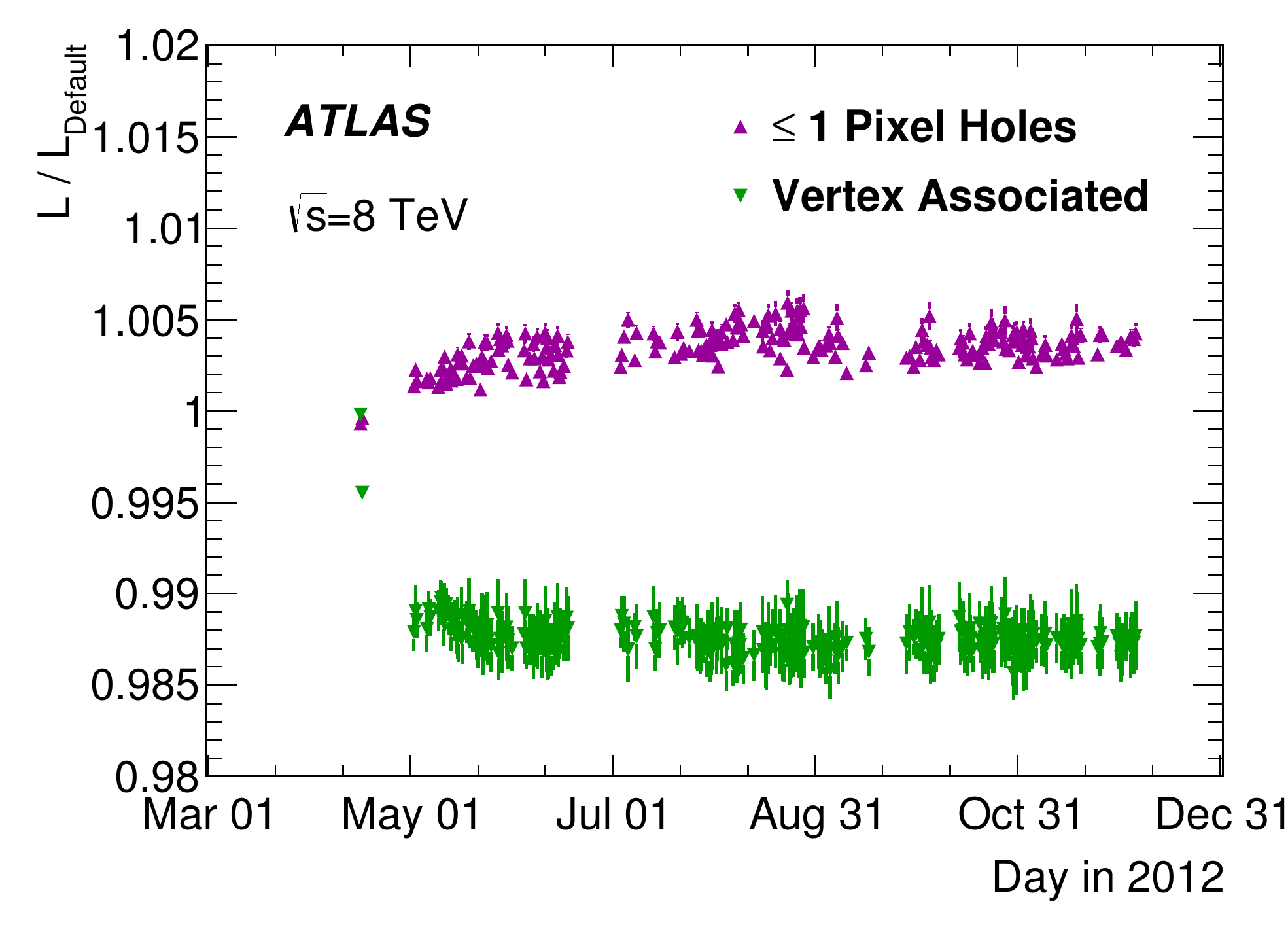} 
   \caption{History of the integrated-luminosity values reported by the two alternative track-counting methods, normalized to that from the default track selection, each as absolutely calibrated by the \vdM method. 
Each point represents the mean over a single ATLAS run. The error bars reflect the systematic uncertainty associated with the simulation-based fake-track subtraction. No track-counting data are available prior to the first \vdM-scan session (16 April 2012).
}
\label{fig:trkWPvsTime}
\end{figure}

Figure~\ref{fig:trkWPvsTime} displays the history of the luminosity reported by the two alternative track-counting working points introduced in Sect.~\ref{subsec:IDalgos}, normalized to that from the default WP. In contrast to what is presented in Fig.~\ref{fig:internStab}, these ratios are not anchored, but directly reflect the relative response of the three algorithms as calibrated in the November 2012 \vdM-scan session. While the three working points are consistent within 0.2\% at the very beginning of the 2012 running period (which corresponds to the April \vdM-scan session), counting vertex-associated tracks results, during most of the year, in a luminosity value lower by about 1.3\% compared to the other two WPs. Comparison with the history of the mean pile-up parameter (Fig.~\ref{fig:muvstime}) suggests that this inconsistency is not time-related but $\mu$-dependent, as further discussed in Sect.~\ref{subsec:mudependence}.

\subsubsection{Consistency between luminometer subsystems}
\label{subsubsec:crossDetStab}

\begin{figure}
\centering
\subfigure[]{
\includegraphics[width=0.485\textwidth]{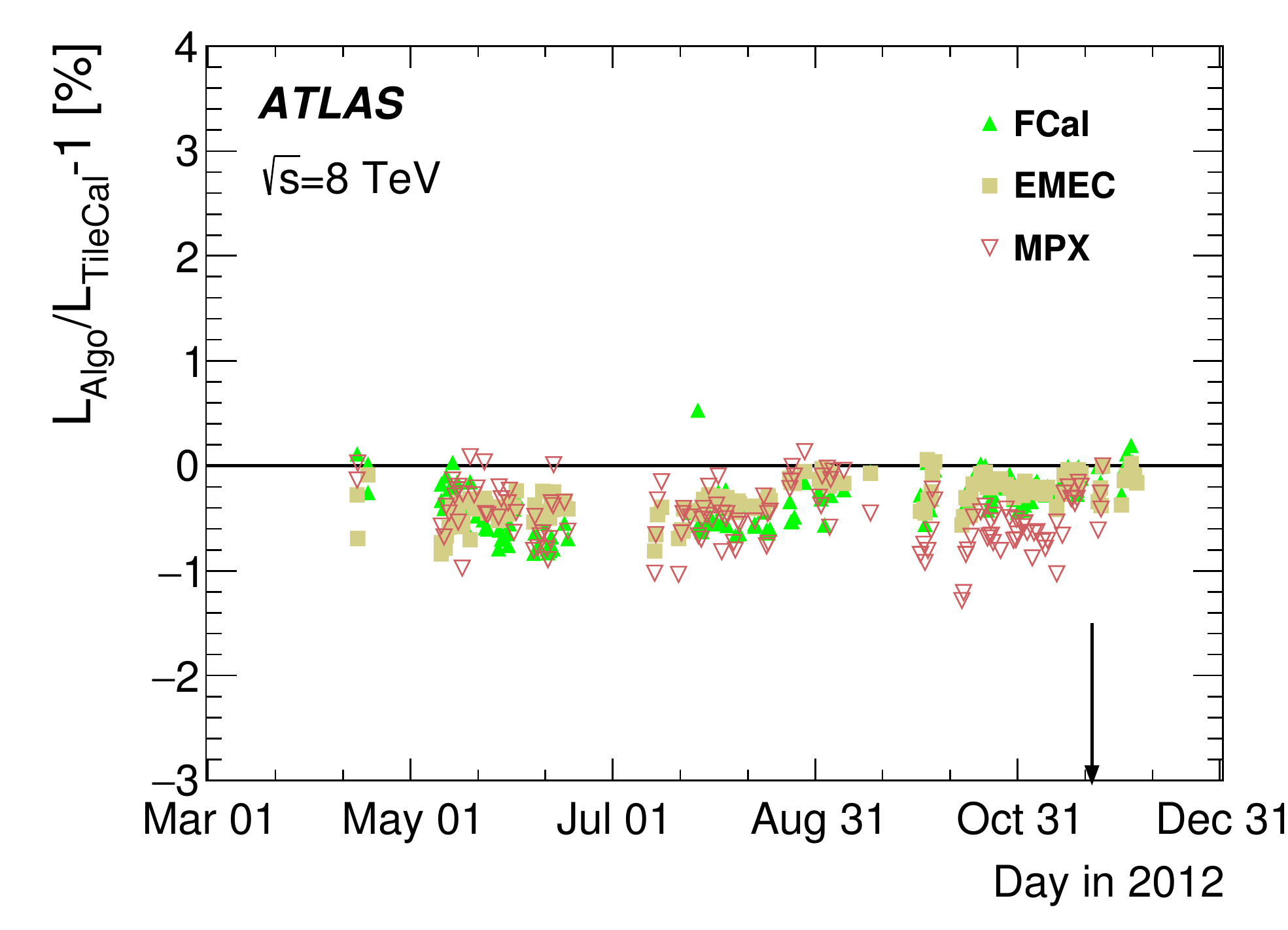}
}
\subfigure[]{
\includegraphics[width=0.485\textwidth]{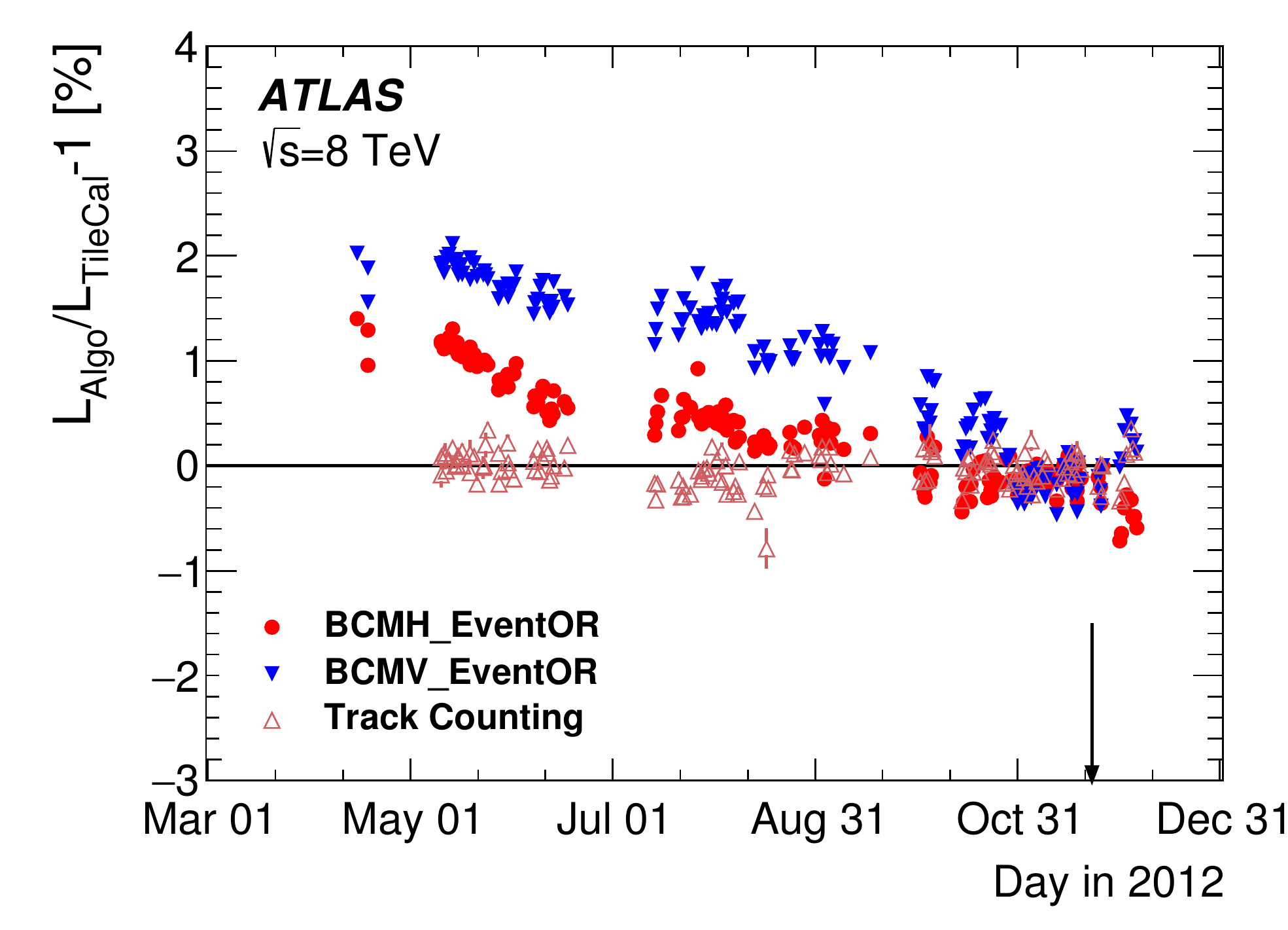}
}
\subfigure[]{
\includegraphics[width=0.485\textwidth]{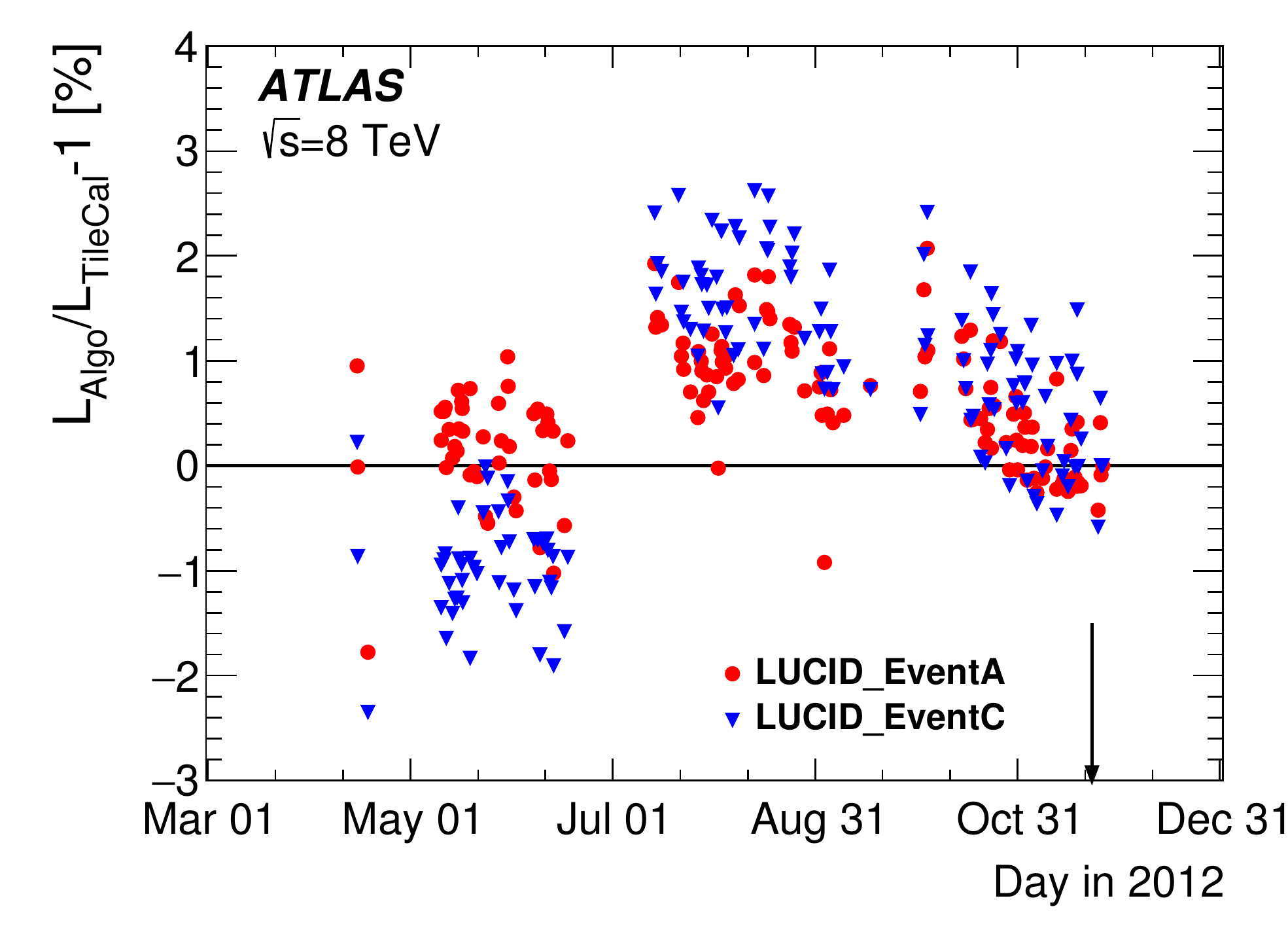}
}
\caption{History of the luminosity per run, compared to the value measured  by TileCal, for (a) bunch-integrating, (b) BCM and track-counting, and (c) LUCID algorithms, during routine physics operation at high luminosity. Each point shows for a single run the mean deviation from a reference run taken on November 25, 2012 (LHC fill 3323). The EMEC, FCal and TileCal values are computed using the average of the luminosities reported by the A and C arms of the corresponding calorimeter; the MPX values reflect the average over the six sensors. 
The step in LUCID response 
is moderate thanks to the use of the April calibration for the LUCID data recorded before July. 
The vertical arrows indicate the time of the November \vdM scan session.
}  
\label{fig:timeStab_uncor}
\end{figure}

Figure~\ref{fig:timeStab_uncor} shows the ratio of the integrated luminosity per ATLAS run as measured by a variety of luminosity algorithms, to that reported by the TileCal. Even though a systematic trend between the LAr and TileCal measurements is apparent, the calorimeter algorithms are consistent to better than $\pm 0.7\%$. The TileCal luminosity is consistent with that from the default track-counting algorithm to within $\pm 0.4$\% or less. 

\par
In contrast, both BCM and LUCID exhibit significant variations in response over the course of 2012, which vary from channel to channel and are attributed to, respectively, radiation-induced lattice defects and PMT aging. Among these, the BCMH\_EventOR algorithm exhibits the least severe deviation from its response at the time of the November \vdM-scan session. Its long-term drift is, however, large enough to warrant a time-dependent response correction that is based on one of the more stable relative-luminosity monitors shown in Fig.~\ref{fig:timeStab_uncor}, and that is described in Sect.~\ref{subsubsec:longTermDrfitCrctn}.
\subsection{\boldmath  $\mu$ dependence}
\label{subsec:mudependence}

As the pile-up response of a given luminosity algorithm is determined by the instrumental characteristics of the luminometer considered, the BCMH\_EventOR and BCMV\_EventOR algorithms are expected to exhibit little $\mu$-dependence with respect to each other, even if both may be affected by a common non-linearity with respect to the actual instantaneous luminosity. The same applies to ratios of luminosity values reported independently by the A and C arms of FCal, EMEC, LUCID and TileCal.

\par
\begin{figure}
\centering
\subfigure[]{
\includegraphics[width=0.77\textwidth]{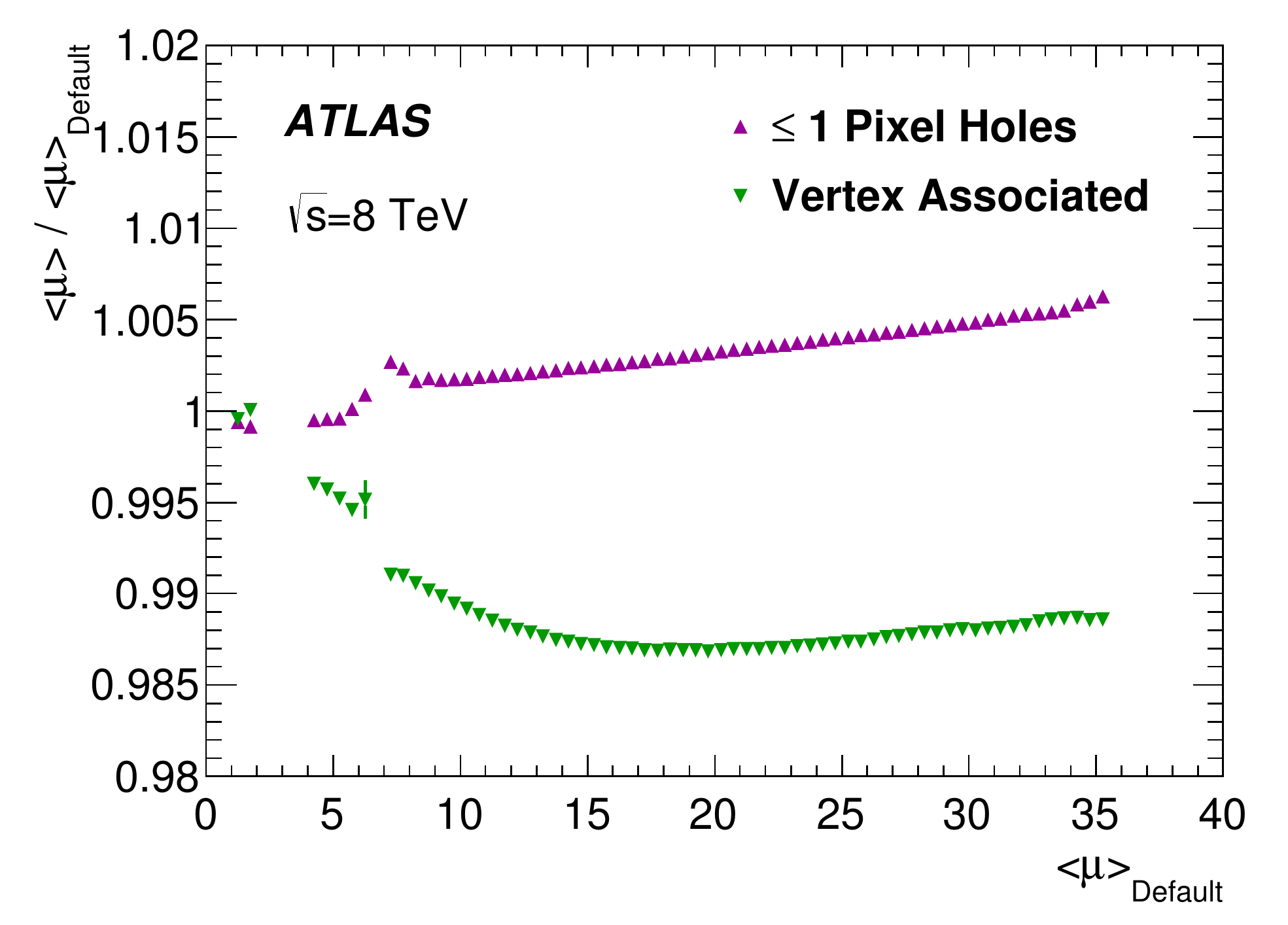}
}
\subfigure[]{
\includegraphics[width=0.77\textwidth]{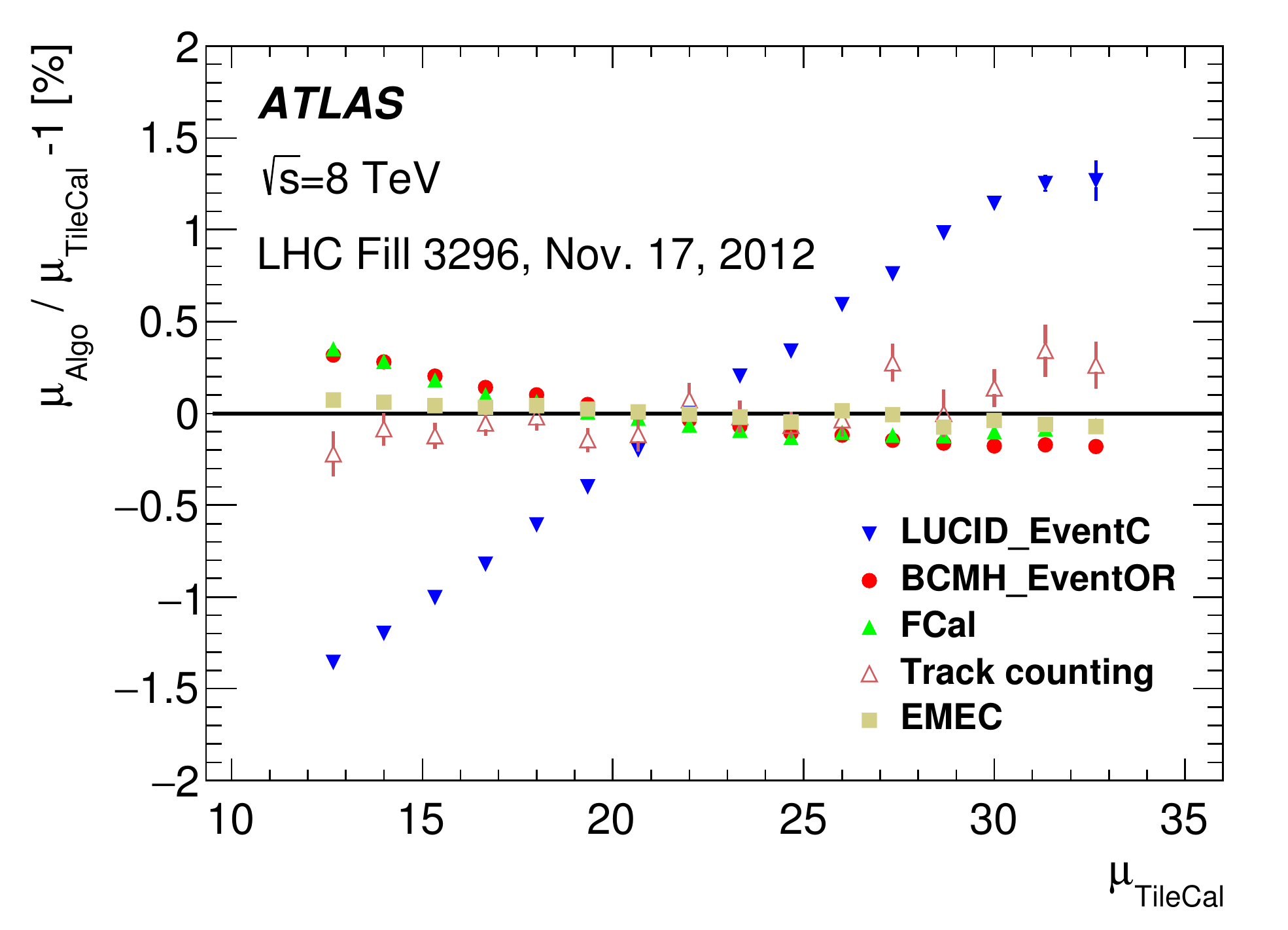}
}
\caption{ (a) Ratio of the bunch-averaged pile-up parameter \muav reported using different track-counting working points, to that from the default WP, as a function of the \muav value obtained using the default WP. 
The data are averaged over all stable-beam runs.
(b) Fractional deviation of the bunch-averaged pile-up parameter \muav, obtained using different algorithms, from the TileCal value, as a function of $\muav_{\mathrm{TileCal}}$, during a physics run selected to cover the widest possible \muav range. The data are normalized such that all algorithms yield the same integrated luminosity in the run considered.
}  
\label{fig:mudep}
\end{figure}

In contrast, the track-counting luminosities obtained using the three track selections defined in Sect.~\ref{subsec:IDalgos} exhibit a noticeable relative non-linearity (Fig.~\ref{fig:mudep}(a)). The pattern is consistent with that observed in Fig.~\ref{fig:trkWPvsTime}. At very low $\mu$, the three working points are fully consistent, as expected from having been \vdM-calibrated  at $ \mu \sim 0.5$. As $\mu$ increases, loosening the pixel-hole requirement on the selected tracks results, after fake-track subtraction, in a residual positive non-linearity of at most 0.7\% in the reported \muav value. In contrast, the vertex-associated track count exhibits, also after fake-track correction, a negative non-linearity with respect to the default WP, which peaks at $-1.3$\% and then decreases in magnitude.
Even though the simulation should account for the pile-up dependence of the fake-track fraction and of the track- and vertex-reconstruction efficiencies, it fails to explain the relative $\mu$-dependence observed in the data between the three track-counting selections. The onset of the discrepancies appears to lie in the range $2<\mu<10$. However, only very limited data, all from a single run with a small number of isolated bunches, are available in that $\mu$ range, so that no firm conclusions can be drawn.  
A conservative approach is therefore adopted: the observed discrepancy between track-counting WPs is used as a data-driven upper limit on a potential bias affecting the absolute track-based luminosity scale in the high-$\mu$ regime. The impact of this systematic uncertainty is discussed in Sect.~\ref{subsubsec:calibTrnsfr}.

\par
In the absence of any absolute linearity reference, potential pile--up-dependent biases in the high-$\mu$ regime can be constrained by the relative $\mu$-dependence of the luminosity values reported by luminometers based on very different technologies (Fig.~\ref{fig:mudep}(b)). 
The relative non-linearity  between the BCMH\_EventOR and the TileCal (the default track-counting) algorithm does not exceed $\pm 0.3$\% ($\pm 0.5$\%)  over the \muav range accessible in this run; the root causes of the relative $\mu$-dependence between these three luminometers remain under investigation. 
An extensive analysis of the more severe LUCID non-linearity indicates that under typical physics operating conditions, the large currents drawn by the LUCID PMTs significantly distort their response.

\par
The run-averaged pile-up parameter changes from one run to the next, because of variations both in the initial luminosity and in the duration of LHC fills. Therefore, the larger the relative $\mu$-dependence between two algorithms, the larger the fill-to-fill fluctuations in the ratio of the run-integrated luminosities reported by these two algorithms. This effect contributes significantly to the point-to-point scatter that is apparent in Fig.~\ref{fig:timeStab_uncor}.

\section{Luminosity determination during physics running}
\label{sec:LDtmPhys}

To determine the integrated luminosity used in ATLAS physics analyses, a single bunch-by-bunch algorithm is selected as the baseline to provide the central value for a certain time range (Sect.~\ref{subsec:prefAlgo}).
The corresponding \vdM-calibrated luminosity values are first background-subtracted (Sect.~\ref{subsec:LBgds}), and then corrected  for rate- and time-dependent biases that impact high-luminosity operation (Sect.~\ref{subsec:highLCrctns}).
The consistency of the various ATLAS luminosity measurements after all corrections is quantified in Sect.~\ref{subsec:finalCstcy}, together with the associated systematic uncertainty.

\subsection{Baseline luminosity algorithm}
\label{subsec:prefAlgo}

The choice of algorithm is determined in part by the reproducibility and long-term stability of its absolute calibration. Figure~\ref{figvdMCalStblty} shows that in this respect, the BCMH\_EventOR and track-counting algorithms perform noticeably better than BCMV\_EventOR and LUCID. Studies of relative stability during physics running (Fig.~\ref{fig:timeStab_uncor}) and of $\mu$ dependence (Fig.~\ref{fig:mudep}(b)) lead to the same conclusion. As track counting is active only during stable-beam operation and is statistically marginal at the luminosity-block level, it is not suitable for use as a baseline algorithm, but it is retained as a reference method to assess systematic biases. The BCMH\_EventOR algorithm supplies the absolute luminosity during most of the 2012 running period; it is supplemented by the LUCID\_EventA algorithm during the few runs where the BCM is not available, and which represent less than 1\% of the 2012 integrated luminosity.
\subsection{Background subtraction}
\label{subsec:LBgds}

During high-luminosity physics running, instrumental noise and single-beam backgrounds become negligible by comparison to the luminosity; only afterglow remains as a significant background. With a 2012 bunch spacing of 50~ns and typically over 1000 colliding bunches, it reaches a fairly stable equilibrium after the first few bunches in a train. It is observed to scale with the instantaneous luminosity and typically amounts to 0.2--0.5\% of the luminosity signal.

\par
The  bunch-by-bunch noise- and afterglow-subtraction procedure described in Sect.~\ref{subsec:scanBgds} is applied to all BCM and LUCID luminosity determinations. Since the afterglow level in the BCID immediately following a colliding-bunch slot may differ from that in the second BCID after this slot (i.e. in the next colliding-bunch slot), BCIDs at the end of a bunch train were used to evaluate a possible bias in the method. This study suggests that the subtraction over-corrects the BCMH\_EventOR luminosity by approximately 0.2\%. A systematic uncertainty of $\pm 0.2\%$ is therefore assigned to the afterglow correction.

\subsection{Corrections to the absolute calibration in the high-luminosity regime}
\label{subsec:highLCrctns}

Extrapolating the curves of Fig.~\ref{fig:mudep}(b) to very low \muav suggests that for some algorithms, the \vdM-based luminosity scale may not be directly applicable in the pile-up regime typical of physics operation. Percent-level corrections are indeed required (Sect.~\ref{subsubsec:calibTrnsfr}) to transfer, at one point in time, the absolute calibration of BCM and LUCID from the low-luminosity regime of \vdM scans  ($\mu \sim 0.5$, $\Lum \sim 2 \times 10^{30} \cms$) to that of routine physics operation ($\mu \sim$ 20--25, $\Lum > 10^{33} \cms$). In addition, a time-dependent correction (Sect.~\ref{subsubsec:longTermDrfitCrctn}) must be applied to the luminosity of the baseline algorithm to compensate for the long-term drifts apparent in Fig.~\ref{fig:timeStab_uncor}. 

\subsubsection{Calibration transfer from the \vdM regime to physics conditions}
\label{subsubsec:calibTrnsfr}

\begin{figure}
\centering
\subfigure[]{
\includegraphics[width=0.77\textwidth]{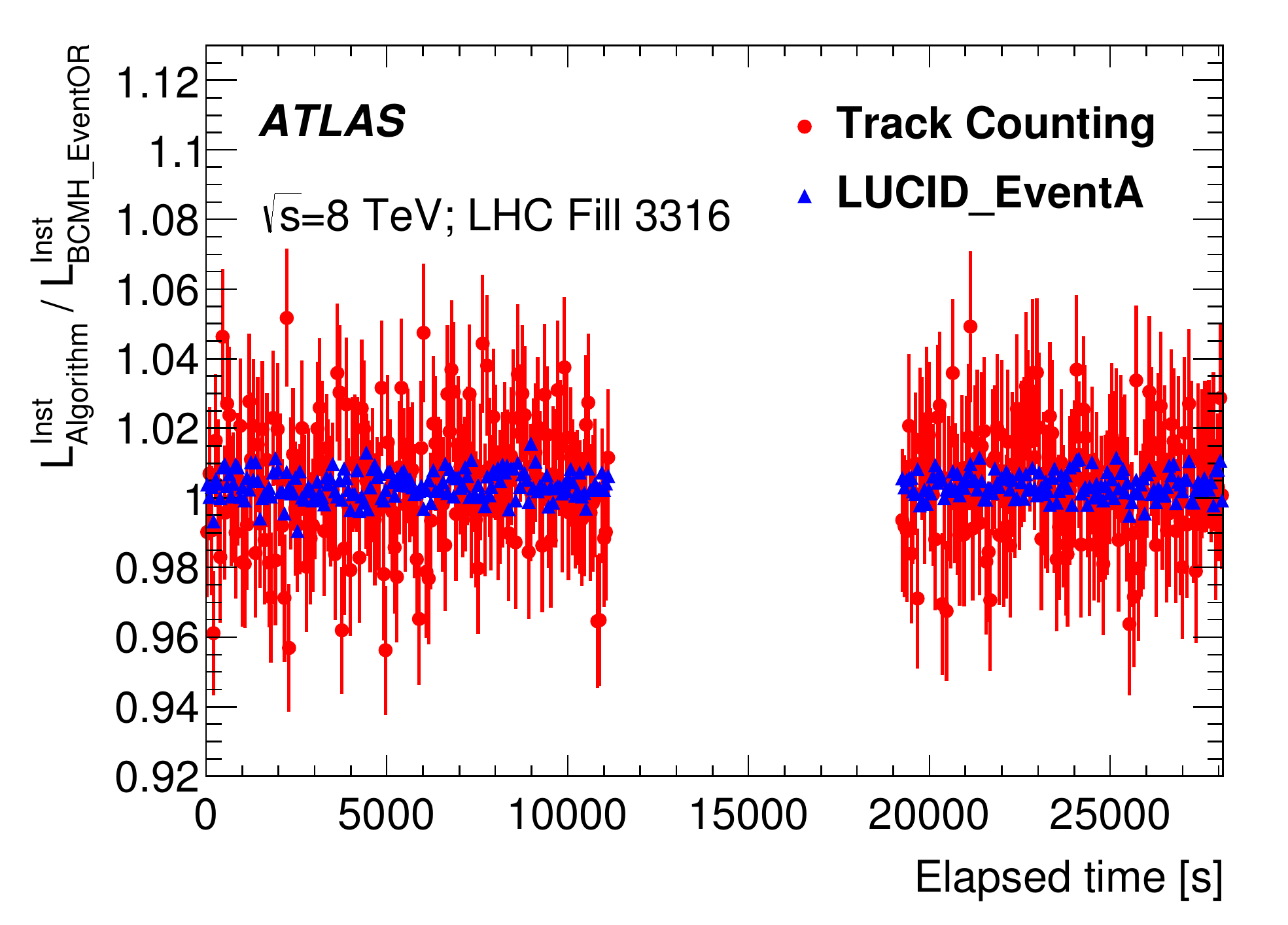}
}
\subfigure[]{
\includegraphics[width=0.77\textwidth]{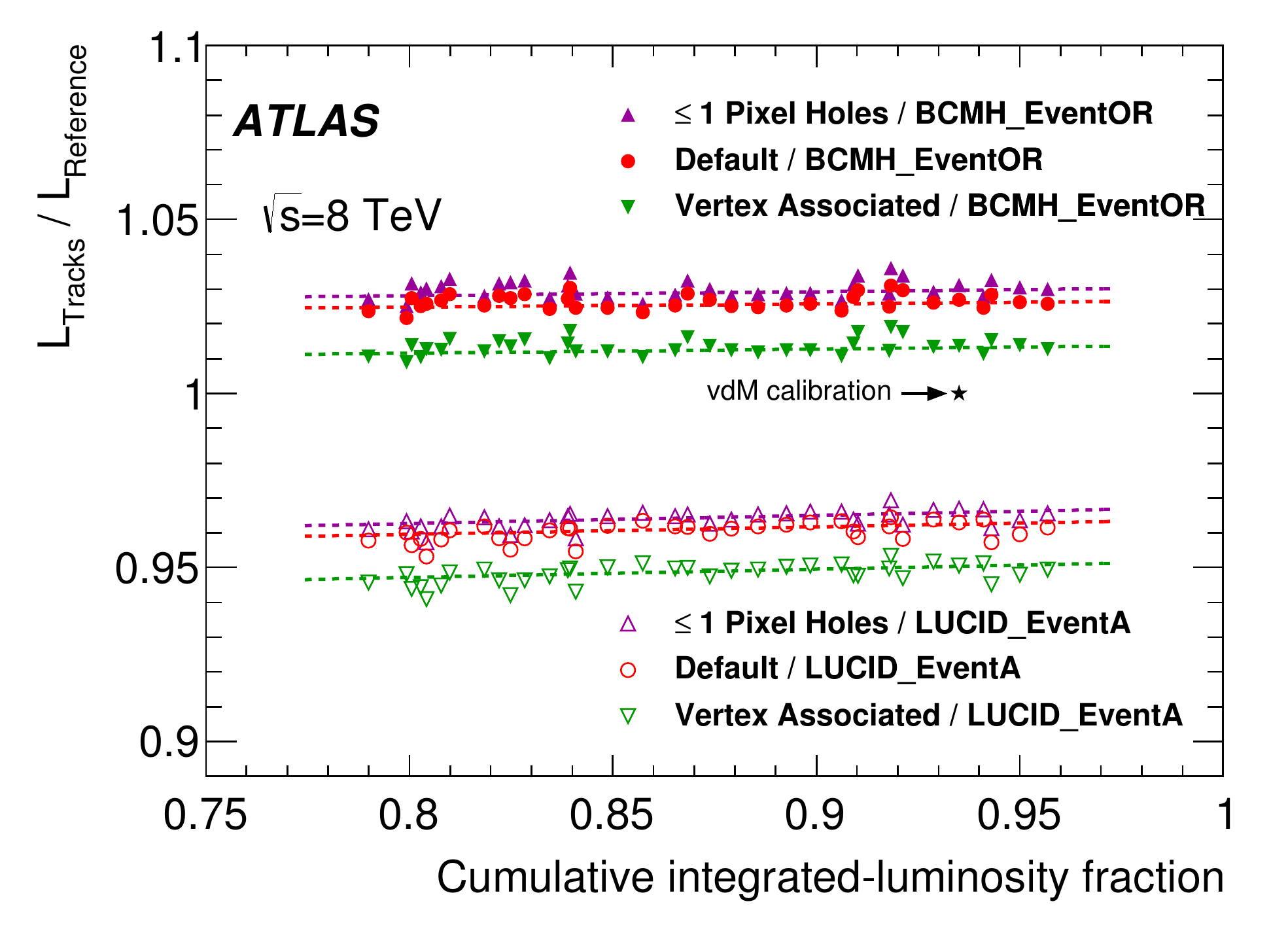}
}
\caption{ (a) History of the ratio of the instantaneous luminosity reported by the default track-counting and LUCID\_EventA algorithms to that from the BCMH\_EventOR algorithm under \vdM-scan conditions, during LHC fill 3316. The gap corresponds to scan set XV. The error bars are statistical.
(b) Evolution of the ratio of the integrated luminosity per run reported by the three track-counting algorithms to that from the BCMH\_EventOR and LUCID\_EventA algorithms, in the few weeks in late 2012 during which the BCM response is approximately constant, as a function of the cumulative delivered luminosity (normalized to the 2012 total). Each point shows the ratio for a single high-luminosity run. The dashed lines are straight-line fits to the data. The reference run (LHC fill 3323) took place the day following the November \vdM-scan session, which is indicated by the star.
}  
\label{fig:calibTrnsfr}
\end{figure}

The history of the instantaneous-luminosity values reported during part of the November \vdM-scan session by the track-counting and LUCID\_EventA algorithms, relative to the BCMH\_EventOR algorithm and using the calibrations listed in Table~\ref{tab:sigvisresult}, is presented in Fig.~\ref{fig:calibTrnsfr}(a). The ratio of the default track-counting (LUCID) luminosity integrated over several hours immediately before and after scan set XV, to that from the BCMH\_EventOR algorithm, is consistent with unity within 0.5\% (0.4\%). The run-integrated luminosity values associated, in that same fill, with the other two track selections (not shown) are consistent with the default track selection within less than one per mille.

\par
However, at high luminosity these ratios differ from unity by several percent (Fig.~\ref{fig:calibTrnsfr}(b)), with all BCM (LUCID) algorithms reporting a lower (higher) luminosity compared to the track-counting method. In addition, the vertex-associated track selection is no longer consistent with the other two, as discussed in Sect.~\ref{sec:stability}.

\par
To provide consistent luminosity measurements, all algorithms must be corrected to some common absolute scale in the high-luminosity regime. As calorimeter-based luminometers lack sensitivity in the \vdM-scan regime, only track counting remains to quantify the relative shifts in response of the BCM and LUCID algorithms between the \vdM-scan and high-luminosity regimes. 
First, the run-to-run fluctuations in Fig.~\ref{fig:calibTrnsfr}(b) are smoothed by parameterizing the luminosity ratios as a linear function of the cumulative integrated-luminosity fraction, used here as a proxy for calendar time. 
Then, for each BCM algorithm and for a given track selection, the difference between the fitted ratio in the high-luminosity reference fill where the calibration transfer is performed (LHC fill 3323), and the corresponding run-integrated luminosity ratio under \vdM conditions (LHC fill 3316), quantifies the shift in the BCM luminosity scale with respect to track counting. The same procedure is applied to LUCID.

\par
The results are summarized in Table~\ref{tab:calibTrnsfr} for the default track selection.
The BCMH\_EventOR efficiency drops by 2.5\% with respect to track counting. Naively extrapolating the relative $\mu$-dependence of these two algorithms from the high-$\mu$ regime (Fig.~\ref{fig:mudep}(b)) to $\mu \sim 0.5$  predicts a shift of 1.3\%, about half of the effect observed.\footnote{Since the mechanisms driving the $\mu$-dependence are neither well characterized nor understood, and in the absence of sufficient data linking the $\mu$ range in routine physics operation (Fig.~\ref{fig:mudep}(b)) to that in the \vdM-scan regime ($\mu \sim 0.5$), such an extrapolation is indicative only: it cannot be relied upon for a quantitative evaluation of the calibration-transfer correction.}
Similarly, the $\mu$-dependence of LUCID\_EventC predicts a 3\% increase in response when going from the \vdM-scan regime to the high-luminosity regime, while the measured step amounts to +3.9\%. These observations suggest that while the measured relative $\mu$-dependence of the three algorithms is consistent with the signs of the calibration shifts and appears to account for a large fraction of their magnitude, other effects also play a role. For instance, studies of the CMS diamond sensors~\cite{bib:CMS_BCM} suggest that the response of the BCM may depend on the total instantaneous collision rate (\ie on the product of \muav and the total number of colliding bunches) through a polarization mechanism associated with radiation-induced lattice defects.

\begin{table}[htbp]
   	\centering
   	\begin{tabular}{p{6 cm}cc} 
		\hline
      		Luminosity algorithm 				& 				Calibration shift w.r.t. track counting [\%]  		\\
		\hline	
		BCMH\_eventOR					& 					$- 2.5 \pm 0.1$		\\
		BCMV\_eventOR					& 					$- 2.9 \pm 0.1$		\\
		LUCID\_eventA					& 					$ + 3.5 \pm 0.1$		\\
		LUCID\_eventC					& 					$ + 3.9 \pm 0.1$		\\
 		\hline
  	 \end{tabular}
	 \caption{Measured fractional shift in luminosity scale between the \vdM-scan regime (LHC fill 3316) and a nearby high-luminosity ATLAS run (LHC fill 3323), using the default track-counting algorithm as the reference. The errors shown are statistical only; they are dominated by track-counting statistics in the \vdM-scan fill, and are therefore fully correlated across the four ratios.}
   	\label{tab:calibTrnsfr}
\end{table}

\par
The track-counting results lie between BCM and LUCID, and using the track scale as a proxy for the true scale is consistent to within 0.5\% with taking the average scale from all the algorithms listed in Table~\ref{tab:calibTrnsfr}. The choice of which track selection to use as reference is somewhat arbitrary. The default working point appears as the natural choice given that it exhibits the smallest relative $\mu$-dependence with respect to TileCal, suffers from the smallest uncertainty arising from the simulation-based fake-track subtraction, and lies between the extremes of the three track selections.

\par 
The systematic uncertainty in the calibration-transfer corrections of Table~\ref{tab:calibTrnsfr} is estimated to be $\pm 1.4$\%. It is dominated by the 1.3\% inconsistency (Figs.~\ref{fig:mudep}(a) and \ref{fig:calibTrnsfr}(b)) between the default and the vertex-associated track selections. Additional contributions arise from the small inconsistency between the BCM-based and track-based luminosity measurements during the \vdM-scan fill (0.5\%), from a small deadtime correction that affects the \vdM-scan track-counting data only (0.2\%), and from the track-counting statistics during the \vdM-scan fill (0.1\%). The slight integrated-luminosity (or time) dependence of the BCM to track-counting luminosity ratio visible in Fig.~\ref{fig:calibTrnsfr}(b) is accounted for as part of the long-term drift correction, discussed next.

\subsubsection{Long-term drift correction}
\label{subsubsec:longTermDrfitCrctn}

\begin{figure}
\centering
\subfigure[]{
\includegraphics[width=0.77\textwidth]{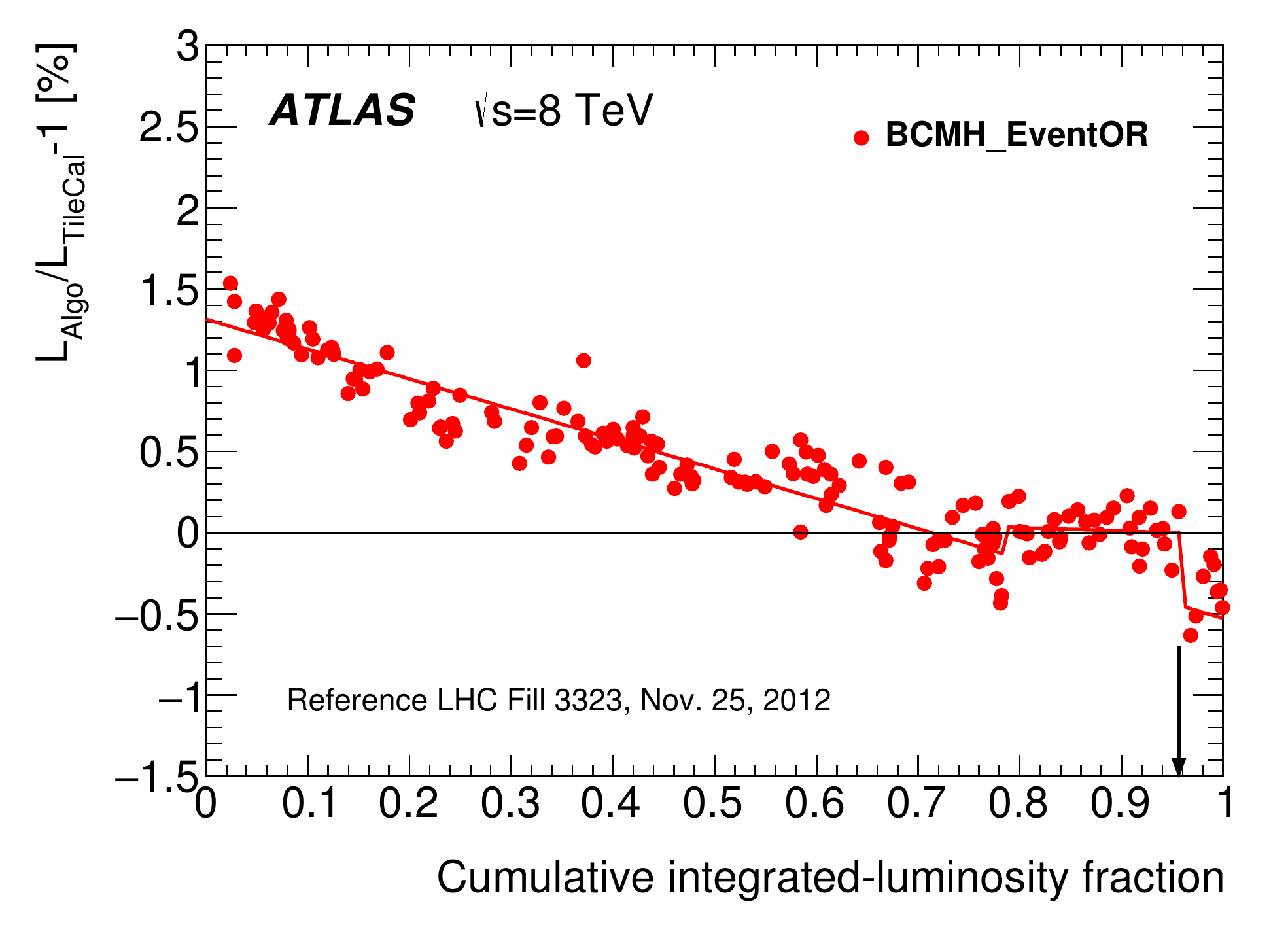}
}
\subfigure[]{
\includegraphics[width=0.77\textwidth]{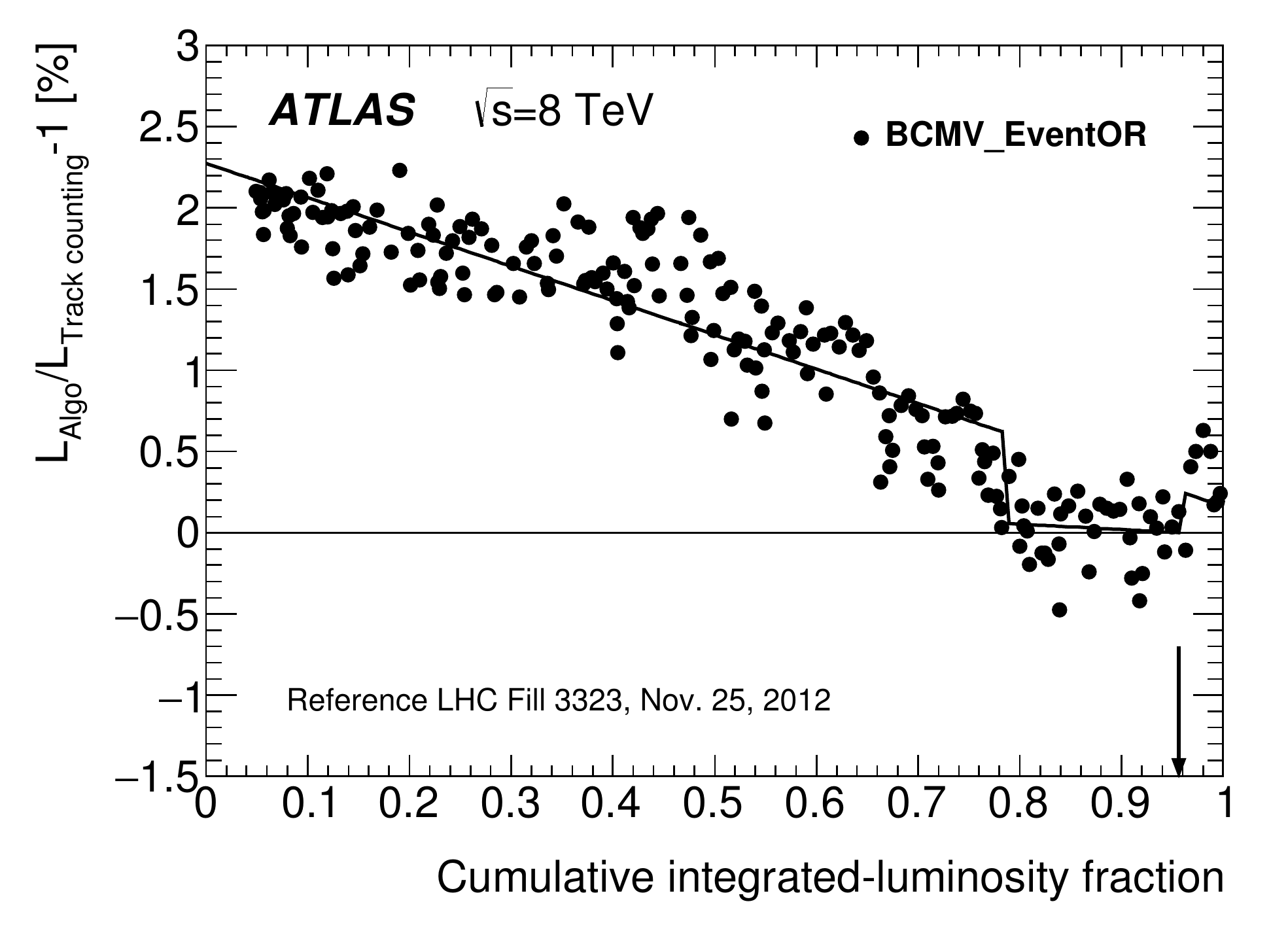}
}
\caption{ History of the fractional difference in integrated luminosity per run  (a) between the BCMH\_EventOR and the TileCal algorithm, and (b) between the BCMV\_EventOR and the default track-counting algorithm. Each point shows the mean difference for a single run compared to that in the reference run (LHC fill 3323) in which the calibration transfer is performed. The lines represent the fit discussed in the text. The vertical arrow indicates the time of the reference run.
}  
\label{fig:timeStabFits}
\end{figure}

The second step in transferring the \vdM-based calibrations to an arbitrary high-luminosity physics run consists in correcting for the long-term drifts apparent in Fig.~\ref{fig:timeStab_uncor}, using one of the more stable monitors (EMEC, FCal, TileCal or track counting) as a reference. The absolute luminosity scale of the selected reference monitor is first anchored to that of BCM (or LUCID) in the high-luminosity reference run where the calibration transfer is performed (LHC fill 3323). The run-by-run luminosity ratio of the considered bunch-by-bunch algorithm to the chosen reference is then parameterized as a function of the cumulative integrated-luminosity fraction. This choice of variable, instead of calendar time, is inspired by (but not dependent upon) the assumption that detector aging increases smoothly with integrated radiation dose; it also simplifies the analysis by eliminating the gaps between running periods (Fig.~\ref{fig:timeStab_uncor}).
A two-segment, piece-wise linear fit is used to smooth the run-to-run fluctuations, with one segment covering the entire year except for the BCM noise period, and the second, shorter segment accounting for the gain shift during that same noise period (Fig.~\ref{fig:timeStabFits}). This empirical parameterization  yields a satisfactory description of the entire data set. It provides a run-by-run correction to the instantaneous luminosity reported by each BCM or LUCID algorithm: a positive (negative) value of the fit function in a given ATLAS run results in a downwards (upwards) luminosity adjustment for every luminosity block in that run. This implies that the absolute luminosity scale in each LHC fill is effectively carried by the reference monitor, while the time- and BCID-dependence of the luminosity during that same fill continues to be provided by the bunch-by-bunch algorithm considered.

\par
The net impact of this procedure on the integrated luminosity for the entire 2012 running period is documented in Table~\ref{tab:driftCrctn}. The TileCal- and track--counting-based corrections are effectively indistinguishable; the former is chosen for the central value because of the slightly smaller run-to-run scatter of the BCM/TileCal luminosity ratio.  The largest difference between reference monitors amounts to 0.3\%, and reflects the relative slope between the FCal and TileCal algorithms in Fig.~\ref{fig:timeStab_uncor}(a). This value is taken as the systematic uncertainty in the long-term drift correction.
\begin{table}
   \centering
   \begin{tabular}{c|cccc}

      	\hline
      	Reference algorithm 	&   	 \multicolumn{4}{c}{        Fractional change in integrated luminosity [\%]}   				\\
      	\hline
	                                	& BCMH\_EventOR		& BCMV\_EventOR	& LUCID\_EventA		& LUCID\_EventC 	\\
	EMEC				& --0.59			  	& --1.26			& --0.70				& --0.49			\\	
	FCal					& --0.70				& --1.36			& --0.68				& --0.52			\\	
	TileCal				& --0.44			  	& --1.09			& --0.54				& --0.26			\\	
	Track Counting			& --0.45			  	& --1.12			& --0.57				& --0.34			\\		  	
	\hline
   \end{tabular}
      \caption{Impact of the long-term drift correction on the 2012 integrated luminosity.}
   \label{tab:driftCrctn}
\end{table}

\subsection{Consistency of ATLAS luminosity measurements after all corrections}
\label{subsec:finalCstcy}

A global check of the consistency of the corrections described in Sects.~\ref{subsec:LBgds} and \ref{subsec:highLCrctns} is provided by the comparison of the 2012 integrated-luminosity values reported by different bunch-by-bunch algorithms. For high-luminosity runs ($\beta^\star = 0.6$\,m and at least 1050 colliding bunches) under stable-beam conditions, after background subtraction, calibration transfer and long-term drift correction of the BCM and LUCID data, the integrated luminosity reported by BCMV\_EventOR  agrees with that from the BCMH\_EventOR baseline within 0.01\%. For the subset of such runs where both LUCID and BCM deliver valid luminosity data, which corresponds to about 91\% of the 2012 integrated luminosity, both single-arm LUCID algorithms agree with the BCMH\_EventOR baseline within 0.5\%. It should be stressed, however, that these BCM- and LUCID-based luminosity determinations are correlated, because they were all drift-corrected to the same reference.

\par
\begin{figure}
\centering
\subfigure[]{
\includegraphics[width=0.77\textwidth]{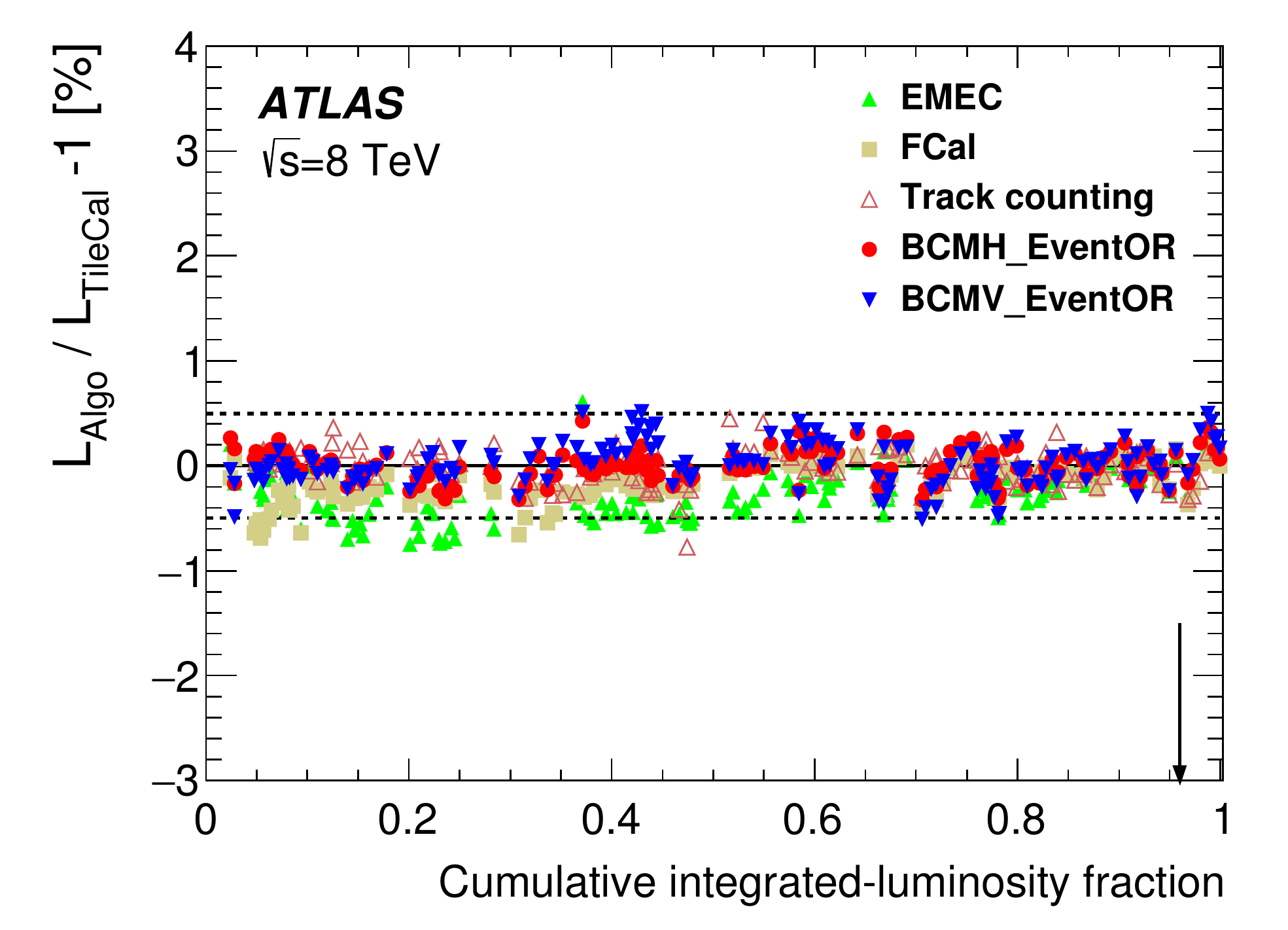}
}
\subfigure[]{
\includegraphics[width=0.77\textwidth]{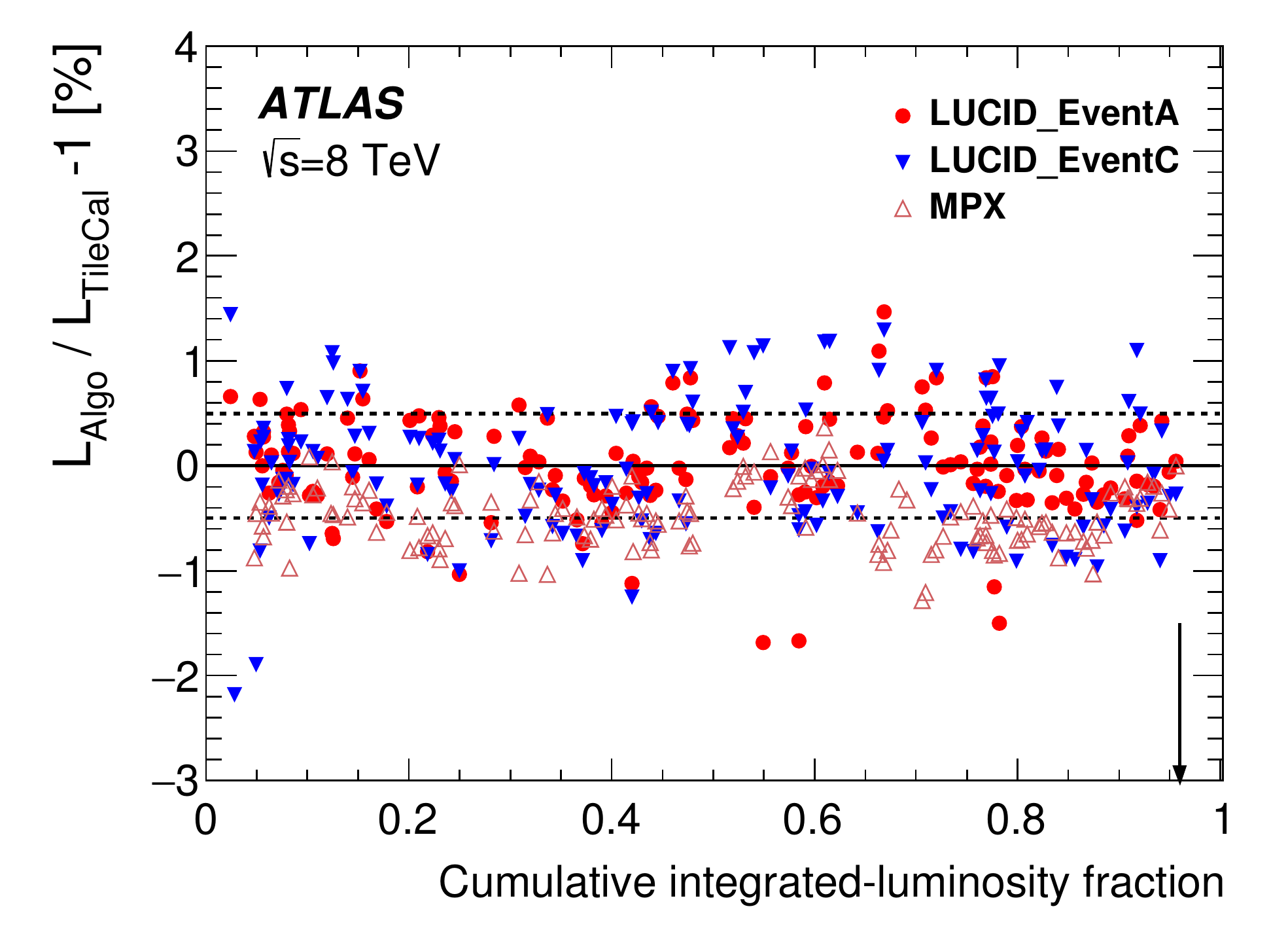}
}
\caption{ History of the fractional difference in run-integrated luminosity between the TileCal algorithm and the drift-corrected (a) BCM and (b) LUCID and MPX algorithms.  The results of the other possible reference monitors (EMEC, FCal and track counting) are taken from Fig.~\ref{fig:timeStab_uncor} and included here for comparison. Each point shows the mean difference for a single run compared to that in the reference fill indicated by the arrow. The dashed horizontal lines delimit a $\pm 0.5$\% window around zero.
}  
\label{fig:timeStabCrctd}
\end{figure}

The internal consistency of the absolute luminosity measurements at $\sqrt{s} = 8$\,TeV in the high-luminosity regime is illustrated in Fig.~\ref{fig:timeStabCrctd}. The run-to-run fluctuations reflect the combined impact of the relative $\mu$-dependence of the various algorithms, of imperfectly corrected medium-term drifts and of other sources of non-reproducibility. With the exception of some of the LUCID data, they remain within a $\pm 0.5$\% band, which provides a measure of the systematic uncertainty associated with the run-to-run consistency of independent luminosity measurements.

\section{Total luminosity uncertainty for the 2012 \pp\ run}
\label{sec:totalSyst}

Table~\ref{tab:totLsyst} regroups the contributions to the total uncertainty in the luminosity values provided for physics analyses. The \vdM-calibration uncertainties are detailed in Tables~\ref{tab:sigvisresult} and \ref{tab:vdMsysts}. The afterglow subtraction, the calibration transfer from the \vdM-scan to the high-luminosity regime and the long-term drift correction applied to the bunch-by-bunch luminometers are described in Sects.~\ref{subsec:LBgds}, \ref{subsubsec:calibTrnsfr} and \ref{subsubsec:longTermDrfitCrctn} respectively. The run-to-run consistency of the ATLAS luminosity measurements is assessed in Sect.~\ref{subsec:finalCstcy}. The resulting total uncertainty amounts to $\pm 1.9$\%.

\begin{table}
   \centering
   \begin{tabular}{cc}
      	\hline
      	Uncertainty source 		& $\delta {\cal L}/{\cal L}$ [\%]	\\
      	\hline
	van der Meer calibration	& 1.2						\\	
	Afterglow subtraction	& 0.2						\\
	Calibration transfer from \vdM-scan to high-luminosity regime	
						& 1.4						\\	
	Long-term drift correction	& 0.3						\\
	Run-to-run consistency	& 0.5						\\
	\hline
	Total					& 1.9						\\
	\hline
   \end{tabular}
   \caption{Relative uncertainty in the calibrated luminosity scale, broken down by source.}
   \label{tab:totLsyst}
\end{table}

\section{Summary}
\label{sec:conclusions}

The ATLAS luminosity scale for the 2012 LHC run has been calibrated using data from dedicated beam-separation scans, also known as van der Meer scans. The \vdM-calibration uncertainty is smaller than for the 2011 data set~\cite{Aad:2013ucp}, thanks to improved control of beam-dynamical effects (beam--beam deflections, dynamic $\beta$, non-factorization) and to a refined analysis of the non-reproducibility of beam conditions (orbit drift, emittance growth). 
The total systematic uncertainty in the delivered luminosity is no longer dominated by \vdM-calibration uncertainties. The largest contribution arises from instrumental effects that require the transfer of the absolute luminosity scale from the low-rate  \vdM-scan regime to the high-luminosity conditions of routine physics operation; residual run-to-run and long-term inconsistencies between independent luminosity measurements also contribute significantly.

\par
The combination of these systematic uncertainties results in a final uncertainty of $\delta {\cal L}/ {\cal L} = \pm 1.9\%$ in the luminosity measured by ATLAS during $pp$ collisions at $\sqrt{s} = 8$~\TeV\   for the 22.7\,$ \mathrm{fb}^{-1}$ of data delivered to ATLAS in 2012. This uncertainty applies to the high-luminosity data sample and any subset thereof, but not necessarily to a few special runs taken under very low pile-up conditions, such as those dedicated to elastic-scattering measurements: the latter require a separate analysis tailored to their specific experimental conditions.

\section*{Acknowledgements}

We thank CERN for the very successful operation of the LHC, as well as the
support staff from our institutions without whom ATLAS could not be
operated efficiently.

We acknowledge the support of ANPCyT, Argentina; YerPhI, Armenia; ARC, Australia; BMWFW and FWF, Austria; ANAS, Azerbaijan; SSTC, Belarus; CNPq and FAPESP, Brazil; NSERC, NRC and CFI, Canada; CERN; CONICYT, Chile; CAS, MOST and NSFC, China; COLCIENCIAS, Colombia; MSMT CR, MPO CR and VSC CR, Czech Republic; DNRF and DNSRC, Denmark; IN2P3-CNRS, CEA-DSM/IRFU, France; GNSF, Georgia; BMBF, HGF, and MPG, Germany; GSRT, Greece; RGC, Hong Kong SAR, China; ISF, I-CORE and Benoziyo Center, Israel; INFN, Italy; MEXT and JSPS, Japan; CNRST, Morocco; FOM and NWO, Netherlands; RCN, Norway; MNiSW and NCN, Poland; FCT, Portugal; MNE/IFA, Romania; MES of Russia and NRC KI, Russian Federation; JINR; MESTD, Serbia; MSSR, Slovakia; ARRS and MIZ\v{S}, Slovenia; DST/NRF, South Africa; MINECO, Spain; SRC and Wallenberg Foundation, Sweden; SERI, SNSF and Cantons of Bern and Geneva, Switzerland; MOST, Taiwan; TAEK, Turkey; STFC, United Kingdom; DOE and NSF, United States of America. In addition, individual groups and members have received support from BCKDF, the Canada Council, CANARIE, CRC, Compute Canada, FQRNT, and the Ontario Innovation Trust, Canada; EPLANET, ERC, FP7, Horizon 2020 and Marie Sk{\l}odowska-Curie Actions, European Union; Investissements d'Avenir Labex and Idex, ANR, R{\'e}gion Auvergne and Fondation Partager le Savoir, France; DFG and AvH Foundation, Germany; Herakleitos, Thales and Aristeia programmes co-financed by EU-ESF and the Greek NSRF; BSF, GIF and Minerva, Israel; BRF, Norway; Generalitat de Catalunya, Generalitat Valenciana, Spain; the Royal Society and Leverhulme Trust, United Kingdom.

The crucial computing support from all WLCG partners is acknowledged gratefully, in particular from CERN, the ATLAS Tier-1 facilities at TRIUMF (Canada), NDGF (Denmark, Norway, Sweden), CC-IN2P3 (France), KIT/GridKA (Germany), INFN-CNAF (Italy), NL-T1 (Netherlands), PIC (Spain), ASGC (Taiwan), RAL (UK) and BNL (USA), the Tier-2 facilities worldwide and large non-WLCG resource providers. Major contributors of computing resources are listed in Ref.~\cite{ATL-GEN-PUB-2016-002}.

%

\printbibliography

\newpage 
\begin{flushleft}
{\Large The ATLAS Collaboration}

\bigskip

M.~Aaboud$^{\rm 136d}$,
G.~Aad$^{\rm 87}$,
B.~Abbott$^{\rm 114}$,
J.~Abdallah$^{\rm 65}$,
O.~Abdinov$^{\rm 12}$,
B.~Abeloos$^{\rm 118}$,
R.~Aben$^{\rm 108}$,
O.S.~AbouZeid$^{\rm 138}$,
N.L.~Abraham$^{\rm 150}$,
H.~Abramowicz$^{\rm 154}$
% ...
\bigskip
\\
% ...
$^{12}$ Institute of Physics, Azerbaijan Academy of Sciences, Baku, Azerbaijan\\
% ...
$^{65}$ University of Iowa, Iowa City IA, United States of America\\
% ...
$^{87}$ CPPM, Aix-Marseille Universit{\'e} and CNRS/IN2P3, Marseille, France\\
% ...
$^{108}$ Nikhef National Institute for Subatomic Physics and University of Amsterdam, Amsterdam, Netherlands\\
% ...
$^{114}$ Homer L. Dodge Department of Physics and Astronomy, University of Oklahoma, Norman OK, United States of America\\
% ...
$^{118}$ LAL, Univ. Paris-Sud, CNRS/IN2P3, Universit{\'e} Paris-Saclay, Orsay, France\\
% ...
$^{136}$ $^{(a)}$ Facult{\'e} des Sciences Ain Chock, R{\'e}seau Universitaire de Physique des Hautes Energies - Universit{\'e} Hassan II, Casablanca; $^{(b)}$ Centre National de l'Energie des Sciences Techniques Nucleaires, Rabat; $^{(c)}$ Facult{\'e} des Sciences Semlalia, Universit{\'e} Cadi Ayyad, LPHEA-Marrakech; $^{(d)}$ Facult{\'e} des Sciences, Universit{\'e} Mohamed Premier and LPTPM, Oujda; $^{(e)}$ Facult{\'e} des sciences, Universit{\'e} Mohammed V, Rabat, Morocco\\
% ...
$^{138}$ Santa Cruz Institute for Particle Physics, University of California Santa Cruz, Santa Cruz CA, United States of America\\
% ...
$^{150}$ Department of Physics and Astronomy, University of Sussex, Brighton, United Kingdom\\
% ...
$^{154}$ Raymond and Beverly Sackler School of Physics and Astronomy, Tel Aviv University, Tel Aviv, Israel\\
% ...
$^{a}$ Also at Department of Physics, King's College London, London, United Kingdom\\
$^{b}$ Also at Institute of Physics, Azerbaijan Academy of Sciences, Baku, Azerbaijan\\
$^{c}$ Also at Novosibirsk State University, Novosibirsk, Russia\\
$^{d}$ Also at TRIUMF, Vancouver BC, Canada\\
$^{e}$ Also at Department of Physics {\&} Astronomy, University of Louisville, Louisville, KY, United States of America\\
% ...
\end{flushleft}



\end{document}